\documentclass{aa}

\usepackage{natbib}
\usepackage{graphicx}
\bibpunct{(}{)}{;}{a}{}{,}

\newcommand{\mic}{\hbox{${\mu}$m}}
\newcommand{\Msun}{$M_{\sun}$}
\newcommand{\Rsun}{$R_{\sun}$}
\newcommand{\Lsun}{$L_{\sun}$}

\newcommand{\Mstar}{$M_{\star}$}

\newcommand{\Lstar}{$L_{\star}$}
\newcommand{\Tstar}{$T_{\star}$}
\newcommand{\Mwd}{$M_{\rm WD}$}

\newcommand{\Lwd}{$L_{\rm WD}$}
\newcommand{\Twd}{$T_{\rm WD}$}
\newcommand{\dustgas}{$\rho_{\rm d}/\rho$}

\newcommand{\Ks}{\mbox{$K_{\rm S}$}}
\newcommand{\Lc}{\mbox{$L_{\rm c}$}}
\newcommand{\Lpah}{\mbox{$L_{\rm PAH}$}}
\newcommand{\HmK}{\mbox{$H{-}K$}}
\newcommand{\KmL}{\mbox{$K_{\rm S}{-}L_{\rm c}$}}
\newcommand{\HmL}{\mbox{$H{-}L_{\rm c}$}}
\newcommand{\irc}{\object{IRC $\!$+10\,216}}
\newcommand{\hlt}{\object{HL Tau}}
\newcommand{\irs}{\object{L1551 IRS\,5}}
\newcommand{\rr}{\object{Red Rectangle}}
\newcommand{\hd}{\object{HD\,44179}}

\defcitealias{Men'shchikovHenning1997}{Men'shchikov \& Henning 1997}
\defcitealias{Jura_etal1997}{Jura et al. 1997}
\defcitealias{Osterbart_etal1997}{Pa\-per I}
\defcitealias{Men'shchikov_etal1998}{Pa\-per II}
\defcitealias{Tuthill_etal2002}{Pa\-per III}
\defcitealias{Cohen_etal1975}{Cohen et al. 1975}
\defcitealias{Perkins_etal1981}{Perkins et al. 1981}
\defcitealias{VanWinckel2001}{Van Winckel 2001}
\defcitealias{Waters_etal1998}{Waters et al. 1998}

\begin{document}

\title
{
Properties of the close binary and circumbinary torus\\
of the {\rr}
}

\author
{
A.\,B. Men'shchikov\inst{1}\and
D. Schertl\inst{1}\and
P.\,G. Tuthill\inst{2}\and
G. Weigelt\inst{1}\and
L.\,R. Yungelson\inst{3}
}

\institute
{
Max-Planck-Institut f{\"u}r Radioastronomie, Auf dem H{\"u}gel 69, D--53121 Bonn, Germany\\
e-mail: sasha, ds, weigelt (@mpifr-bonn.mpg.de)
\and
Astronomy Department, School of Physics, University of Sydney, NSW 2006, Australia\\
e-mail: gekko@physics.usyd.edu.au
\and
Institute of Astronomy, Russian Academy of Sciences, Pyatnitskaya 48, Moscow, Russia\\
e-mail: lry@inasan.rssi.ru
}

\offprints{A.\,B.\,Men'shchikov}

\date{Received / Accepted}

\titlerunning{Close binary and circumbinary torus of the {\rr}}

\authorrunning{Men'shchikov et al.}


\abstract
{
New diffraction-limited speckle images of the {\rr} in the wavelength range
2.1--3.3\,{\mic} with angular resolutions of 44--68 mas
\citep{Tuthill_etal2002} and previous speckle images at 0.7--2.2\,{\mic}
\citep{Osterbart_etal1997,Men'shchikov_etal1998} revealed well-resolved bright
bipolar outflow lobes and long {\sf X}-shaped spikes originating deep inside
the outflow cavities. This set of high-resolution images stimulated us to
reanalyze all infrared observations of the {\rr} using our two-dimensional
radiative transfer code. The high-resolution images imply a geometrically and
optically thick torus-like density distribution with bipolar conical cavities
and are inconsistent with the flat disk geometry frequently used to visualize
bipolar nebulae. The new detailed modeling, together with estimates of the
interstellar extinction in the direction of the {\rr} enabled us to more
accurately determine one of the key parameters, the distance $D \approx$
710\,pc with model uncertainties of 70\,pc, which is twice as far
as the commonly used estimate of 330\,pc. The central binary is surrounded by
a compact, massive ($M \approx$ 1.2\,{\Msun}), very dense dusty torus
with hydrogen densities reaching $n_{\rm H} \approx 2.5 \times
10^{12}$\,cm$^{-3}$ (dust-to-gas mass ratio {\dustgas} $\approx 0.01$).
The model implies that most of the dust mass in the dense torus is in very
large particles and, on scales of more than an arcsecond, the polar
outflow regions are denser than the surrounding medium. The bright
component of the spectroscopic binary {\hd} is a post-AGB star with mass
{\Mstar} $\approx 0.57$\,{\Msun}, luminosity {\Lstar} $\approx 6000$\,{\Lsun},
and effective temperature {\Tstar} $\approx 7750$\,K. Based on
the orbital elements of the binary, we identify its invisible component with a
helium white dwarf with {\Mwd} $\approx$ 0.35\,{\Msun}, {\Lwd} $\sim$
100\,{\Lsun}, and {\Twd} $\sim 6 \times 10^{4}$\,K. The hot white
dwarf ionizes the low-density bipolar outflow cavities inside the dense torus,
producing a small H\,II region observed at radio wavelengths. We propose an
evolutionary scenario for the formation of the {\rr} nebula, in which the
binary initially had 2.3 and 1.9\,{\Msun} components at a separation of $\sim$
130\,{\Rsun}. The nebula was formed in the ejection of a common envelope after
Roche lobe overflow by the present post-AGB star.
}
\maketitle


\section{Introduction}
\label{Introduction}

The {\rr} is a spectacular bipolar reflection nebula around an evolved close
binary star (also known as {\hd}, \object{AFGL\,915},
\object{IRAS\,06176--1036}). The object has been extensively studied for more
than two decades \citep[see, e.g., references
in][]{Waters_etal1998,Men'shchikov_etal1998}. Recent diffraction-limited
speckle images of the object with 62--76\,mas resolution in the optical
\citep[0.6--0.8\,{\mic},][ hereafter Paper~I]{Osterbart_etal1997} and near
infrared \citep[0.7--2.2\,{\mic},][ hereafter Paper~II]{Men'shchikov_etal1998}
displayed a compact, highly symmetric bipolar nebula with pronounced {\sf
X}-shaped spikes, implying a toroidal distribution of the circumstellar
material. No direct light from the completely obscured central binary could be
seen.

New diffraction-limited images of the {\rr} in the near-IR (2.1--3.3\,{\mic})
with unprecedented angular reso\-lutions of 46--68\,mas were recently presented
by \cite{Tuthill_etal2002} (hereafter Paper~III). The images were reconstructed
from the Keck telescope speckle data using the bispectrum speckle
interferometry method. The highest-resolution images clearly show two bright
lobes above and below the midplane of an inclined torus or geometrically very
thick disk. {\sf X}-shaped spikes along the surfaces of the conical outflow
cavities contribute to the intensity distribution of the two bright lobes,
making them appear widened and even double-peaked. A striking feature of the
{\rr} bipolar nebula is its self-similar appearance on scales from 80\,mas to
1{\arcmin} and from the red light to at least 10\,{\mic}, implying that large
grains of at least several microns in size dominate scattering.

\begin{figure*}
\hspace{0.075\hsize}
\resizebox{0.85\hsize}{!}
   {
    \hspace{-5mm}
    \includegraphics{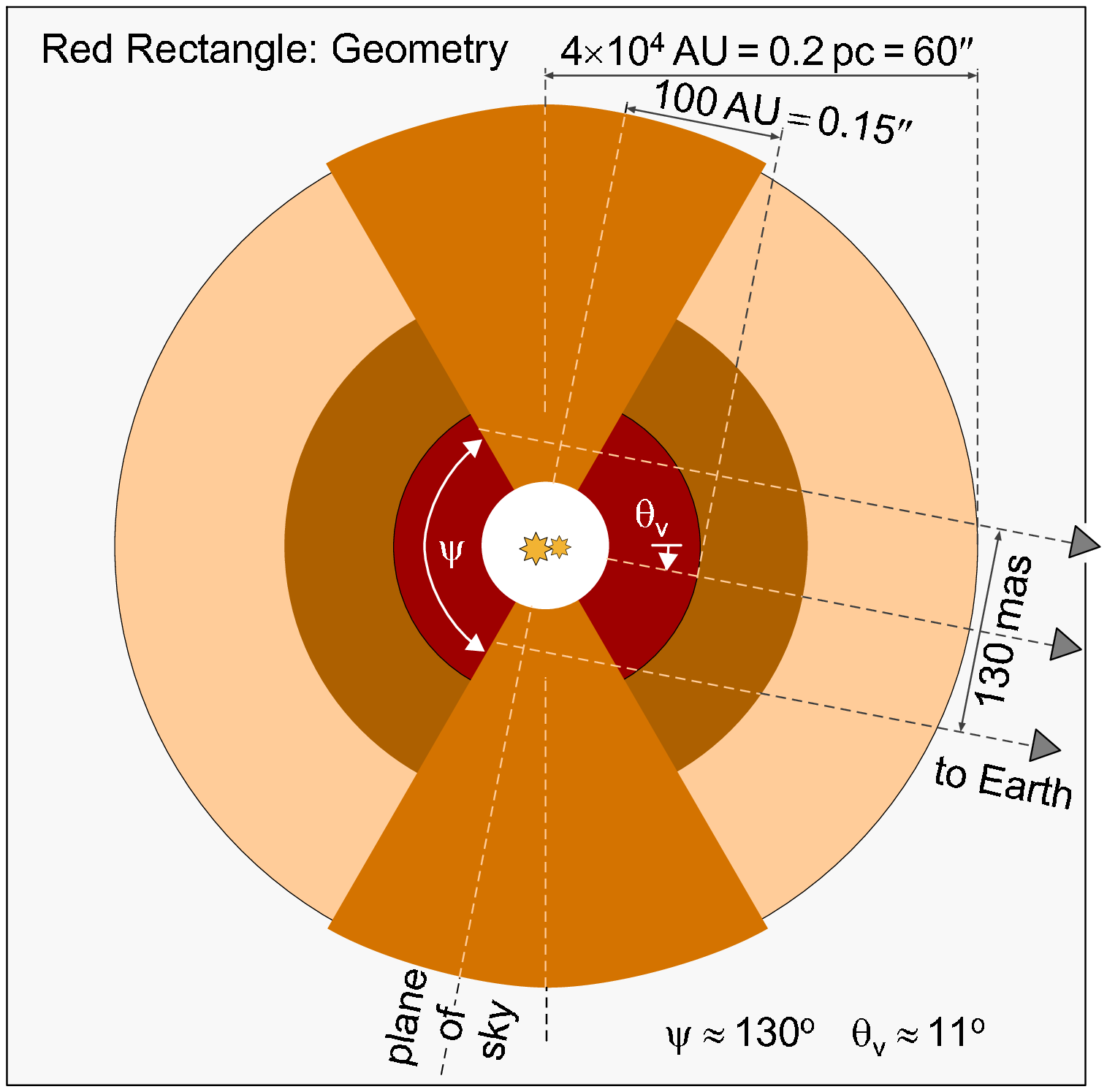}
    \hspace{10mm}
    \includegraphics{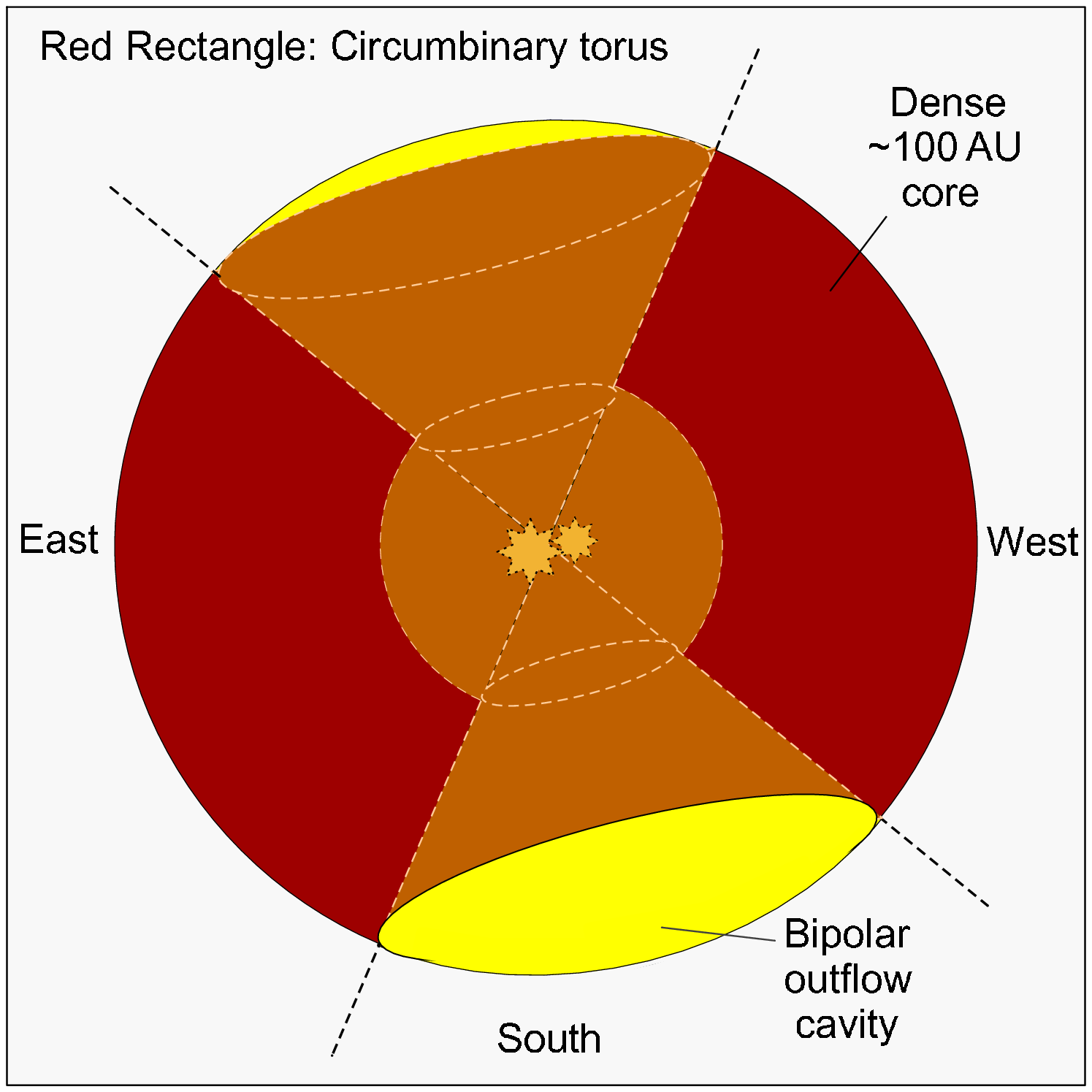}
   }
\caption
{
Geometry of the circumstellar environment of the close binary {\hd} ({\em left
panel}) and three-dimensional representation of the massive circumbinary torus
of the {\rr} ({\em right panel}) as it appears in the near-IR images of
\citetalias{Osterbart_etal1997}, \citetalias{Men'shchikov_etal1998}, and
\citetalias{Tuthill_etal2002} in projection onto the sky plane. Schematically
shown are four regions of the model -- the innermost dense torus with bipolar
cavities (100\,AU radius; dark color), the less dense envelope with a ${\rho}
\propto r^{-1.5}$ density profile (400\,AU radius; medium color), the bipolar
outflow cavities (lighter color), and the outer extended envelope with a steep
${\rho} \propto r^{-4}$ density gradient (4$ \times 10^{4}$\,AU; the lightest
color). The geometry is defined by the opening angle of the cavities, $\omega =
\pi - \psi \approx$ 50{\degr} ($\psi \approx$ 130{\degr}) and the viewing
angle, $\theta_{\rm v} \approx$ 11{\degr}, between the equatorial plane and the
line of sight.
}
\label{Geometries}
\end{figure*}

Only a few two-dimensional (2D) radiative transfer calculations of the bipolar
envelope of the {\rr} have been published to date. \citet{Yusef-Zadeh_etal1984}
first simulated the well-known optical images of the nebula with {\sf X}-shaped
spikes using a Monte-Carlo scattering method. Varying the density distribution
and scattering properties of dust grains, they found that a torus-like
configuration with a $\rho \propto r^{-2}$ radial density profile and with
biconical cavities having a full opening angle of 70{\degr} is able to
reproduce the shape of the nebula. These calculations gave support to the
previously suggested idea \citep{Cohen_etal1975,Morris1981,Perkins_etal1981}
that a quasi-spherical envelope with bipolar cavities can reproduce the shape
of the {\rr}.

Extending the previous modeling, \citet{Lopez_etal1997} applied a Monte-Carlo
technique in which they were able to not only simulate scattering of the
stellar radiation at a selected wavelength, but also calculate the radiative
equilibrium temperature and emission of dust. With multi-wavelength radiative
transfer computations, they aimed to constrain the model by comparing it with
the observed spectral energy distribution (SED) and a deconvolved 0{\farcs}2
resolution intensity map at 2.2\,{\mic} \citep{Cruzalebes_etal1996}. This work
represented the first step in the direction of a more realistic modeling of the
{\rr} capable of explaining a larger set of observational data and to
reconstruct reliable properties of the object.

In the previous 2D modeling presented in \citetalias{Men'shchikov_etal1998}, we
applied a frequency-dependent ray tracing method
\citep{Men'shchikovHenning1997} to construct a detailed model of the {\rr}
consistent with a much larger number of observational constraints. For the
first time, the model reproduced reasonably well the entire SED of the {\rr}
from the ultraviolet to centimeter wavelengths and the highest-resolution (76
mas) speckle-interferometry images at 0.656\,{\mic}, 0.8\,{\mic}, 1.65\,{\mic},
and 2.2\,{\mic}. The extensive modeling allowed us to derive the geometry of a
compact circumbinary torus-like structure, such as the opening angle $\omega =
70${\degr} of the bipolar cavities and the inclination angle $\theta_{\rm v} =
7${\degr} of the symmetry axis. For an assumed distance of 330 pc, the model
reconstructed physical parameters of the object, such as the total luminosity
$L_{\star} \approx 3000$ {\Lsun}, the radius $R \sim 30$ AU and the mass $M
\approx 0.25$ {\Msun} of the opaque ($A_V \approx 30$) torus, the density
distribution $\rho \propto r^{-2}$ for $r < 16$ AU and $\rho \propto r^{-4}$
for $r > 16$ AU, and very large sizes of dust particles ($a \sim 0.2$ cm).

Although the model describes reasonably well the large number of constraints,
our new Keck telescope images of the {\rr} with unprecedented resolutions
\citepalias[44--68 mas,][]{Tuthill_etal2002} have shown that the model is not
consistent with the longest-wavelength 3.1\,{\mic} and 3.3\,{\mic} images. As
we have already demonstrated in \citetalias{Tuthill_etal2002}, the model
predicts a single elongated peak at this wavelength, whereas the Keck image
clearly displays two lobes divided by a dark lane, very similar to the
shorter-wavelength images. In the model, there is too much emission from the
hot grains close to the inner boundary of the torus, which implies insufficient
optical depths. This finding stimulated us to recompute the model taking into
account the new constraints in addition to the old data.

This paper presents a new detailed study of the {\rr} based on
our previous model \citepalias{Men'shchikov_etal1998} and on the new
constraints provided by the Keck telescope images
\citepalias{Tuthill_etal2002}. In Sect.~\ref{RadTraModel} we describe our
assumptions and radiative transfer model of the dusty circumbinary torus. In
Sect.~\ref{ModelResults} we present the model results and compare them with
available observational data. In Sect.~\ref{Discussion} we discuss the
parameters of the {\rr} and evolution of its close binary. In
Sect.~\ref{Conclusions} we summarize conclusions of this work.


\section{Radiative transfer model}
\label{RadTraModel}

\subsection{General formulation}
\label{Formulation}

The detailed 2D radiative transfer model described below is similar to our
recent model presented in \citetalias{Men'shchikov_etal1998}. This new modeling
was stimulated by the speckle images of the {\rr} with the unprecedented,
diffraction-limited resolutions of 44--68\,mas, based on data collected with the
10\,m Keck telescope \citepalias{Tuthill_etal2002}. The images extend
our previous series of high-resolution optical and near-IR images of the object
\citepalias{Osterbart_etal1997,Men'shchikov_etal1998} to 3.3\,{\mic} and place
stronger constraints on models. The new images gave us the opportunity to
reanalyze the wealth of existing data on the {\rr} in the widest wavelength
range from the far-UV to radio wavelengths, using accurate radiative transfer
calculations. Following \citetalias{Tuthill_etal2002}, we denote the Keck
telescope near-infrared camera filters at 2.2, 3.08, and 3.31\,{\mic} as {\Ks},
{\Lc}, and {\Lpah}, respectively.

Our aim was to derive from the observations reliable information on the
structure and physical properties of the {\rr} and on the evolutionary status
of its central binary, {\hd}. We utilized our 2D radiative transfer code based
on a ray-tracing method \citep{Men'shchikovHenning1997}, which solves the
frequency-dependent radiative transfer problem in axially-symmetric dusty
envelopes for a number of dust components and grain sizes. A large parameter
space was explored in hundreds of models by comparing the resulting spectral
energy distribution and images to the available observational
constraints. For more discussion of our modeling approach, see
\citetalias{Men'shchikov_etal1998}, \cite{Men'shchikovHenning2000}, and
\cite{Men'shchikov_etal2001}.


\subsection{Geometry and assumptions}
\label{GeomAss}

The axially-symmetric model torus of the {\rr} is shown in
Fig.~\ref{Geometries}. The central binary {\hd} is embedded in a very large
dusty envelope of an axially-symmetric geometry defined by the biconical
outflow cavities in the otherwise spherical distribution of material. For
brevity, we call such a structure ``torus'', although it displays obvious
differences from a mathematician's torus.

The density distributions ${\rho}\,(r)$ of the massive torus and its outer
envelope depend only on the radial distance $r$ from the center. The density
$\rho_{\rm o} (r)$ of the bipolar outflow cavities is much lower than
${\rho}\,(r)$ in the central regions of the massive torus ($r <$ 800\,AU),
whereas in its outer parts the outflow is denser than the surrounding envelope.
Processes of dust condensation or sublimation put the inner boundary of the
dusty torus at $R_1 \approx$ 14 AU, where temperatures of dust grains $T_{\rm
d} \approx$ 1000 K. The location of the outer boundary $R_2$ can be constrained
from below by a deep coronagraphic image of the {\rr} tracing the nebula up to
1{\arcmin} from the center \citep{VanWinckel2001}. Although the envelope can
well be even more extended, we have fixed the model boundary at that angular
distance. In contrast to other mass-losing evolved stars \citep[e.g.,
{\irc},][]{Men'shchikov_etal2001}, any more precise location of $R_2$ in this
object is unimportant: with the steep outer density distribution of our model
($\rho \propto r^{-4}$, Sect.~\ref{DensTemp}), 99\,{\%} of the mass is
contained in the compact dense torus ($r <$ 100\,AU).

As usual, only the transfer of {\em dust} radiation was computed in this work.
The gas component, present in the model only implicitly, is described by a
dust-to-gas ratio {\dustgas} = $0.01$ within the dense torus.
Instead of assuming an arbitrary value, we used constraints from a detailed
analysis of the binary evolution in the {\rr} (see Sect.~\ref{Evolution} for
details) to estimate and fix the total mass of the circumbinary material at $M
\approx 1.2$\,{\Msun}, therefore determining {\dustgas}. Since there are no
reliable constraints for the outer regions of the envelope, we adopted
{\dustgas} = 0.004 outside the dense torus, which is close to the usual ratio
for the mass-losing evolved stars.

Since the semimajor axis of the spectroscopically observed component of the
close binary (Sect.~\ref{Binary}) is significantly smaller than $R_1$, we can
replace the binary at the center of the system with a single star (see also
Sects.~\ref{EffTem}, \ref{Luminosity}) for the purpose of the radiative
transfer calculations. Furthermore, we assume that the plane of the binary's
orbit is parallel to the midplane of the circumbinary torus, i.e., that the
inclination angle of the orbit is $i = {{\pi}/2 - \theta_{\rm v}}$ (cf.
Fig.~\ref{Geometries}).

Extending the previous modeling \citepalias{Men'shchikov_etal1998}, we used a
more realistic approach to describe dust grains, the main ingredient of the
radiative transfer models. Instead of having dust particles of a single (very
large) radius, we assume that the grains have a wide distribution of sizes. The
radiative transfer code can handle any (reasonable) number of dust components,
each of them with an arbitrary number of grain sizes. In practice, the number
of the dust components needs to be kept at a minimum, balanced with available
observational constraints. Details of the spatial distribution of chemical
compositions, shapes, sizes, and other parameters of real dust grains in this
object are poorly known. Large uncertainties of the real properties of dust
enormously increase the (free) parameter space of the models to explore. Since
our purpose was to reconstruct global properties of the {\rr}, we had to
simplify the dust model (see Sect.~\ref{Dust}) and ignore in this work the
presence of polycyclic aromatic hydrocarbon molecules (PAHs) and of crystalline
silicate dust \citep{Waters_etal1998}. These are only minor components of the
dust and our experience shows that they would not alter results of the
modeling.


\subsection{Effective temperature}
\label{EffTem}

It is impossible to observe the intrinsic energy distribution of the central
source in the {\rr} directly since the close binary is obscured by the
optically thick torus. The radiation of both components of the binary is
reprocessed by dust grains, which makes our understanding of what is going on
in the system very difficult. \cite{Cohen_etal1975} classified the radiation
source as a B9--A0 III star with {\Tstar} $\approx 10^4$\,K. From
high-resolution spectroscopy, \cite{Waelkens_etal1992} favored lower effective
temperatures of {\Tstar} $\approx$ 7500\,K (or 8000\,K) and gravity $\log\,g =
0.8$ (or 1.4). \cite{Knapp_etal1995} and \cite{Jura_etal1997} measured radio
emission from the {\rr}, concluding that it is produced by a small H\,II region
(radius of $\sim$ 17--170\,AU for our $D$ = 710 pc). \cite{Knapp_etal1995}
deduced a much higher temperature of {\em at least} $2.5 \times 10^{4}$\,K for
the ionizing star and \cite{Jura_etal1997} estimated an electron density of
$\sim 10^{6}$\,cm$^{-3}$ in the region. The UV spectrum of the {\rr}
does not show such a hot continuum in scattered light, therefore we can
safely assume that the other star, a cooler supergiant with {\Tstar} $\approx$
8000\,K, is responsible for the entire SED of the {\rr} (see also
Sect.~\ref{Luminosity}).

In this work, we have used several models of stellar atmospheres with effective
temperatures between 7500 and 8000\,K \citep{Kurucz1993}. In the final model,
described in Sect.~\ref{ModelResults}, we adopted an atmosphere parameterized
by {\Tstar} = 7750\,K, gravity $\log\,g = 1.5$, metallicity [M/H] = $-$3, and
microturbulent velocity of 2\,km\,s$^{-1}$. The high-resolution spectrum of the
atmosphere had been smoothed to a resolution of ${\lambda}/\Delta{\lambda}
\sim$ 100 and extrapolated to wavelengths $\lambda >$ 10\,{\mic}. Although our
final model's parameters correspond to a slightly lower value of $\log\,g \approx
1.2$, this difference does not have any impact on the results of our continuum
radiative transfer calculations.


\subsection{Distance}
\label{Distance}

Accurate estimates of distances $D$ are important for reliable derivation of
parameters of individual objects. The most frequently cited distance to {\hd}
of 330 pc \citep{Cohen_etal1975} was obtained by assuming that the star is
a visual binary, that it has the same bolometric luminosity as
\object{LkH${\alpha}$\,208} and as \object{R Mon}, and averaging the two
distances. Since now we know that \object{LkH${\alpha}$\,208} and \object{R
Mon} are of a different spectral type and nature (massive pre-main-sequence
Herbig Ae/Be stars), this derivation does not seem reliable. The same
conclusion can be made with respect to another estimate of 280 pc
\citep{Schmidt_etal1980}, which assumed that the binary is visual and has
identical components of spectral type A0\,III, whose bolometric luminosities
are $\sim$ 360 {\Lsun} each. An upper limit of 800--1400 pc
\citep{Waelkens_etal1992} was based on the measurements of photospheric
abundances, accurate assessment of the effective temperature and gravity
(Sect.~\ref{EffTem}), and the assumption that circumstellar dust produces a
2.7 mag extinction. Although the latter may not be reliable, the distance limit
seems reasonable. An estimate of 940 pc \citep{Knapp_etal1995} was based on the
upper end of the luminosity range of 2300--7500 {\Lsun} suggested by
\citet{Waelkens_etal1992} and the bolometric flux of the {\rr}. Since the
luminosity was chosen somewhat arbitrarily, it implies an uncertainty
by a factor of 1.8 toward shorter distances. Parallax measurements with the
Hipparcos satellite gave the value of 2.62$\,\pm\,$2.37\,mas
\citep{Perryman_etal1997}, which amounts to a distance of 380\,pc with quite
large uncertainties making the whole range of 200--4000\,pc plausible.

The widely adopted distance of 330\,pc is definitely inconsistent with the
well-established post-AGB status of the {\rr}. If the value were accurate, the
total luminosity of the central binary (estimated from the SED assuming
spherical symmetry of the nebula) would only be 920\,{\Lsun}
\citep{LeinertHaas1989}. This luminosity is too low for a single post-AGB star
(by a factor of $\sim$ 3--10) and it seems to be impossible to account for such
a difference in the framework of the stellar evolution theory for single stars.
Although binarity does modify stellar evolution to some extent, we still
believe that the most natural and likely cause of this discrepancy is invalid
distance.

\begin{figure}
\begin{center}
\resizebox{0.9\hsize}{!}{\includegraphics{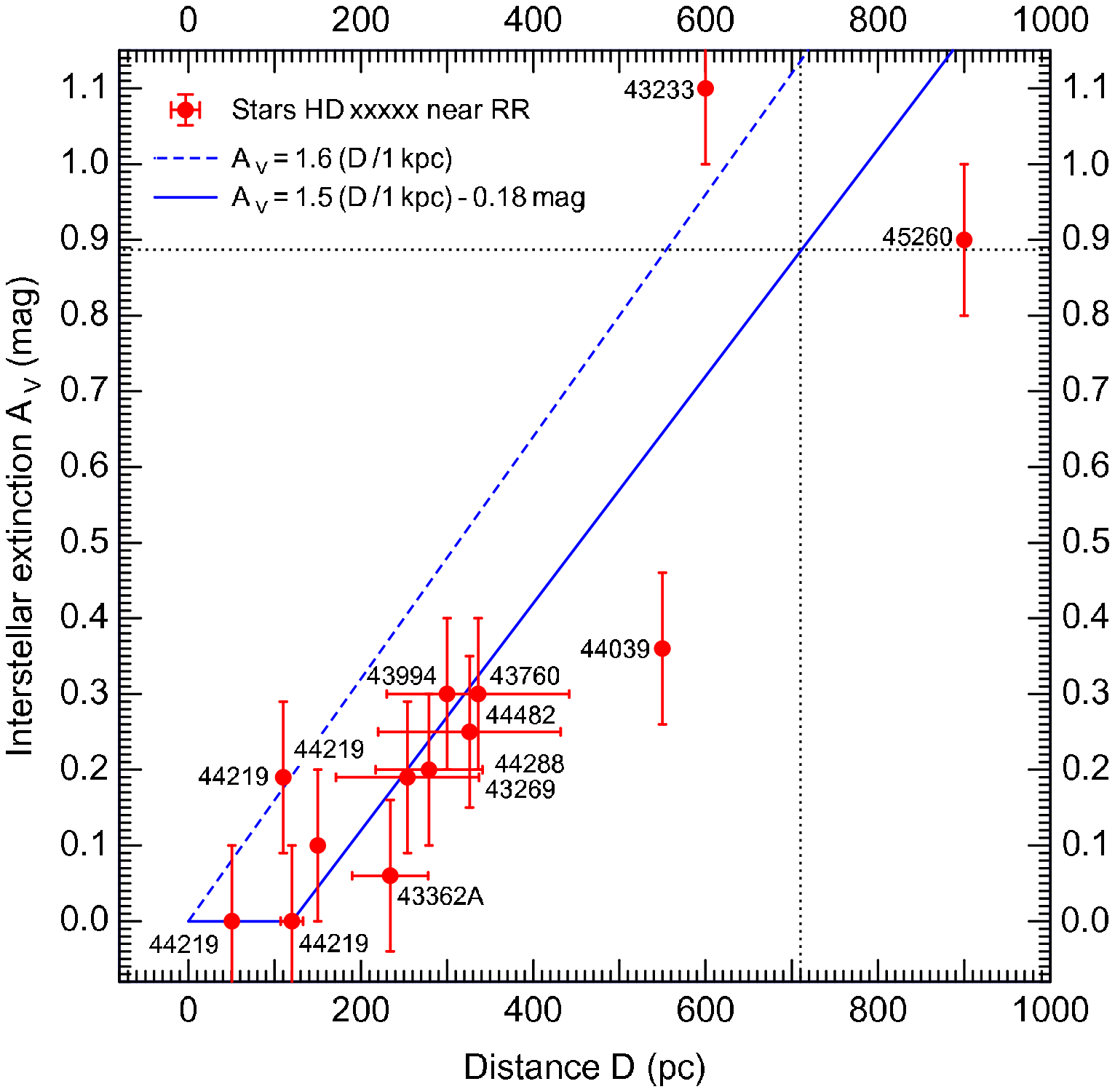}}
\caption
{
Observationally determined interstellar visual extinction $A_V$ for stars
with known distances $D$ in the direction of the {\rr} (within an angular distance
of $\sim$ 1{\degr} from the object). The stars are labeled with their
respective HD names. The dashed line shows an average dependence $A_V(D)$
for our Galaxy. The solid line displays the relation used in our modeling,
which takes into account the well-known fact that within roughly 100 pc from
the Sun there is almost no extinction. The dotted lines indicate the distance
and extinction derived in this study. Horizontal error bars, drawn only
for the Hipparcos data, correspond to the uncertainties of the measured
parallaxes. The vertical bars are approximate uncertainties associated with the
color excess estimates.
}
\label{Extinc}
\end{center}
\end{figure}

The results of our modeling (Sect.~\ref{SpEnDi}) confirm this idea,
demonstrating that an almost perfect fit to the SED of the {\rr} can be
obtained only if the interstellar extinction (proportional to the distance) is
properly taken into account. The interstellar reddening varies with direction
in the Galaxy, and the variations increase closer to the galactic plane. The
{\rr} is located at a galactic latitude of $b = -$11{\fdg}8, in a region
without noticeable dense interstellar dust clouds. The published photometric
data for the {\rr} \citep{Cohen_etal1975,Kilkenny_etal1985} show no significant
color variations with an average $B{-}V$ = 0.42 mag. The stellar gravity and
effective temperature correspond to an intrinsic color index $(B{-}V)_{0}$ =
0.18 mag \citep{Straizhis1977}. As the extinction by the dense circumbinary
torus is gray, a color excess $E_{B{-}V}$ = 0.24 mag is due to the interstellar
dust. This results in an observational estimate of the interstellar $A_V$ =
0.74 mag, which is close to the value of $A_V$ = 0.89 mag we obtained in our
model fitting of the observed SED (Sect.~\ref{SpEnDi}).

From these data one can estimate $D$ toward the {\rr} using the photometry and
spectral classification of nearby stars available in the literature, as
well as the Hipparcos parallaxes. Our search for the stars was facilitated by the
SIMBAD and the Two Micron All Sky Survey databases. Although not many stars
around the {\rr} have the estimates of spectral types, the existing data allow
us to put some constraints on the relationship between $A_V$ and $D$ in this
direction (Fig.~\ref{Extinc}). The stars located closer than 350 pc have $A_V
\le$ 0.3 mag, demonstrating that a distance of 330 pc \citep{Cohen_etal1975} is
inconsistent with $A_V$ of 0.74--0.89 mag derived in this paper. Data on two
B-type stars suggest that $A_V$ reaches 1 mag at $\sim 800 \pm 100$ pc,
although there is a very large gap in the diagram, in the region of our
interest. A rough interpolation between the few available points seems to
suggest a distance between 600 and 800 pc, however, there are no stars with
known spectral types in the region. Photometric and spectroscopic observations
of $V \sim$ 9--10 mag A- and B-type stars in the neighborhood of the {\rr} on
the sky would be very useful in obtaining a better estimate.

In this study we adopted an average interstellar extinction per hydrogen column
number density $A_V / N({\rm H}) = 5.3 \times 10^{-22}$ mag\,cm$^2$
\citep{Bohlin_etal1978} and a hydrogen number density $n_{\rm H} =
1$\,cm$^{-3}$. Assuming also that there is no extinction within 120 pc from the
Sun, this resulted in $A_V = 1.5\,(D / 1\,{\rm kpc}) - 0.18$ mag
(Fig.~\ref{Extinc}). Using the value of $A_V$ = 0.89 mag derived in this
modeling and the dependence $A_V(D)$ shown in Fig.~\ref{Extinc}, we can
determine a distance $D \approx$ 710 pc, with possible uncertainties of $\sim$
100 pc due to unknown Galactic extinction beyond 350 pc from the Sun.

\begin{table*}
\caption{Model parameters of the post-AGB star and circumbinary torus of the {\rr}.}
\label{ModelParams}
\smallskip
\begin{tabular}{lcrlll}
\hline
\noalign{\smallskip}
Parameter                & Symbol             & Value   & Units   & Uncert.       & Comment\\
\noalign{\smallskip}
\hline
\noalign{\smallskip}
Distance                 & $D$                &   710   & pc      & $\pm$10{\%} & derived (from SED and reddening, Sects.~\ref{Distance}, \ref{SpEnDi})\\
Total luminosity         & $L_{\star}$        & 6\,050  & {\Lsun} & $\pm$20{\%} & derived (from SED and distance, Sects.~\ref{Luminosity}, \ref{SpEnDi})\\
Effective temperature    & $T_{\star}$        &  7\,750 & K       & $\pm$10{\%} & derived (from observations, Sect.~\ref{EffTem})\\
Stellar radius           & $R_{\star}$        &    43   & {\Rsun} & ---         & derived ($R_{\star} = {(4\pi\sigma)^{-1/2}}{L_{\star}^{1/2}}\,{T_{\star}^{-2}}$)\\
Inner boundary           & $R_{1}$            &    14   & AU      & ---         & derived ($R_1 \approx$ 71\,$R_{\star}$, Sect.~\ref{DensTemp})\\
Outer boundary           & $R_{2}$            & 43\,000 & AU      & ---         & assumed (from deep optical images, Sect.~\ref{GeomAss})\\
Total mass of envelope   & $M$                &  1.2    & {\Msun} & ---         & derived (corresponds to {\dustgas} = $0.01$, Sects.~\ref{GeomAss}, \ref{Evolution})\\
Gaussian density profile & FWHM               &    26   & AU      & ---         & model (14--100\,AU, Sect.~\ref{DensTemp})\\
1st density exponent     & ${\alpha}_1$       &  $-$1.5 & ---     & ---         & model (100--430\,AU, Sect.~\ref{DensTemp})\\
2nd density exponent     & ${\alpha}_2$       &  $-$4.0 & ---     & $\pm$20{\%} & model (430--43\,000\,AU, Sect.~\ref{DensTemp})\\
Cavity opening angle     & ${\omega}$         &    50   & {\degr} & $\pm$10{\%} & derived (from high-resolution speckle images, Sect.~\ref{Images})\\
Viewing angle            & ${\theta}_{\rm v}$ &    11   & {\degr} & $\pm$10{\%} & derived (from SED and model images, Sects.~\ref{SpEnDi}, \ref{Images})\\
\noalign{\smallskip}
\hline
\end{tabular}
\end{table*}


\subsection{Luminosity}
\label{Luminosity}

With the above distance of 710\,pc, the object's bolometric luminosity
would be 4260\,{\Lsun }, if the nebula were spherically-symmetric.
Although the value would better agree with the AGB status of the star,
this simple estimate based on the assumption of spherical geometry is
incorrect, because the {\rr} is strongly {\em non-}spherical. Large deviations
from spherical geometry in optically thick environments usually lead to large
errors (by a factor of several), depending on the optical depths, viewing
angle, and degree of non-sphericity \citep[see
e.g.,][]{Men'shchikovHenning1997,Men'shchikov_etal1999}. Our present modeling
shows that the actual total luminosity of the binary is {\Lstar} $\approx$
6050\,{\Lsun} (Sect.~\ref{SpEnDi}). This entire luminosity seems to be
produced by the post-AGB component of the binary. Luminosities, masses, and
other parameters of both components, not related directly to the radiative
transfer calculations, are discussed in more detail in Sect.~\ref{Discussion}.


\subsection{Dust particles}
\label{Dust}

As we have shown in the previous modeling \citepalias{Men'shchikov_etal1998},
the dominant particles in the dense dusty torus of the {\rr} are very large.
Much smaller amounts of small dust grains and PAHs seem to exist somewhere in
the nebula, perhaps in its outer dilute parts and in the outflow regions. In
the final model we assumed that there are two (spatially separate) components
of dust grains. Following \citetalias{Men'shchikov_etal1998}, the first
component is made of very large particles of an unspecified chemical
composition, with sizes of 0.2\,cm. This component exists only in the compact,
extremely dense circumbinary torus, where most of the grains are assumed to
have coagulated into the large particles. The other component is the amorphous
carbon grains with radii $a$ in the range of 0.005--600\,{\mic} and with the
standard interstellar size distribution ${\rm d}n/{\rm d}a \propto a^{-3.5}$.
This second component exists beyond the boundary of the compact torus (100\,AU
$< r < R_2$), where the density drops by several orders of
magnitude. This particular choice of the dust parameters is somewhat arbitrary,
as there are no observational constraints to better restrict the freedom. We
should note, however, that it was impossible to make our model explain all
observations when considerable amounts of small dust grains were put in the
dense torus. We can conclude that a difference between the dust
properties of the compact dense torus, made up of very large particles
and of those of the surrounding low-density envelope, made up of normal
grains, is a firm result of the modeling.


\section{Results of the modeling}
\label{ModelResults}

\subsection{Model parameters}
\label{ModParams}

The final parameters of our detailed radiative transfer model and their
approximate uncertainties are presented in Table~\ref{ModelParams}. The model
results are described in Sects.~\ref{SpEnDi}--\ref{IntensProf} in detail.
The density distribution is assumed to be a broken power law $\rho \propto
r^{{\alpha}_i}$, the exponents ${\alpha}_i$ ($i$ = 1, 2) being allowed to
change in different radial zones. The third column lists the values adopted in
the final model. The fifth column contains our estimates of the uncertainties
associated with the final model parameters. They indicate very approximately
the maximum changes of the parameters, such that the fit to the observations
can still be restored by adjusting some other model parameters. The
uncertainties should not be interpreted as absolute error bars. If some of our
general assumptions turn out to be insufficiently realistic, this would
possibly affect the results more than these uncertainties imply.

It may be useful to recall here that the results of our radiative transfer
modeling are invariant with respect to the distance $D$, except for several
parameters which are scaled in the following simple way:
\begin{equation}
L{\,\propto\,}D^2, \,\,M{\,\propto\,}D^2, \,\,R{\,\propto\,}D,
\,\,\rho{\,\propto\,}D^{-1},
\end{equation}
where mass and density refer to the properties of dust; for conversion to the
gas parameters, {\dustgas} must be specified (Sects.~\ref{GeomAss},
\ref{Evolution}). These relations make it easy to scale the model results to
another distance, if necessary.

\begin{figure*}
\begin{center}
\resizebox{0.65\hsize}{!}
   {
    \includegraphics{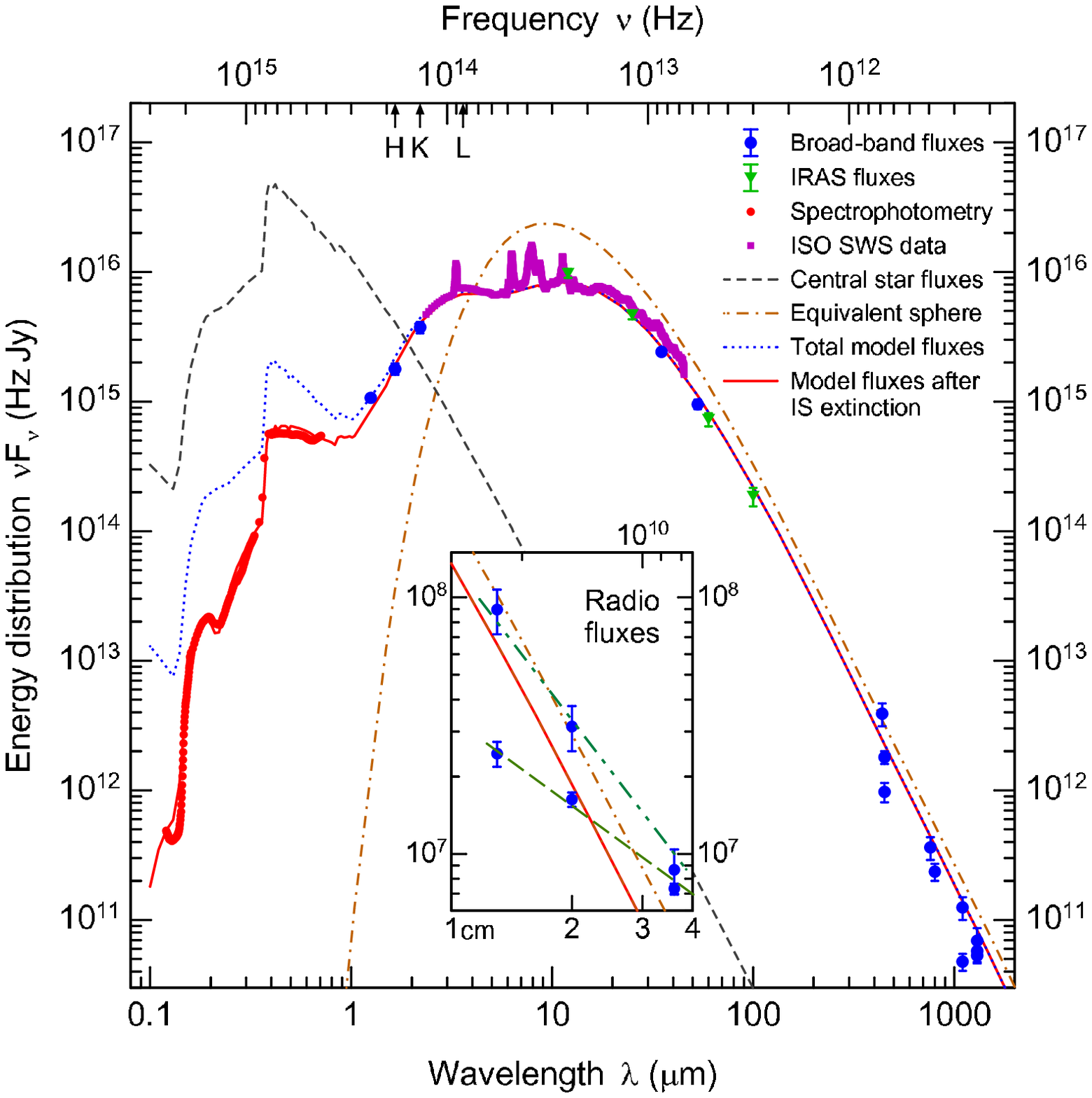}
   }
\caption
{
Observed SED of the {\rr} compared to our model with $L_{\star} =
6050$\,{\Lsun} (solid line). The stellar continuum \citep{Kurucz1993},
the total model fluxes without interstellar reddening, the model fluxes
reddened by the interstellar extinction (for $D$ = 710\,pc), and SED for
an equivalent spherical envelope are plotted. The latter emphasizes the major
influence of the bipolar outflow cavities on the emerging continuum of an
optically thick torus. The observed fluxes are shown by different symbols to
differentiate between broad-band fluxes, IRAS fluxes, spectrophotometry, and
ISO SWS data. If available, error bars for the fluxes are drawn, whenever they
are larger than the symbols. The model assumes that we observe the torus close
to edge-on orientation, at $\theta_{\rm v}$ = 11{\degr}. Three arrows at the
top axis indicate the location of the $H, K$, and $L$ photometric bands. The
insert compares the centimeter-wavelength segment of the model SED with
available radio observations. The dashed line is fitted to the compact emission
fluxes from ionized hydrogen \citep{Jura_etal1997}. Dash-double-dotted line
through the extended fluxes is the sum of the model fluxes (solid line) and the
compact H\,II emission. The far-UV model fluxes rising toward shorter
wavelengths at $\lambda < 0.14$\,{\mic} and a bump on the reddened model curve
show the upper limit for the possible contribution of a hot companion with a
temperature of $6 \times 10^{4}$ K (see Sect.~\ref{Binary}).
}
\label{SED}
\end{center}
\end{figure*}


\subsection{Spectral energy distribution}
\label{SpEnDi}

In addition to the fluxes used in \citetalias{Men'shchikov_etal1998}, in our
new modeling we included data collected by the Infrared Space Observatory's
short-wavelength spectrometer \citep[ISO SWS,][]{Waters_etal1998} as well as
new (sub-)millimeter fluxes
\citep{GreavesHolland1997,JuraTurner1998,SanchezContreras_etal1998}. To make
sure that the observed SED of the {\rr} is not influenced by observations at
different epochs, we ignored older broad-band fluxes (in the wavelength range
covered by the ISO data) containing essentially no additional information on
the spatial distribution of intensity. As our model does not contain any
species producing unidentified nebular red emission between 0.55 and
0.75\,{\mic} \citep{Schmidt_etal1980}, we included an optical spectrum of the
innermost few arcseconds around {\hd} \citep{ReeseSitko1996} that shows no
extended red emission.

Figure~\ref{SED} compares the model SED to the broad-band flux distribution and
spectrophotometry of the {\rr}. To emphasize the influence of the bipolar
geometry, the SED of an equivalent spherical envelope is also shown. The latter
differs from the torus model only by the absence of the outflow cavities (i.e.
${\psi}$ = 180{\degr}, $\omega$ = 0, Fig.~\ref{Geometries}). The model of the
{\rr} fits the observed continuum underlying the emission features almost
perfectly, from the far UV to centimeter wavelengths. Three lower radio fluxes
in small beams with a spectral index of $1.5$ are most likely produced by a
compact H\,II region \citep{Jura_etal1997}, whereas three higher fluxes in
larger beams are mostly due to dust emission. When the compact radio emission
of the H\,II region, which has not been included in our modeling, is added to
the total model fluxes, the large-beam fluxes become perfectly fitted
(Fig.~\ref{SED}). There is no discernible effect of different apertures (larger
than a few arcseconds) at any wavelengths in the model SED, which implies that
the dusty torus has density distribution highly concentrated toward the center
(see Sects.~\ref{DensTemp}, \ref{Images}).

The {\rr} differs from other dusty objects in that the dominant dust particles
are very large, most likely millimeter-sized
\citepalias{Jura_etal1997,Men'shchikov_etal1998}. The new modeling presented
here confirms the idea that the dense torus contains most of its circumstellar
dust mass in large particles having gray extinction. We found no model with
significant amounts of small grains that would be able to reproduce all
high-resolution images and the SED. It seems to be impossible, however, to fit
the observed fluxes with only very large grains.

It was clear also from our previous model \citepalias{Men'shchikov_etal1998}
that gray circumstellar extinction alone cannot reproduce the SED of the {\rr}
in the optical and UV. The steep peak of the scattered intrinsic stellar energy
distribution of the post-AGB component in the visible sharply contrasts the almost
horizontal SED in that region. At wavelengths shorter than the Balmer jump, the
observed fluxes drop much deeper and much steeper than simple gray scattering
of stellar radiation would predict (Fig.~\ref{SED}). This puzzle can be solved
in a simple and natural way just by taking into account interstellar
extinction.

Usually it is difficult to discriminate between the absorption by circumstellar
and interstellar dust: small circumstellar dust grains absorb radiation the
same way the interstellar grains do. In contrast to the latter, dust in
stellar environments is embedded within and plays an active role in modifying
the circumstellar radiation field. On the other hand, the interstellar dust is
just passively absorbing and scattering photons on their way to the observer.
Since our modeling has shown that the dust particles in the circumbinary torus
of the {\rr} are predominantly very large, we are left with only one
possibility to explain the discrepancy at short wavelengths, namely, that the
SED is altered by the selective extinction of small {\em interstellar} dust
grains.

The distance to the {\rr} is large enough for the interstellar extinction
(Sect.~\ref{Distance}) to affect the observed SED. For the wavelength
dependence of the interstellar extinction, we used the analytic fits of
$A_{\lambda}/A_V$ given by \cite{Cardelli_etal1989}. The extinction law
parameter $R_V$ was fixed at the generally accepted value of 3.1 corresponding
to the diffuse interstellar medium. The pronounced effect of interstellar
extinction is visible in Fig.~\ref{SED} as the increasingly larger difference
toward shorter wavelengths (${\lambda} \la$ 1\,{\mic}) between the
total (unreddened) model fluxes and the model fluxes reddened by $A_{\lambda}$
($A_V \approx 0.89$ mag). After the interstellar reddening has been taken into
account, the model SED became almost indistinguishable from the observed flux
distribution.

In contrast to the transfer of radiation in circumstellar envelopes,
interstellar extinction does depend explicitly on the distance $D$ to the
object. The accurate fit of the short-wavelength part of the SED enabled us to
derive a distance $D \approx 710$\,pc to the {\rr}, with approximate model
uncertainties of about 10\,{\%}, which is consistent with both the interstellar
reddening data in this direction and the object's luminosity. In fact, this
larger distance implies the stellar luminosity is more than four times
higher, reducing by the same factor the large discrepancy between the post-AGB
nature of {\hd} and its low apparent luminosity (Sect.~\ref{Luminosity}). The
actual luminosity {\Lstar} $\approx 6050$\,{\Lsun} obtained in our
modeling is even higher than the value of 4260\,{\Lsun} one would expect
for a spherical envelope. This is not surprising, given the obvious
non-sphericity of the optically thick circumbinary torus
(Sect.~\ref{Luminosity}).

The above $\sim$ 40\,{\%} increase of the derived luminosity is a result
of accurate 2D radiative transfer calculations. In the optically thick
non-spherical configurations like the torus of the {\rr}, a large fraction of
the central star's radiation can escape in the optically thin polar cavities,
thus changing the value of the luminosity an observer would derive under the
(invalid) assumption of spherical symmetry. The observer would underestimate
the luminosity if the object's orientation is close to edge-on, whereas one
would significantly overestimate the luminosity if one sees the object almost
pole-on. Corrections for converting the observed ``spherical'' luminosities
into the real luminosities of the central energy sources may reach a factor of
several, depending on the distribution of optical depths, opening angle of the
cavities, and the viewing angle of the object \citep[see,
e.g.,][]{Men'shchikovHenning1997,Men'shchikov_etal1999}. It seems to be
difficult (if possible at all) to determine such factors once and for all;
they have to be derived for each individual object in a detailed modeling.

\begin{figure}
\resizebox{\hsize}{!}
   {
    \includegraphics{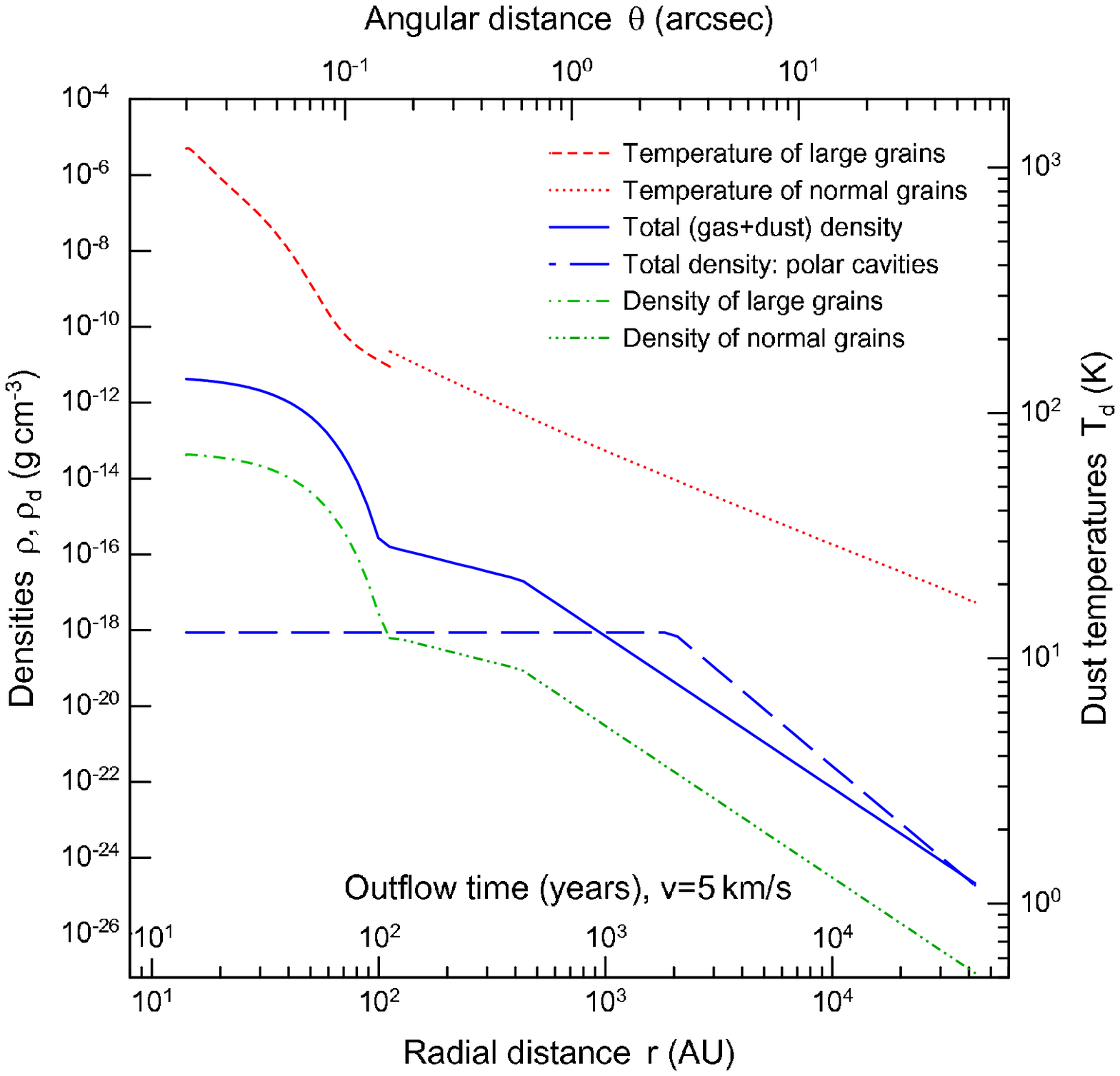}
   }
\caption
{
Temperatures and densities of dust grains in the envelope of the {\rr}. The
distributions are displayed for the smallest grains of the dust components.
Also shown are the total (gas+dust) density profiles in the midplane and in
bipolar outflow cavities along the symmetry axis. The total mass of the
circumstellar material is 1.2\,{\Msun}, corresponding to {\dustgas} =
$0.01$ within the dense torus. Additional labeling above the lower
abscissa shows possible expansion times if the outflow velocity were constant
and equal to 5\,km\,s$^{-1}$ \citep[similar to the observed
velocities,][]{Jura_etal1995,Knapp_etal2000}.
}
\label{DenTem}
\end{figure}


\subsection{Densities and temperatures}
\label{DensTemp}

The density and dust temperature distributions of our final model of the {\rr}
are shown in Fig.~\ref{DenTem} for the smallest grains; the total gas plus
dust density is also plotted assuming a spatially uniform dust-to-gas mass
ratios {\dustgas} = $0.01$ within the dense torus and 0.004
outside (Sects.~\ref{GeomAss}, \ref{Evolution}). Although {\dustgas} may
be a more complex function of the position inside the nebula, we assumed
this simple form in the absence of better constraints. This parameter
is only a means to approximately convert the dust densities into the total
density.

A unique property of the {\rr} is that even the {\Lc} and {\Lpah} images show
almost the same intensity distribution as the images at shorter wavelengths
\citepalias{Osterbart_etal1997,Men'shchikov_etal1998,Tuthill_etal2002}.
Moreover, the biconical shape is preserved across a wide range of wavelengths
(0.6--10\,{\mic}) and spatial scales (0.001--1{\arcmin})
\citepalias[cf.][]{Cohen_etal1975,Perkins_etal1981,
Waters_etal1998,Men'shchikov_etal1998,VanWinckel2001}. This fact provides a
strong observational constraint on the density structure and dust properties of
the circumbinary environment of the {\rr}.

The invariant biconical shape of the nebula at such different scales and even
at the mid-infrared wavelengths suggests that the shape is produced by
scattering. The 10\,{\mic} image presented by \citet{Waters_etal1998}, in
which the bipolar outflow can be traced to 10{\arcsec} from the central stars
rules out thermal emission from dust: in our model, grains in the torus and
in the cavities have temperatures of 32 K and 95 K, respectively, at these
distances. At optical as well as at near-IR wavelengths, the biconical shape is
caused by the radiation scattered by dust grains in the bipolar outflow
regions. The particles must be very large compared to the normal (small)
interstellar grains since the latter have negligible albedo in the
mid-infrared. Dominant grains must have sizes of at least several microns in
order to efficiently scatter the mid-IR radiation.

The present modeling has shown that the density distribution of the
toroidal structure of the {\rr} has two distinct regions: the compact dense
torus and the extended low-density envelope (Figs.~\ref{Geometries},
\ref{DenTem}). The model density distribution displayed in Fig.~\ref{DenTem}
follows a Gaussian profile in the very dense torus, with a full width at half
maximum of 26 AU and a maximum total density of $\rho \approx 4.2 \times
10^{-12}$ g\,cm$^{-3}$ ($n_{\rm H} \approx 2.5 \times 10^{12}$\,cm$^{-3}$).
The dense torus contains 99\,{\%} of the total mass of the circumbinary material.
The density of the toroidal core drops by 5 orders of magnitude at $r \approx
100$ AU (0{\farcs}15), where it changes to a much flatter $\rho \propto
r^{-1.5}$ radial profile, which in the outermost regions at $r \ge$ 430
AU (0{\farcs}6) steepens to a $\rho \propto r^{-4}$ distribution. Constrained
by our new highest-resolution speckle images \citepalias[][ see also
Sect.~\ref{Images}]{Tuthill_etal2002}, the radial density profile differs from
that of the previous model \citepalias{Men'shchikov_etal1998} by the much
steeper, exponential fall-off in the outer regions of the torus and by the
flatter density distribution just outside it. No simpler power-law density
profile has been found to reproduce the well-defined bright bipolar lobes which
are almost invariant between the $H$ and {\Lpah} bands.

The bipolar appearance of the {\rr} implies much lower density in the outflow
cones compared to the torus over the central subarcsecond region
(Fig.~\ref{DenTem}). The inner dense torus is optically thick ($\tau \approx
47$ up to $\lambda \sim$ 1\,cm) and our modeling demonstrated that the
observed appearance of the {\rr} requires the outflow cavities to be optically
thin ($\tau \approx 0.2$). On the other hand, the model shows that the outer
regions of the circumbinary material (at angular distances $\theta \ga$
0{\farcs}5) are optically thin in all directions. This implies that the strong
biconical shape visible in the outer parts of the {\rr} is produced by dust
scattering in the outflow regions which are much {\em denser} there than the
rest of the torus at the distances of $r \ge$ 800 AU from the center. This
conclusion implicitly assumes that the dust properties in the outflow cones and
in the rest of the toroidal envelope are the same. If dust grains in the torus
and in the outflow cones differ in such a way that the outflow regions scatter
significantly more radiation to the observer, there may be no need for a
density contrast to explain the observations.

In the absence of reliable constraints, we modeled the radial density
distribution in the outflow cavities by a constant density $\rho_{\rm o} \approx
9 \times 10^{-19}$ g\,cm$^{-3}$ ($n_{\rm Ho} \approx 5 \times
10^{5}$\,cm$^{-3}$) at $r \le$ 2000 AU (2{\farcs}8) and by a steep
$\rho_{\rm o} \propto r^{-5}$ profile outside (Fig.~\ref{DenTem}). The density
profile of the outer regions is constrained by the long-wavelength fluxes of
the observed SED, which must be produced largely by the dense inner torus with
very large dust grains. At the distance of 2000 AU, the density in the
outflow cavities is larger by a factor of 14 than that of the toroidal
envelope; the density contrast is vanishing toward the outer boundary of the
model. At the inner boundary, the torus is denser by a factor of $\sim 5
\times 10^{6}$ than the outflow cavities. Such a density structure can be
interpreted as a result of higher mass loss and velocity in the outflow along
the symmetry axis during the last $\sim 4 \times 10^{4}$ years, as
compared to the early mass loss that made up the rest of the outer toroidal
envelope.

\begin{figure}
\resizebox{\hsize}{!}
   {
    \includegraphics{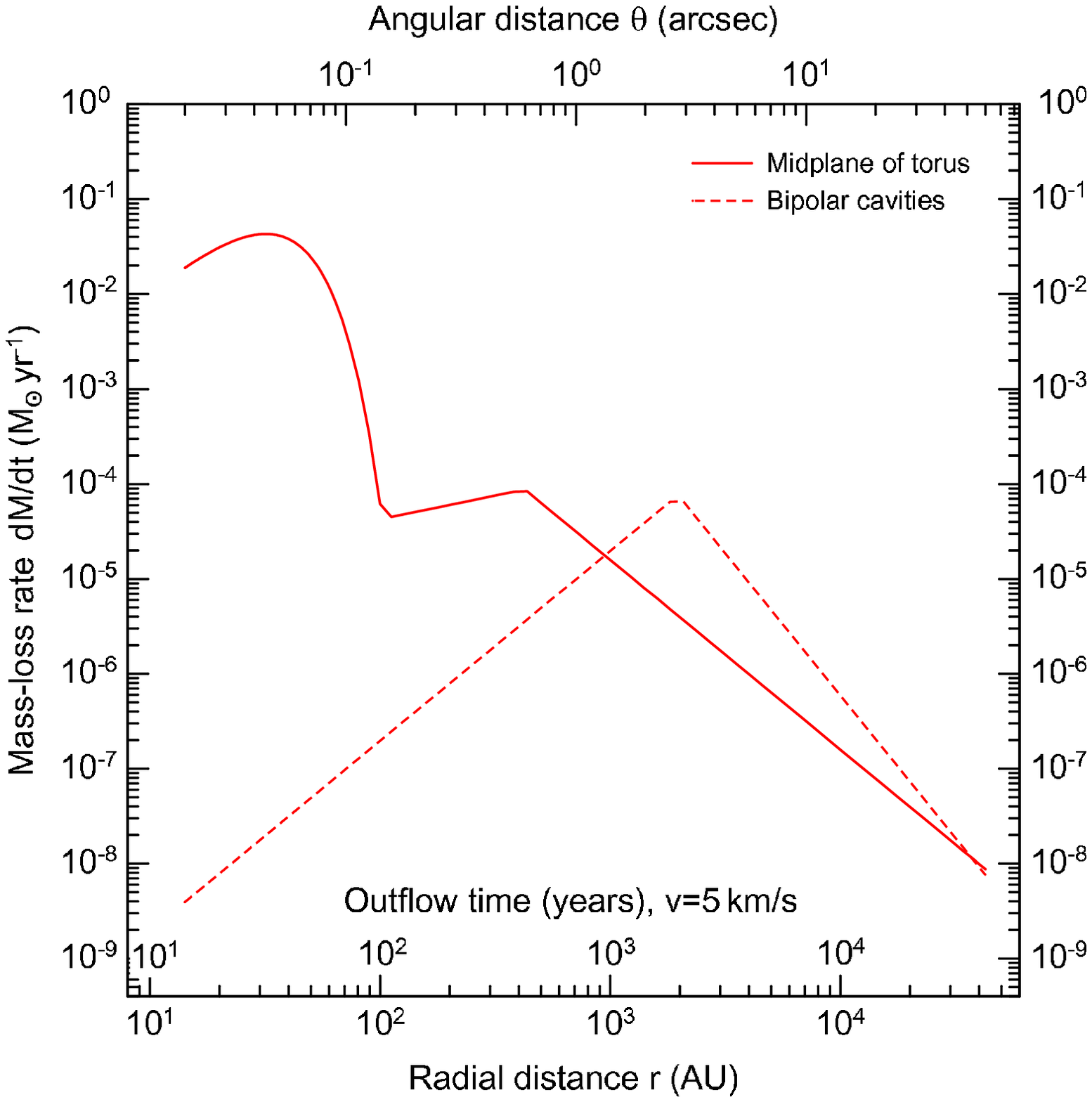}
   }
\caption
{
Mass-loss history of {\hd} within the torus' midplane and in the bipolar
cavities, reconstructed from the density distribution of the {\rr} assuming a
spherical outflow with constant velocity $v = 5$ km\,s$^{-1}$. Although these
assumptions may not precisely describe the entire evolution of the envelope,
the plot gives an approximate overview of the mass-loss history.
}
\label{MLoss}
\end{figure}

With the assumption of a constant-velocity spheri\-cally-symmetric mass
loss, one can derive the mass-loss history of the {\rr} from the density
profiles shown in Fig.~\ref{DenTem}. The reconstruction presented in
Fig.~\ref{MLoss} can provide approximate but useful information on the global
properties and temporal evolution of the mass loss. The very high density
contrast of the toroidal core of our model with respect to the outer envelope
(Fig.~\ref{DenTem}) is reflected in the mass-loss rates that seem to reach
$4 \times 10^{-2}$ {\Msun}\,yr$^{-1}$ within the dense core, very high
values compared to the usual spherical mass loss from AGB stars. However, the
dense core is most likely the result of a complex, time-dependent
hydrodynamical evolution of the self-gravitating envelope ejected from the
central binary. Both expansion and infall are likely to have played a role in
the accumulation of the circumbinary material and in the compression of the
compact torus to very high densities. For an illustration, note that 90\,{\%}
of the dense torus' mass resides within $r = 57$ AU. If the mass is
redistributed over a radius of 430 AU (where ${\rm d}{\dot{M}}/{\rm d}{r}$ and
$\ddot{M}$ change their sign, Fig.~\ref{MLoss}), then the peak densities and
mass-loss rates we derive would be lowered by a factor of 430 to the values of
$n_{\rm H} \sim 10^{10}$\,cm$^{-3}$ and $\dot{M} \sim 10^{-4}$
{\Msun}\,yr$^{-1}$. It seems plausible that the hydrodynamical interplay
between the gravitation, gas and radiation pressure, and bipolar outflow have
led to the present-day density distribution inside the circumbinary torus and
within the bipolar outflow cavities (see also Sect.~\ref{Stability}).

\begin{figure*}
\begin{center}
\resizebox{\hsize}{!}{\hspace{1mm}}
\vspace{-1mm}
\resizebox{\hsize}{!}
   {
    \includegraphics{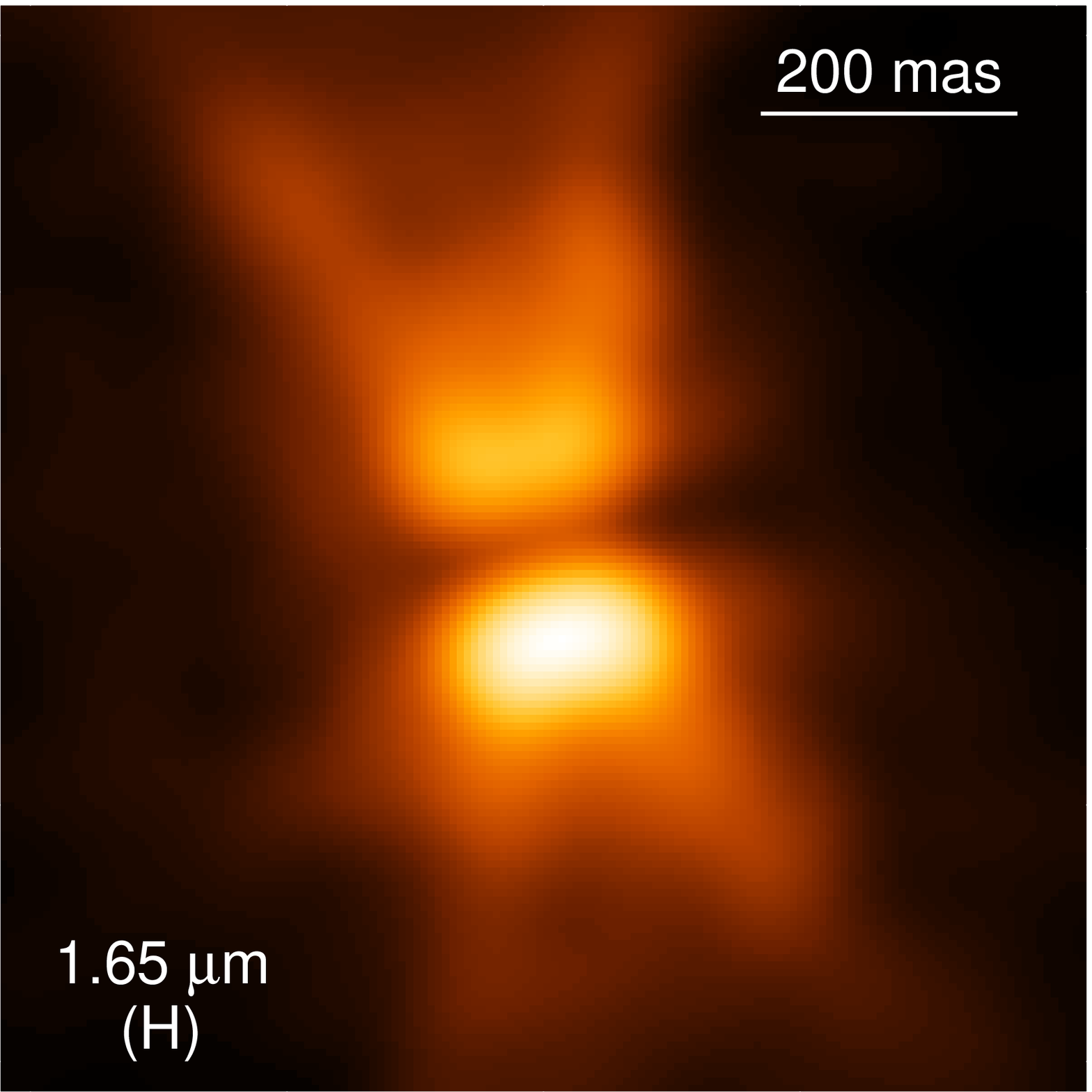}
    \hspace{-3mm}
    \includegraphics{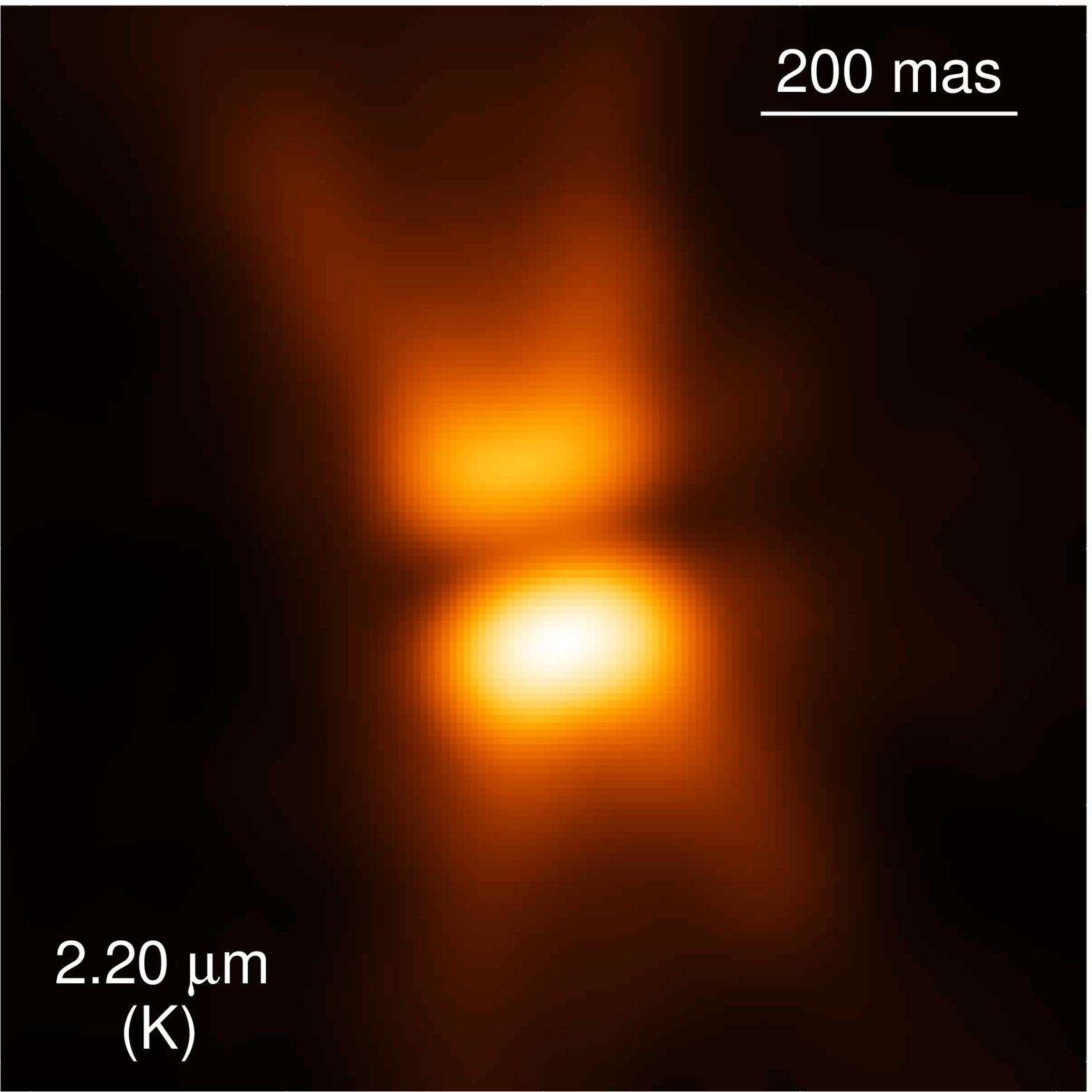}
    \hspace{-3mm}
    \includegraphics{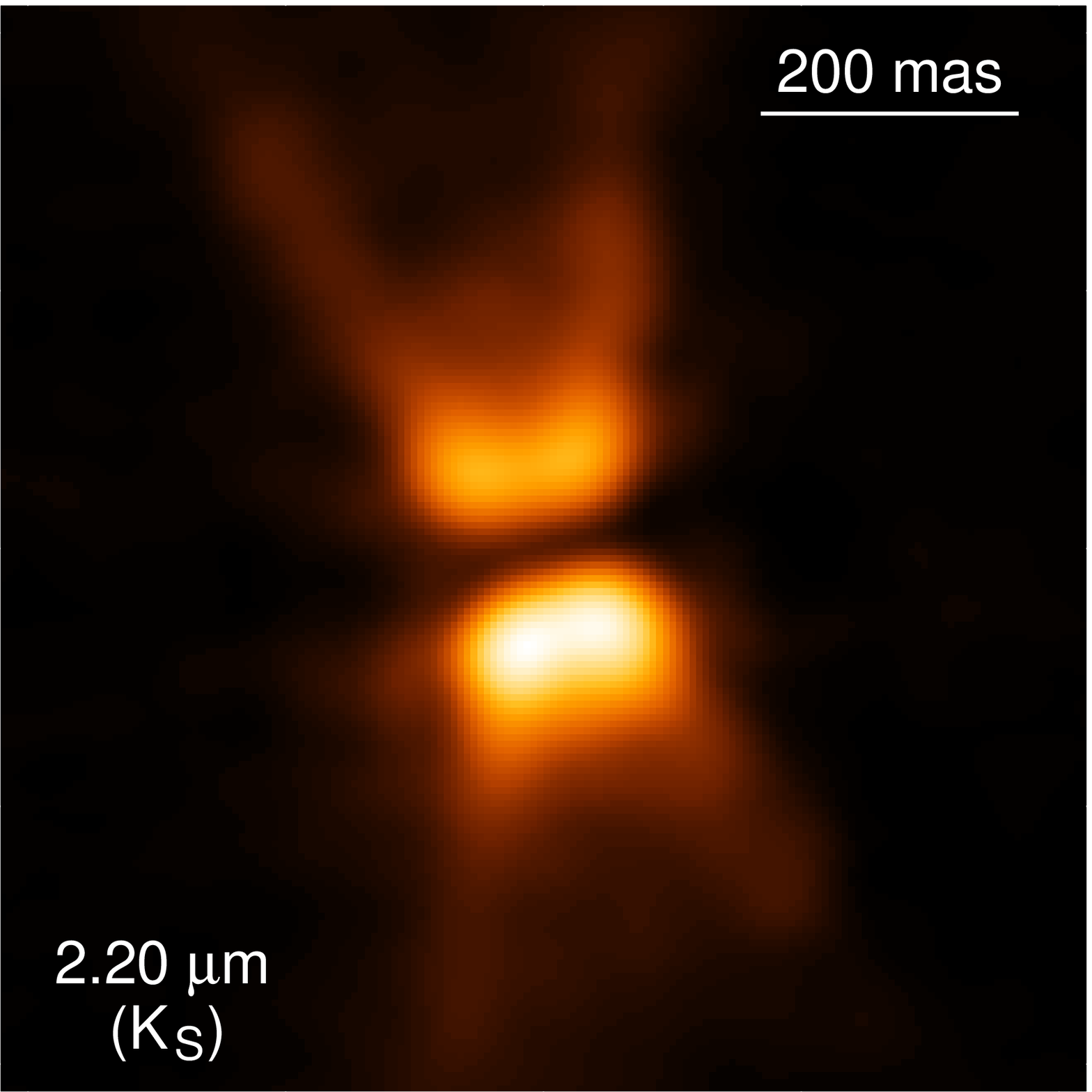}
    \hspace{-3mm}
    \includegraphics{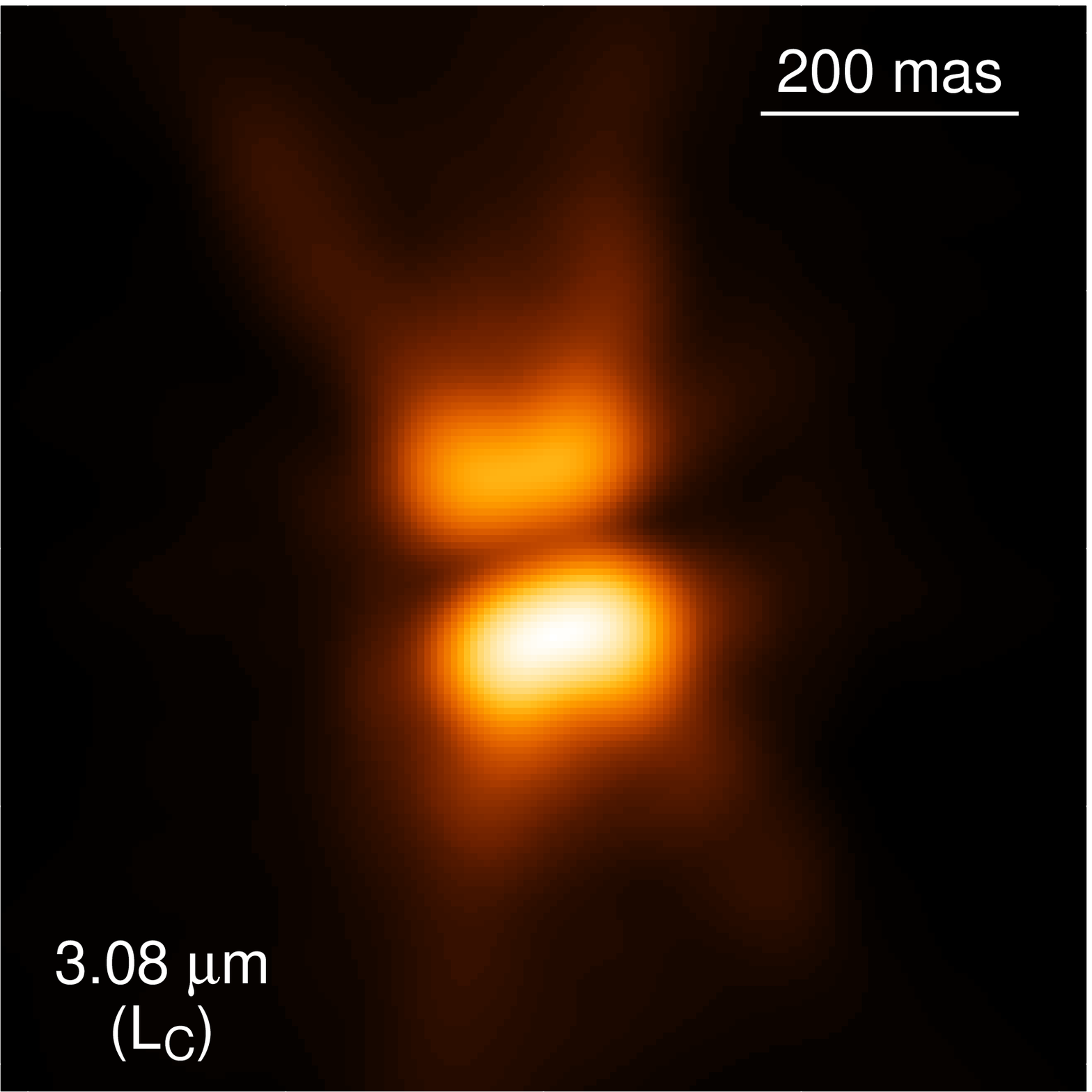}
   }
\resizebox{\hsize}{!}
   {
    \includegraphics{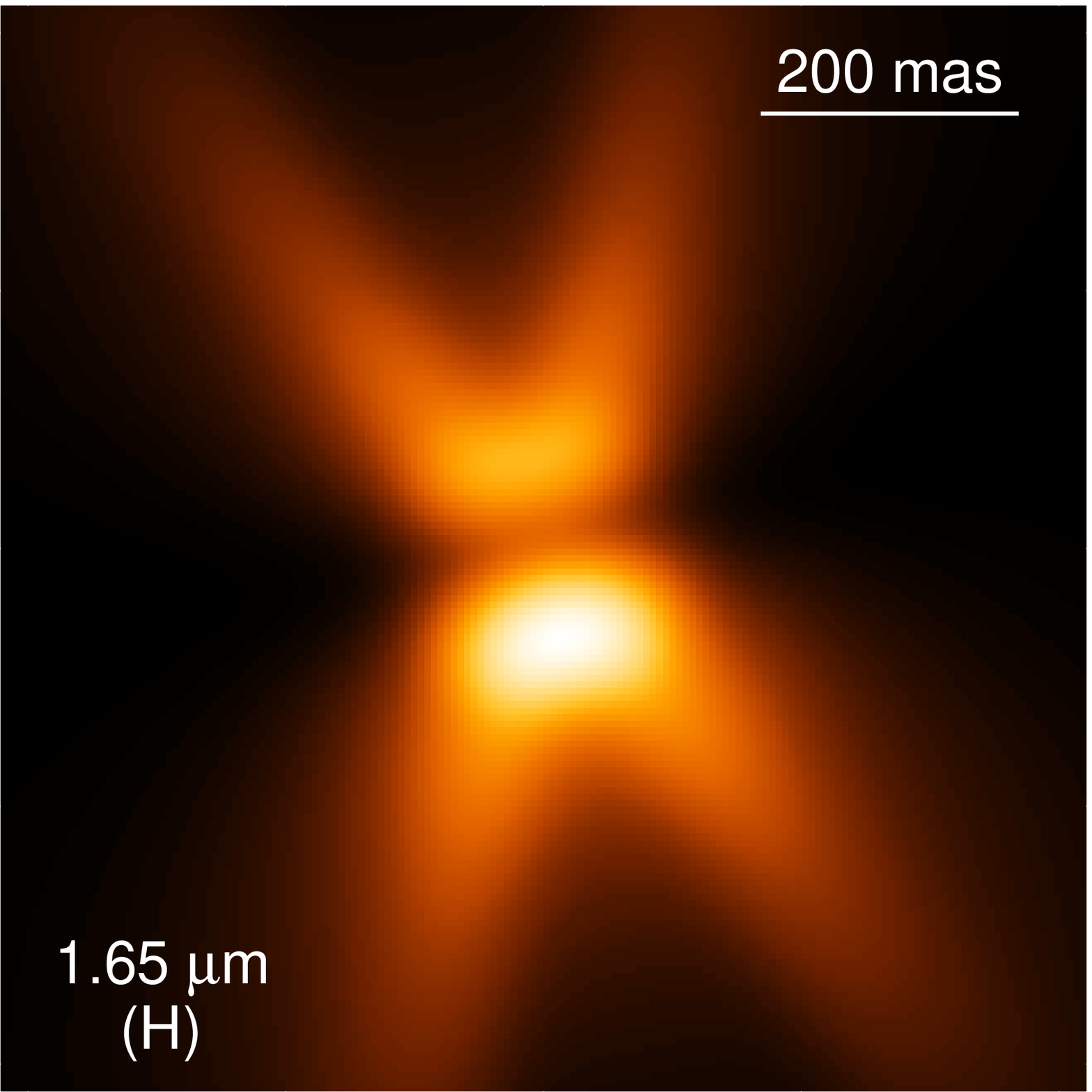}
    \hspace{-3mm}
    \includegraphics{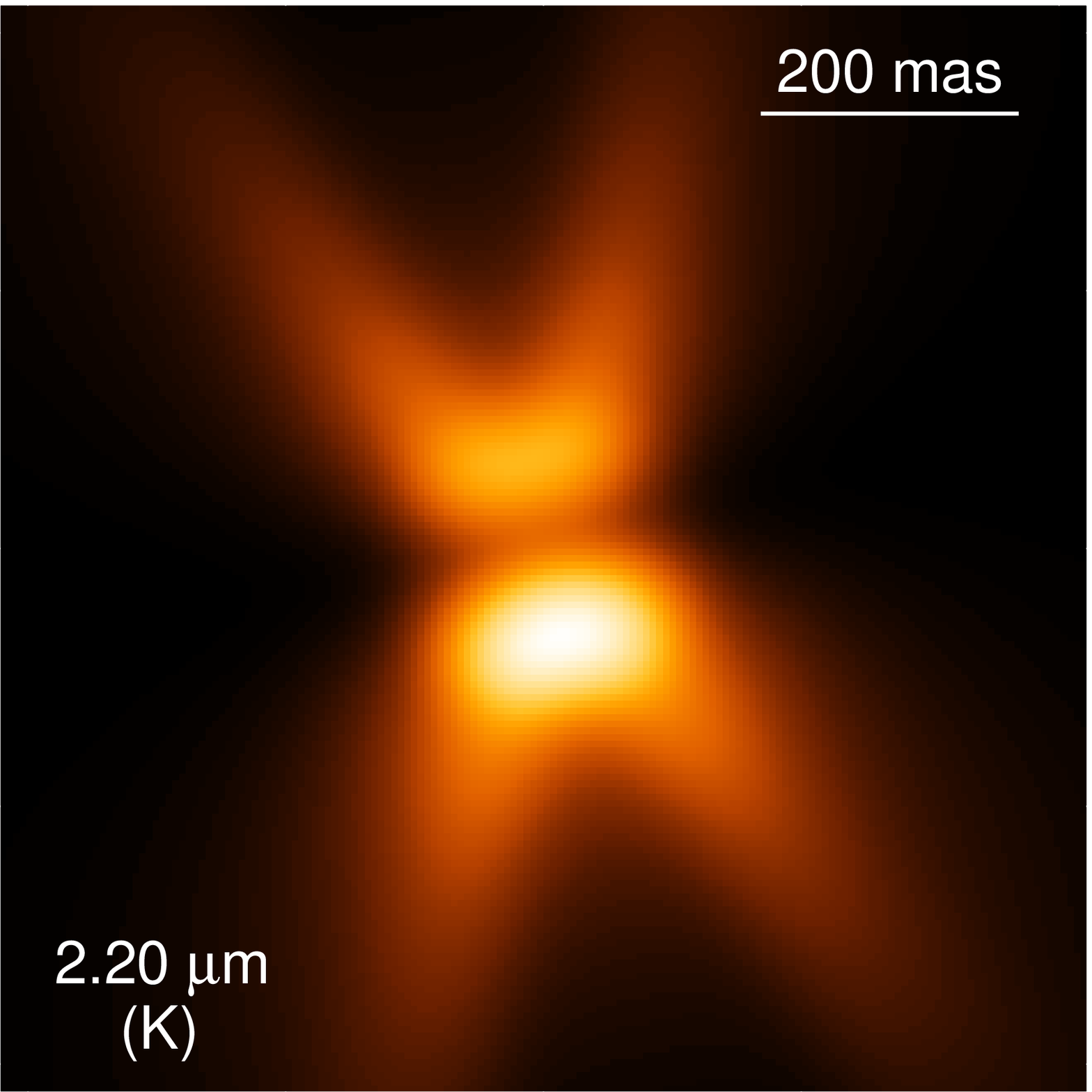}
    \hspace{-3mm}
    \includegraphics{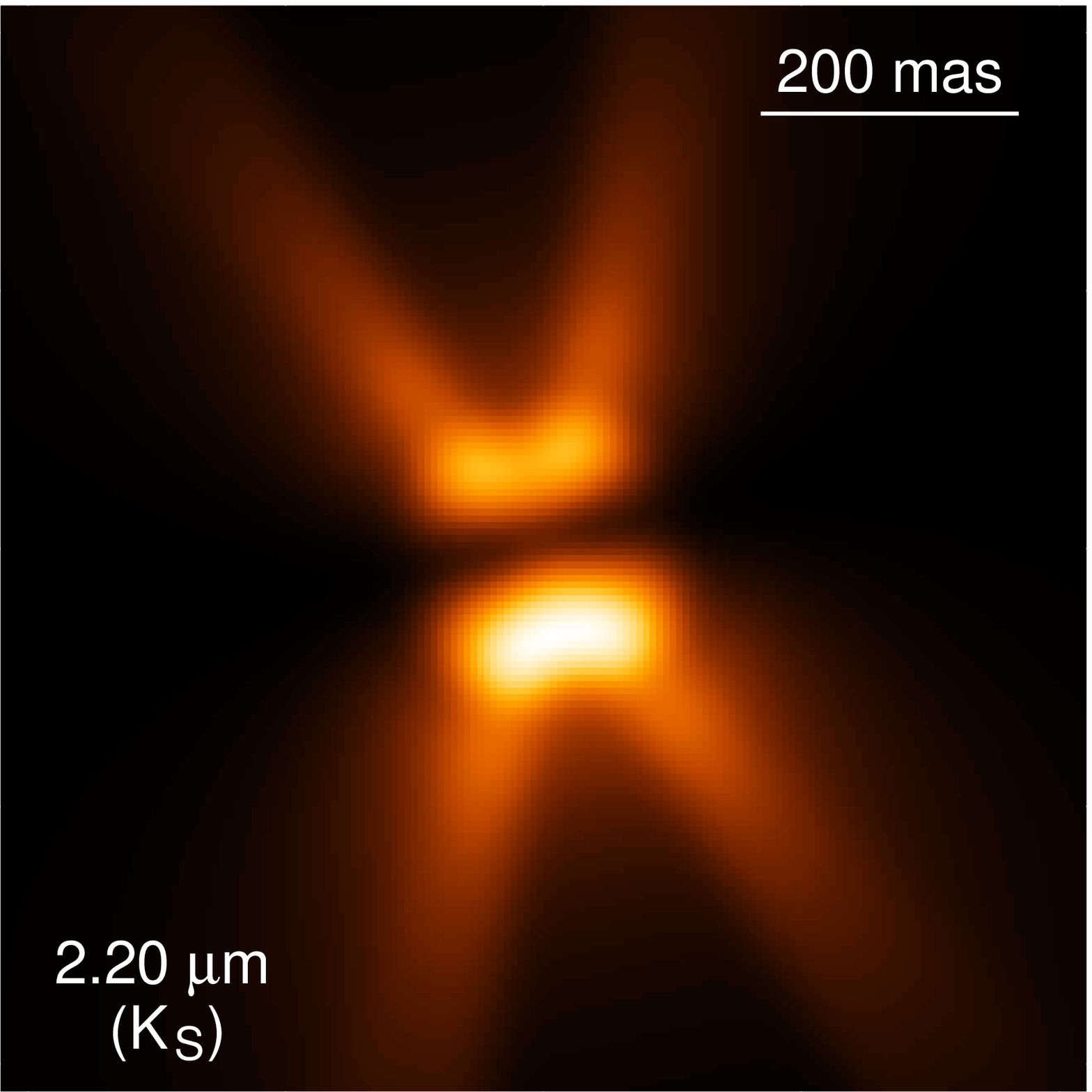}
    \hspace{-3mm}
    \includegraphics{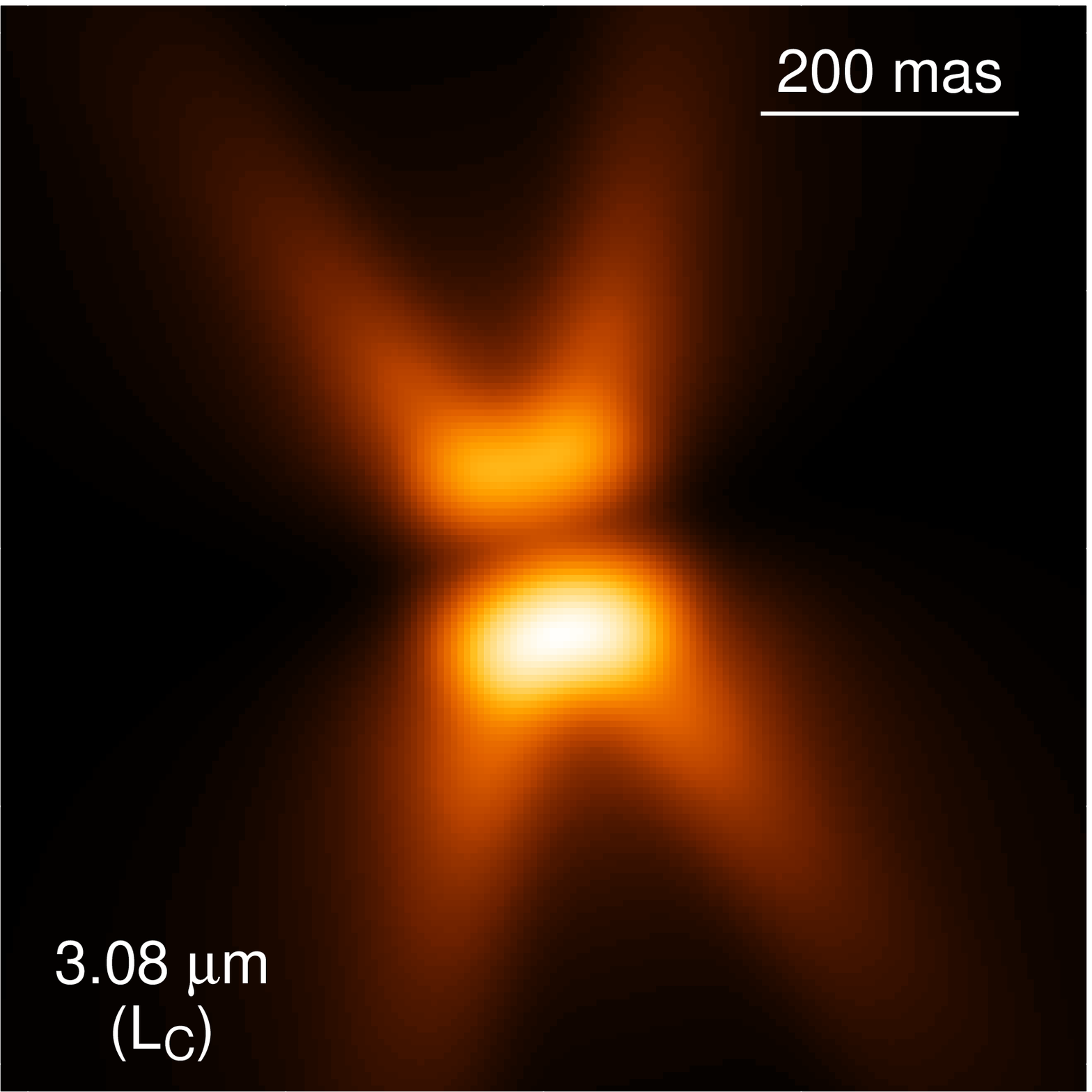}
   }
\caption
{
Comparison of the $H$, $K$ \citepalias[76\,mas
resolution,][]{Men'shchikov_etal1998} and {\Ks}, {\Lc} \citepalias[47 and 64\,mas
resolutions,][]{Tuthill_etal2002} speckle images of the {\rr} ({\em upper row})
with our model images at the same wavelengths ({\em lower row}). The model
images were convolved with Gaussian PSFs having full widths at half-maximum
equal to the resolutions of the corresponding speckle images. North is up and
east is to the left in all panels.
}
\label{Images1}

\resizebox{\hsize}{!}{\hspace{1mm}}
\vspace{-1mm}
\resizebox{\hsize}{!}
   {
    \includegraphics{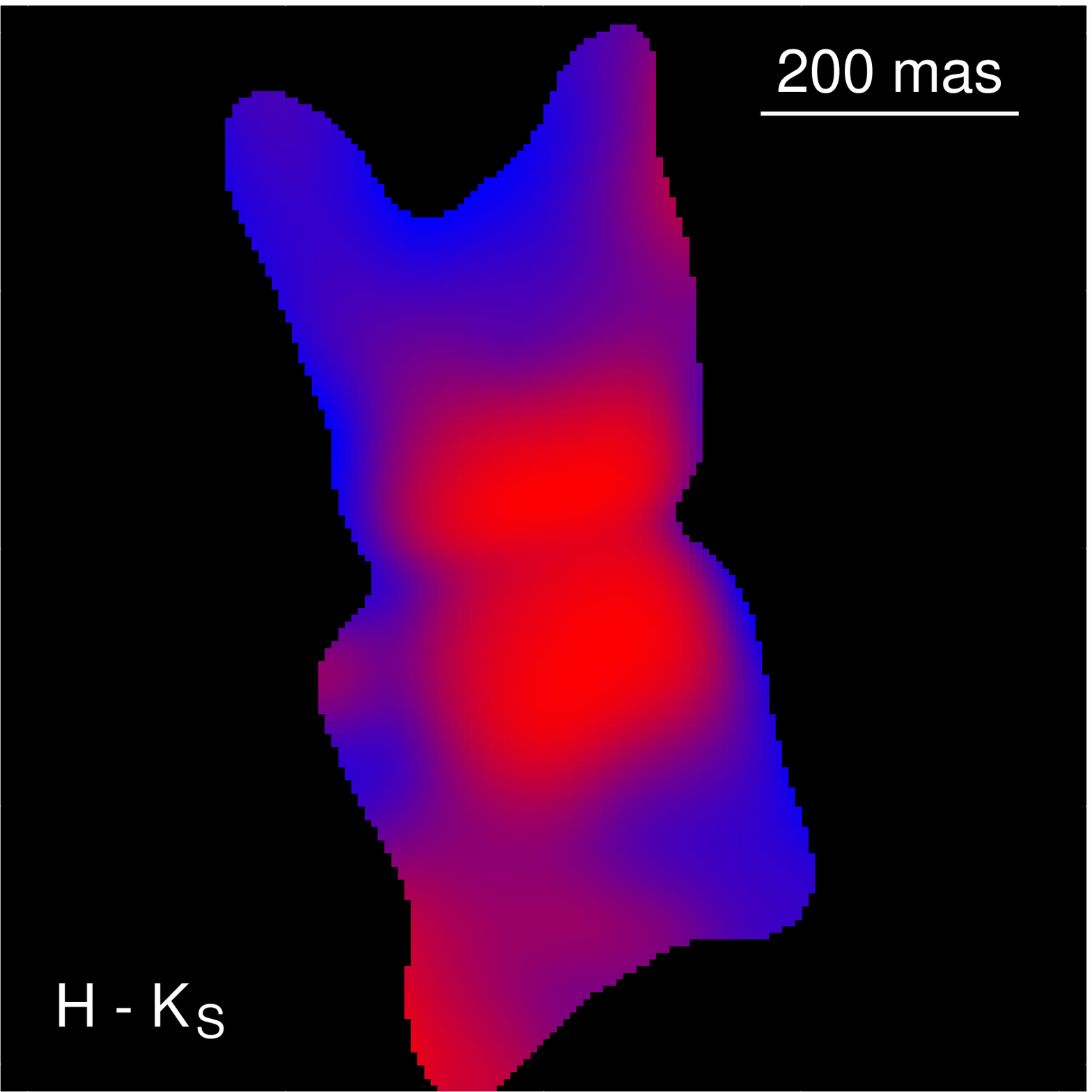}
    \hspace{-4mm}
    \includegraphics{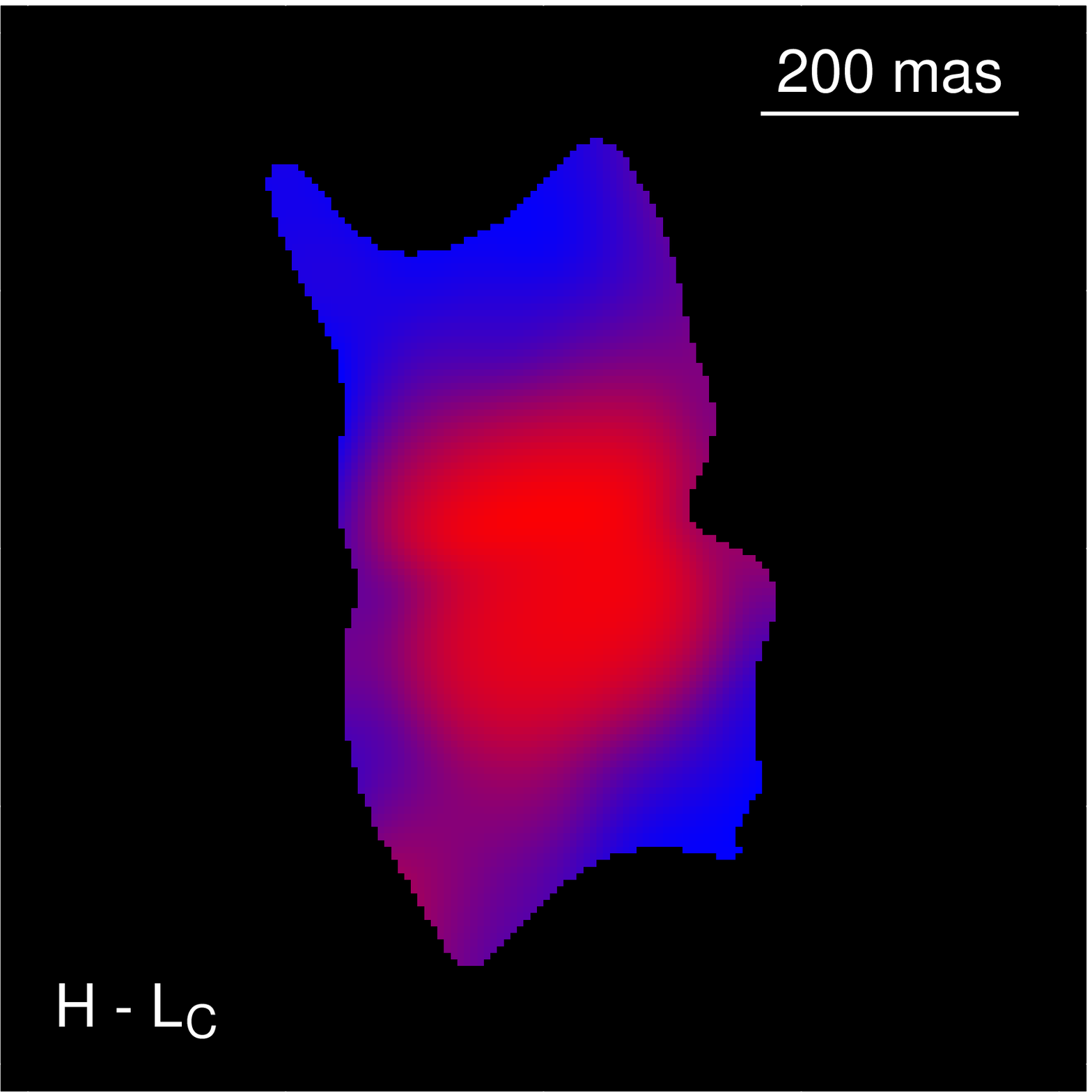}
    \hspace{-4mm}
    \includegraphics{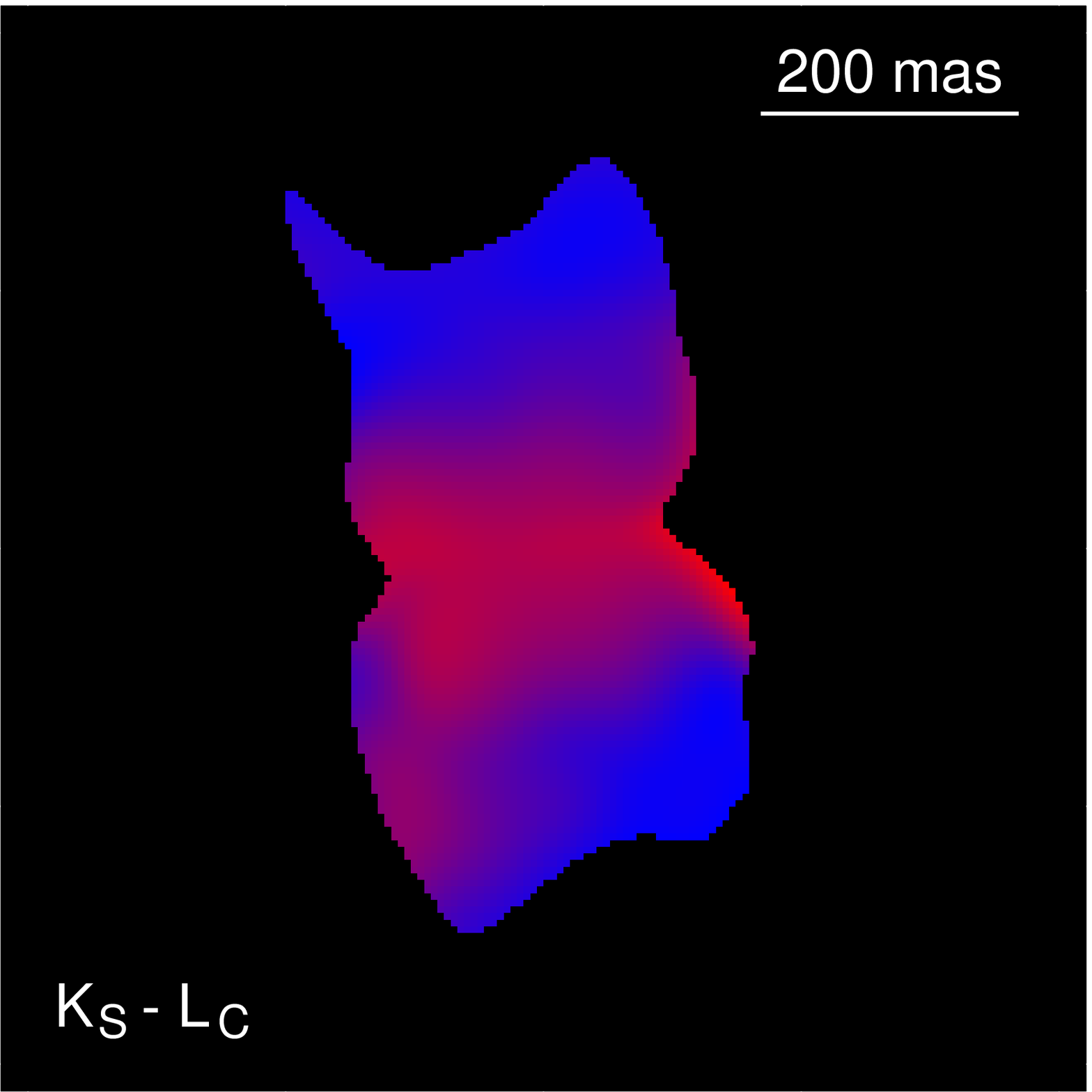}
    \hspace{-1mm}
    \includegraphics{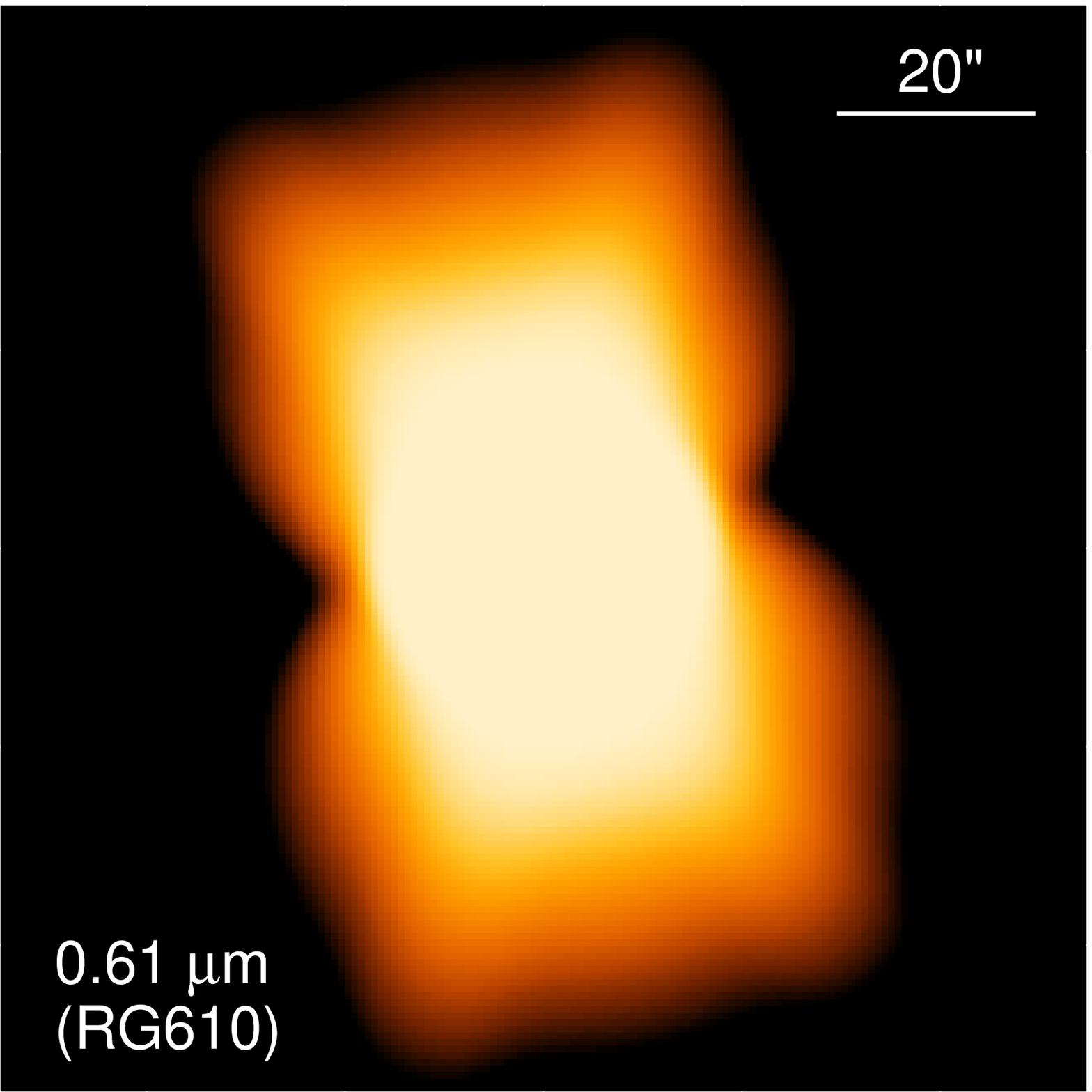}
   }
\resizebox{\hsize}{!}
   {
    \includegraphics{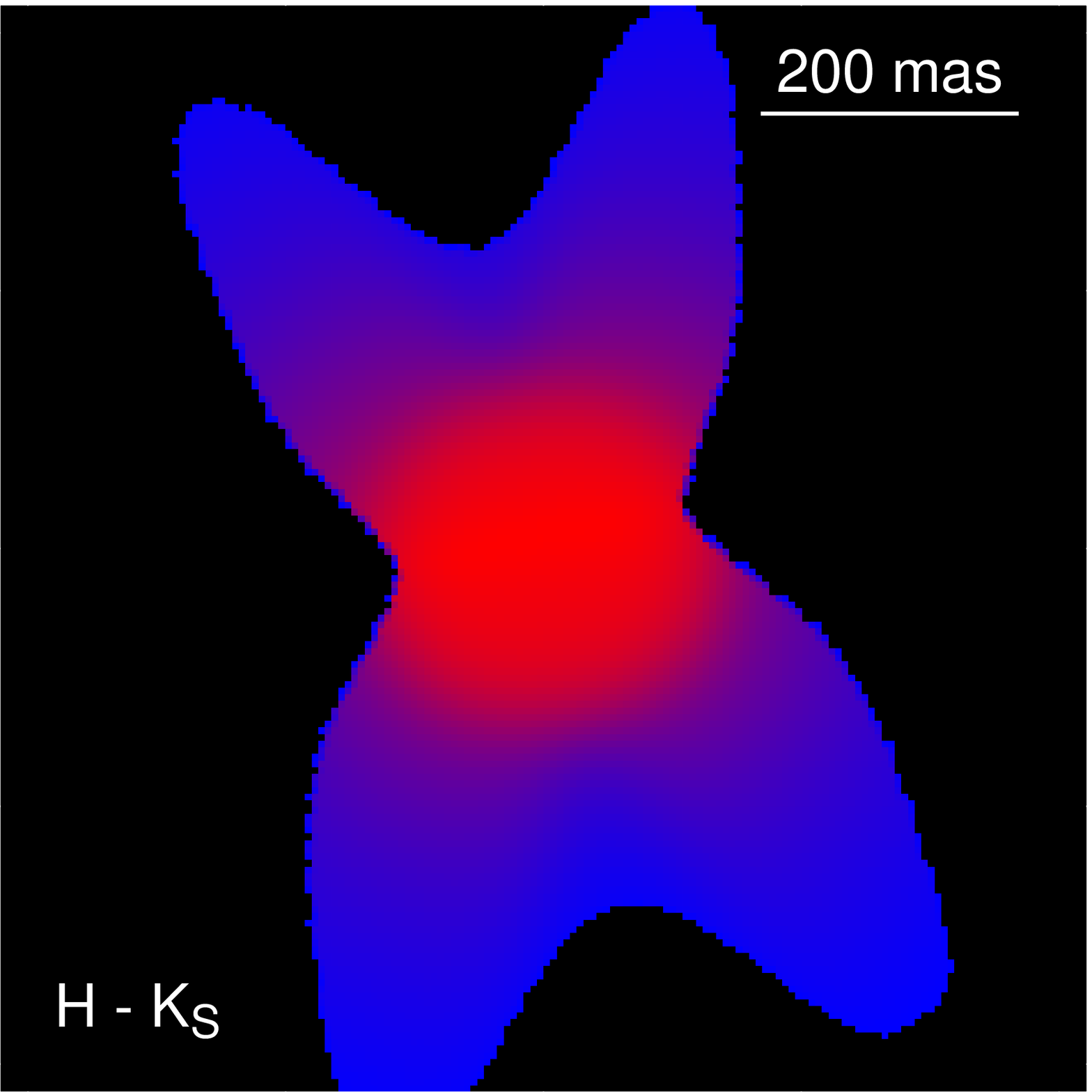}
    \hspace{-4mm}
    \includegraphics{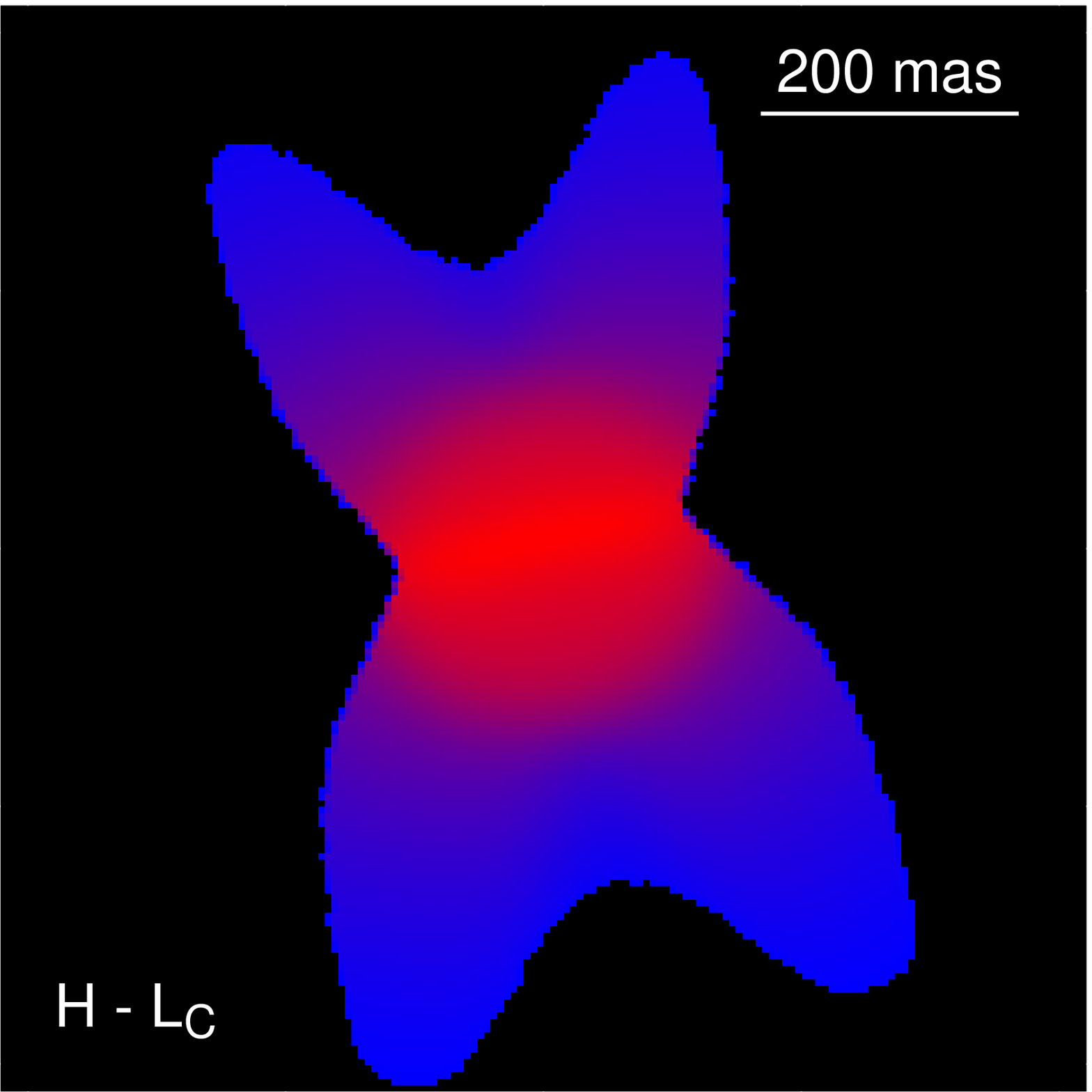}
    \hspace{-4mm}
    \includegraphics{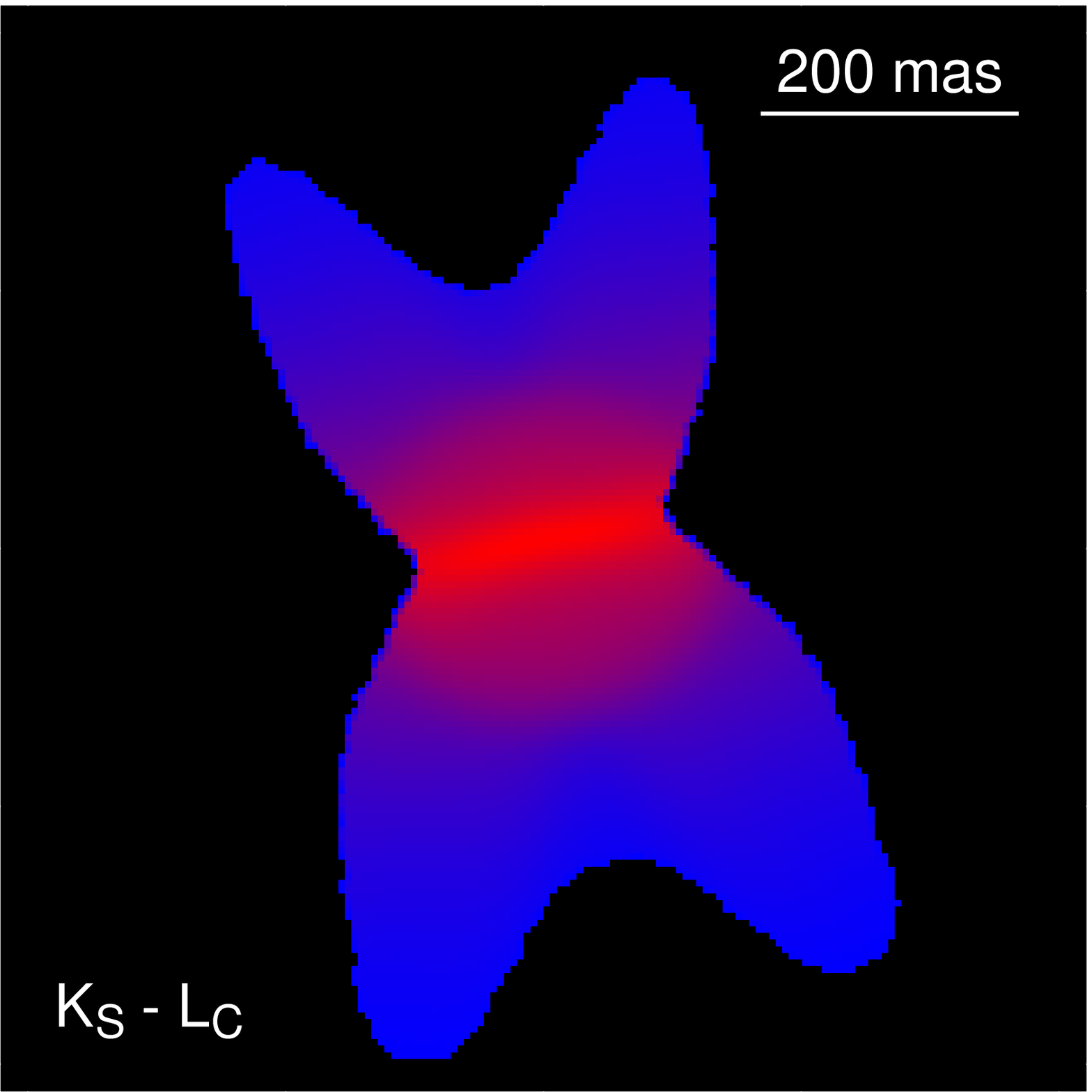}
    \hspace{-1mm}
    \includegraphics{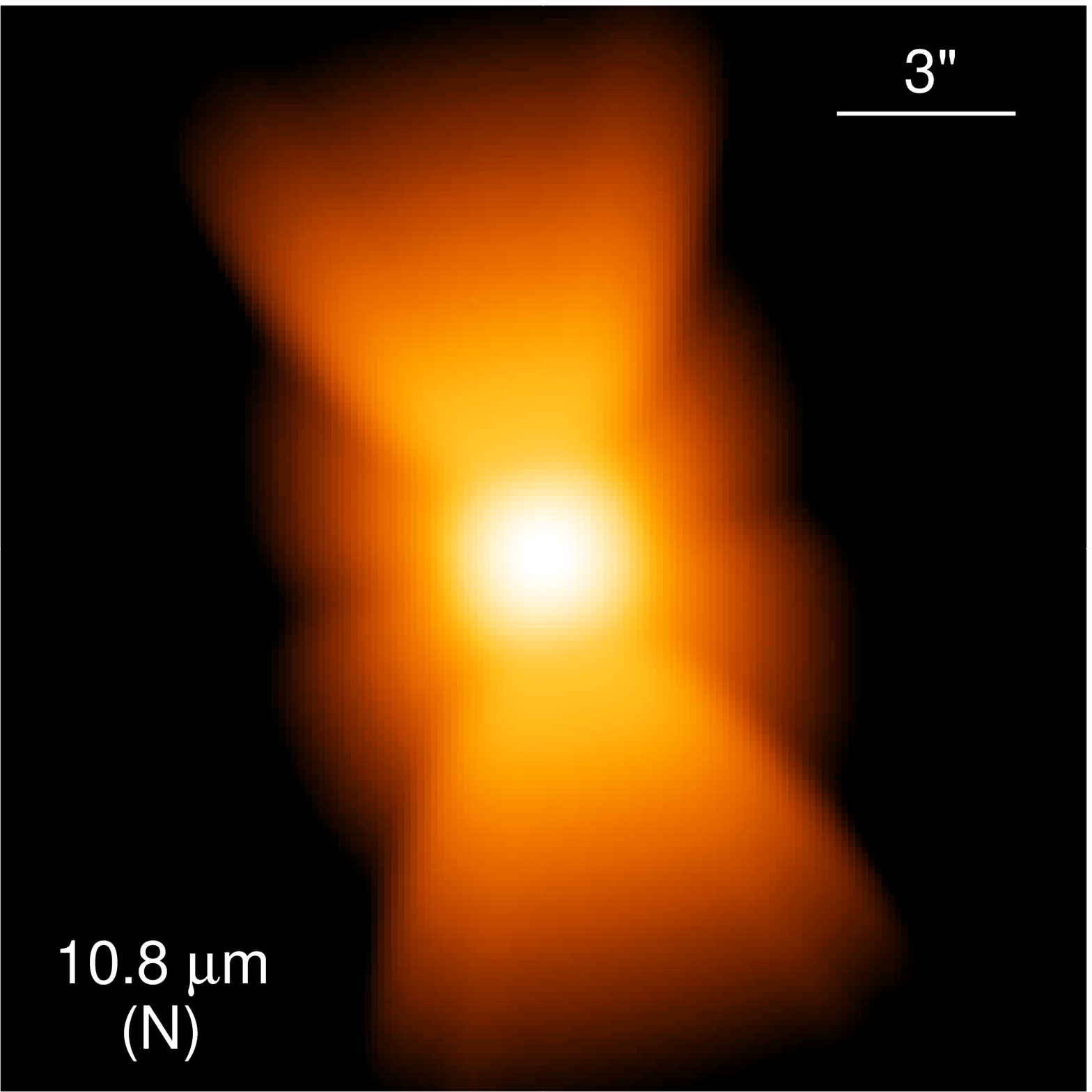}
   }
\caption
{
Comparison of the {\HmK} (76\,mas resolution), {\HmL} (76\,mas), and {\KmL}
(68\,mas) color images of the {\rr} ({\em 3 upper left panels}) with the
corresponding model images ({\em 3 lower left panels}). The color images were
computed from only those parts of the respective images, where intensities are
higher than 5\,{\%} of the peak. Model image of the {\rr} at 0.61\,{\mic} ({\em
upper right panel}) can be compared to a photograph in \cite{Cohen_etal1975}
and to an H$_{\alpha}$ image in \cite{VanWinckel2001}. Model image at
10\,{\mic} ({\em lower right panel}) convolved with a PSF of 0{\farcs}7 can be
compared to an $N$ image in \cite{Waters_etal1998}. North is up and east is to
the left in all panels.
}
\label{Colors}
\end{center}
\end{figure*}


\subsection{Near-infrared images}
\label{Images}

In Fig.~\ref{Images1}, our model images are compared side by side with our
high-resolution images of the {\rr} in the $H$, $K$, {\Ks}, and {\Lc} bands
\citepalias{Men'shchikov_etal1998,Tuthill_etal2002}. The model intensity
distributions at these wavelengths were convolved with circular Gaussian
point-spread functions (PSF) of half-maximum widths of 47--76\,mas
corresponding to the resolutions of the reconstructed speckle images. In
Fig.~\ref{Colors}, we also present the model images at 0.61\,{\mic} and
10\,{\mic}, which can be qualitatively compared to an optical RG\,610
photograph \citep{Cohen_etal1975}, a deep coronagraphic H$_{\alpha}$ image
\citep{VanWinckel2001}, and an $N$-band image \citep{Waters_etal1998}. The
effects of saturation from long exposures in the optical images was roughly
simulated by convolving the model intensity distribution with a circular PSF of
8{\arcsec}, whereas the 10\,{\mic} model image was convolved with a PSF of
0{\farcs}7 which takes into account probable seeing conditions during the
observations. Missing details on the PSFs and intensity scales in the papers
mentioned above prevented us from using the images as constraints in the
modeling and from making more accurate, quantitative comparisons.

Two well-resolved bright lobes above and below the inclined torus of the {\rr}
are seen in the observed images of Fig.~\ref{Images1}, along with the {\sf
X}-shaped spikes originating deep inside the biconical outflow cavities. The
spikes are bright enough to contribute to the intensity distribution of the
bright lobes (outflow cavities), making them appear broadened and even
double-peaked in the highest-resolution {\Ks} image (see also
Sect.~\ref{IntensProf}). The {\sf X}-shaped spikes are caused most likely by
a combination of the limb brightening and scattering by dust in the dense
boundaries of the biconical outflow cavities, as first indicated by
polarization maps \citep{Perkins_etal1981}. Unidentified optical emission
bands, extended red emission, and the 3.3\,{\mic} PAH emission line also
originate close to the biconical surfaces along the {\sf X} spikes
\citep{Kerr_etal1999}.

Extending our previous modeling of \citetalias{Men'shchikov_etal1998}, in
this work we attempted to simulate the {\sf X} spikes appearing in the conical
walls of the outflow cavities along the planes tangential to their surfaces.
To accurately and consistently model the spikes, one needs a more sophisticated
radiative transfer code than we have; moreover, the properties of dust and
other species producing the {\sf X} shape are very poorly known. For these
reasons, we did not try to incorporate the spikes in our 2D code in a
self-consistent way as part of the radiative transfer iterations; instead, we
decided to simulate the spikes for an illustrative purpose when computing the
observable model fluxes. The spikes in the model images were produced by
artificially increasing the scattering term of the source function in the
cavities by 20\,{\%} between the plane tangential to the cavity surface and
another plane closer by 5{\degr} to the symmetry axis.

The spikes are clearly seen in all images of Fig.~\ref{Images1}, both observed
and modeled, traced along the surface of the biconical outflow cavities to
within the distances of 0{\farcs}4 from the center. It is clear that the length
of the spikes depends on the density distribution and dust properties, as
well as on the sensitivity and resolution of observations. A higher
signal-to-noise composite near-IR image of the {\rr}
\citepalias{Tuthill_etal2002} traces them to within 0{\farcs}8 from the central
binary. The spikes can be approximated by straight lines passing through the
east and west peaks of the split northern and southern lobes and intersecting
at the position of the central binary. The opening angle of the conical outflow
cavities, measured between the spikes, is ${{\omega}\approx\,}$50{\degr}.

The near-IR images of the {\rr} (Fig.~\ref{Images1}) clearly show the bright
lobes which our model associates with a geometrically and optically thick torus
inclined toward us in the south by ${\sim\,}$11{\degr}. The images exhibit a
compact, highly symmetric brightness distribution, very similar in shape and
extent between the $H$ and {\Lpah} bands. As is becoming more apparent with
highest resolution of the {\Ks} image, the bright lobes are divided by a very
dark narrow lane, having a similar width in all the images. Even at the longest
wavelengths there are no signs of either the direct light from the central
binary, or an increasing contribution of hot emission ($T_{\rm
d}{\,\sim\,}$1000\,K) from the inner dust boundary of the torus. As discussed
also in \citetalias{Tuthill_etal2002}, such an appearance suggests a strong
concentration of circumstellar matter toward the central binary in an
axially-symmetric density distribution and a high gray optical depth due to
large dust particles (Sect.~\ref{Dust}).

\subsection{Near-infrared color images}
\label{ColorImages}

High-resolution color images provide a sensitive measure of the relative
intensity distribution at selected wavelengths and a comparison of the observed
and model color images gives a better idea of how closely the model resembles
reality. Such a comparison is given in Fig.~\ref{Colors}, where we display
side by side the {\HmK}, {\HmL}, and {\KmL} color images of the {\rr} with
the corresponding model images at the same wavelengths. The corresponding
near-IR speckle images (with resolutions of 68--76\,mas) used to produce the
color images were presented in \citetalias{Men'shchikov_etal1998} and
\citetalias{Tuthill_etal2002}.

The images in Fig.~\ref{Colors} exhibit a broad, relatively flat red plateau
roughly 200\,mas in size, covering the entire region of the bright
lobes and the dark lane between them. The model color distributions in the
near-IR bands are clearly consistent with observations. Quantitatively, the
reddest model colors of 3.0 mag, 5.4 mag, and 2.5 mag (in {\HmK}, {\HmL},
and {\KmL}) are similar to those of 2.0 mag, 4.5 mag, and 2.8 mag in the
observed images. As the color images were computed from only those parts of the
respective $H$, $K$, {\Ks}, and {\Lc} images, where intensities are higher than
5\,{\%} of the peaks, the shapes of the outer edges of the color distributions
essentially outline the brightest parts of the images. The blue {\sf X}-shaped
spikes are also clearly delineated in the color images. The red plateau marks
the position of the very dense, optically thick, inclined torus. The bluer
colors outside the dense torus indicate optically thin regions in and around
the outflow cavities which scatter the hot radiation of the central stars
toward an observer.

\begin{figure*}
\begin{center}
\resizebox{0.945\hsize}{!}
   {
    \includegraphics{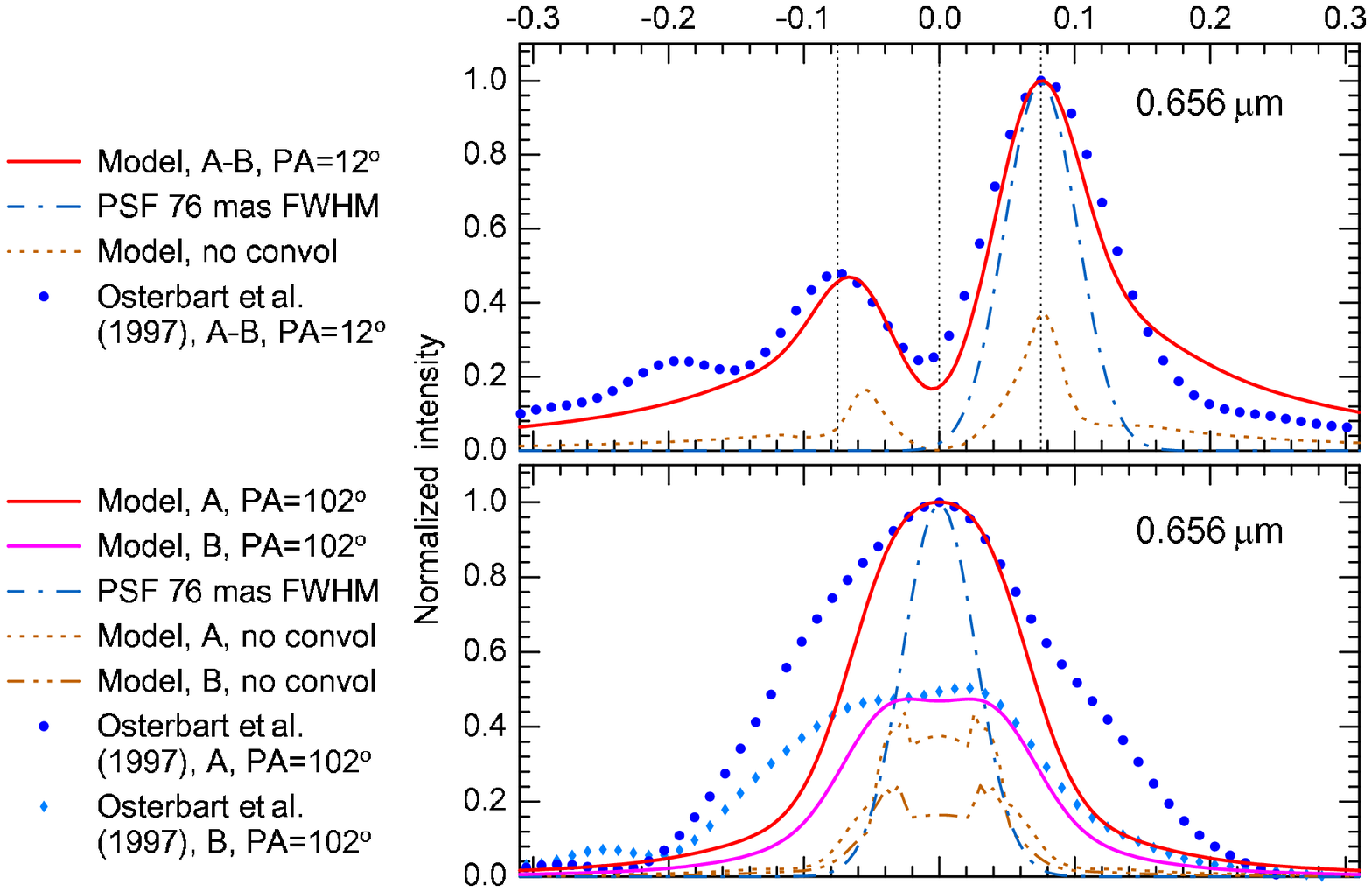}
    \hspace{-3mm}
    \includegraphics{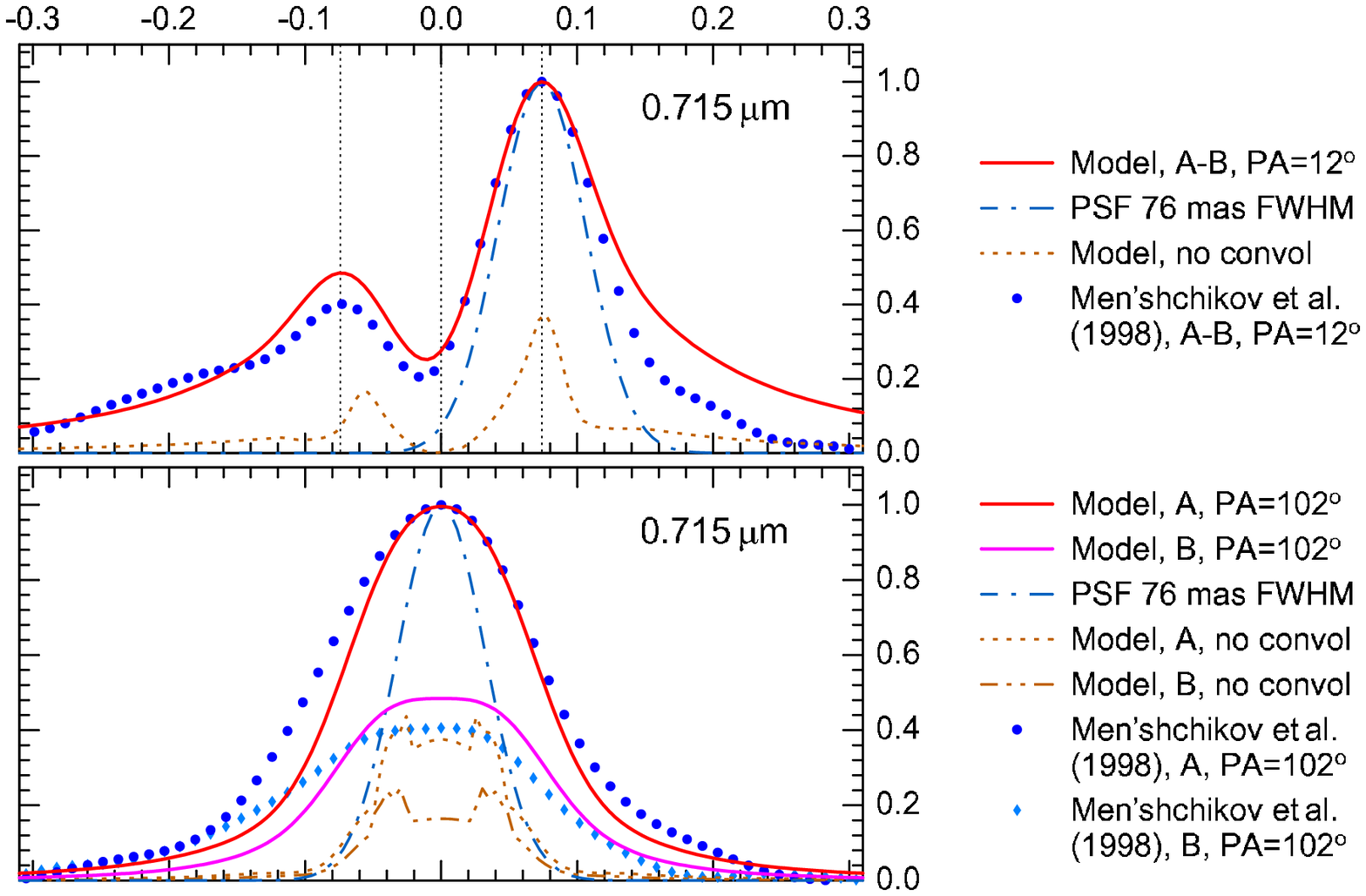}
   }
\resizebox{0.945\hsize}{!}
   {
    \includegraphics{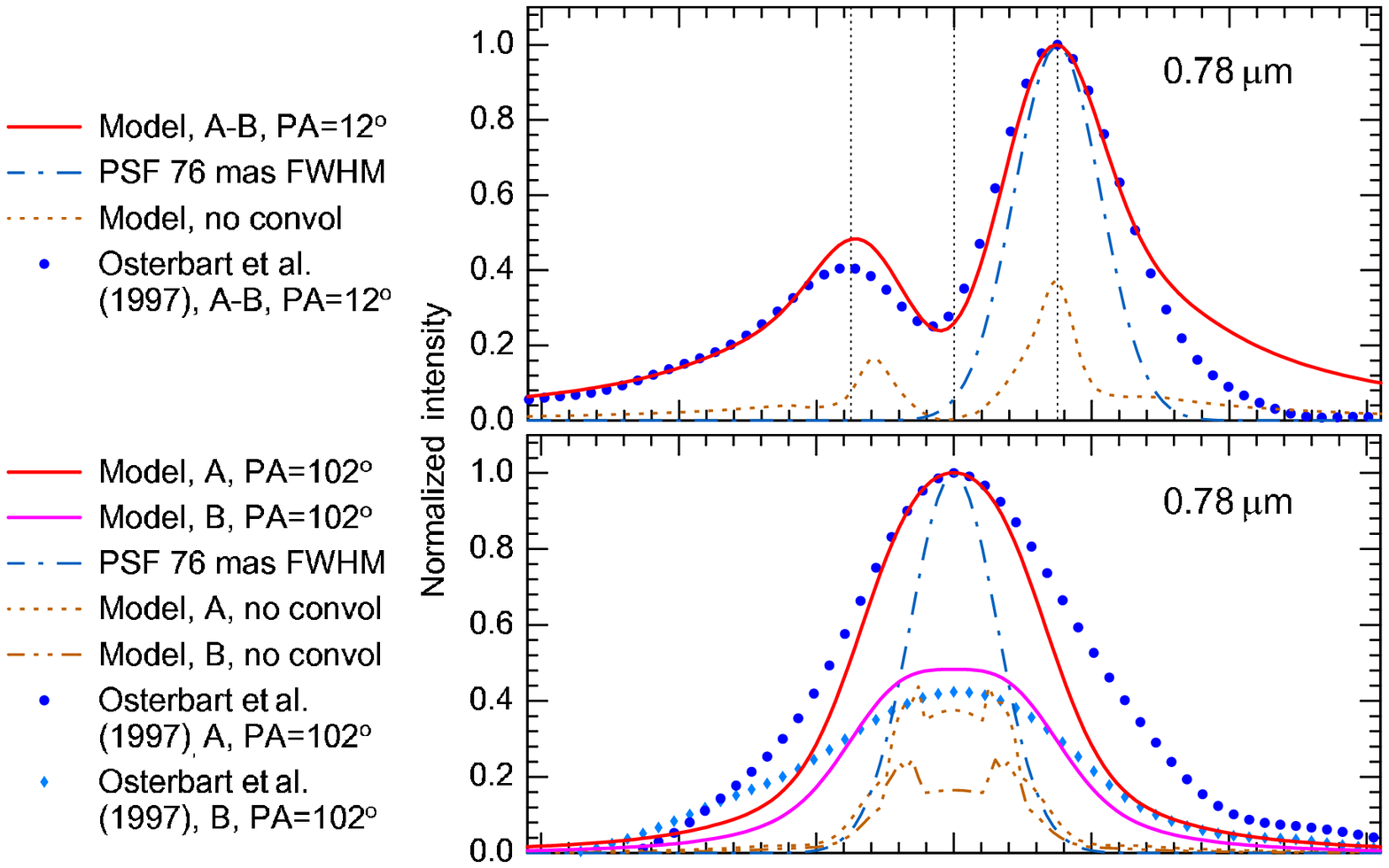}
    \hspace{2.25mm}
    \includegraphics{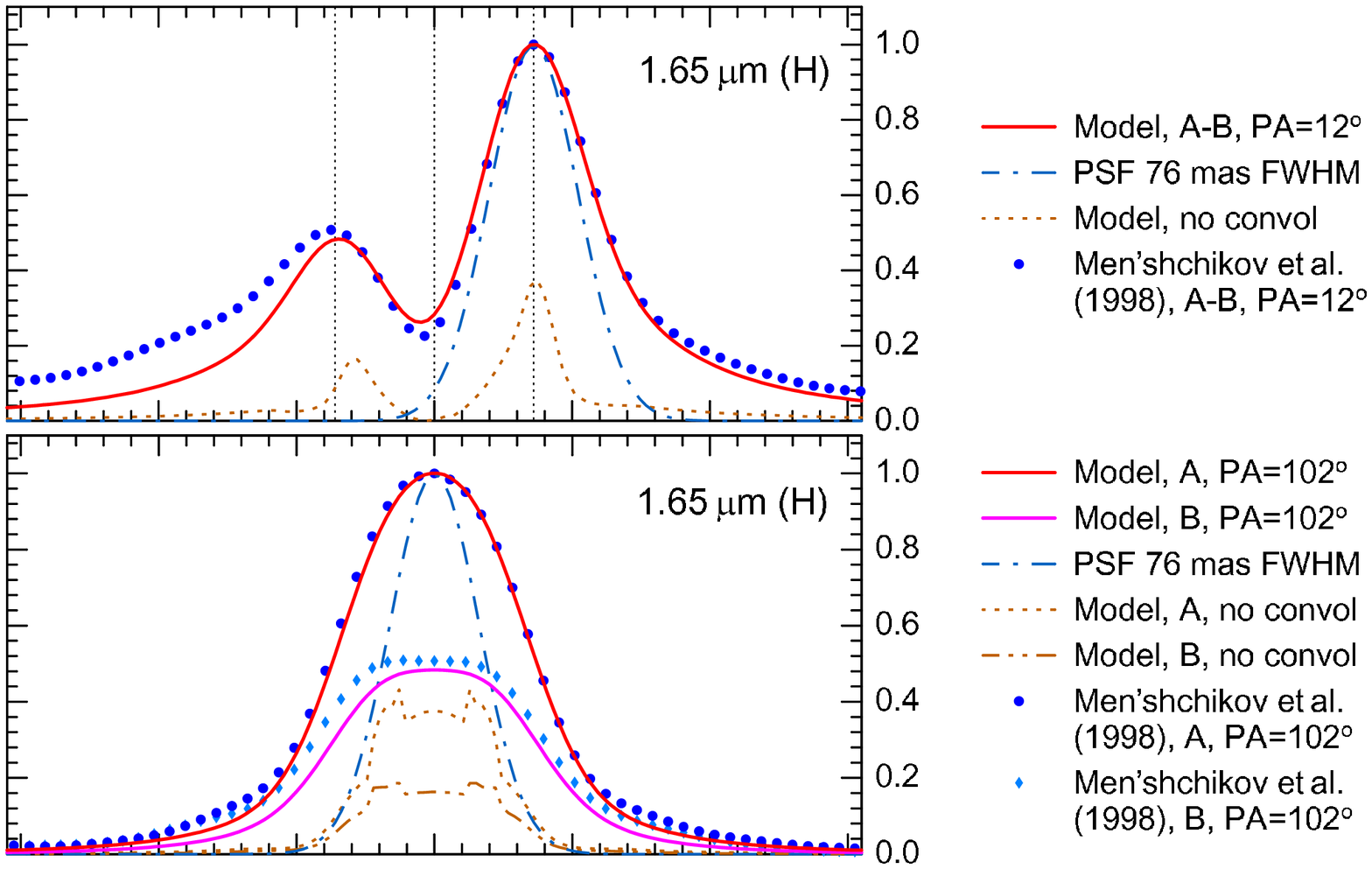}
   }
\resizebox{0.945\hsize}{!}
   {
    \includegraphics{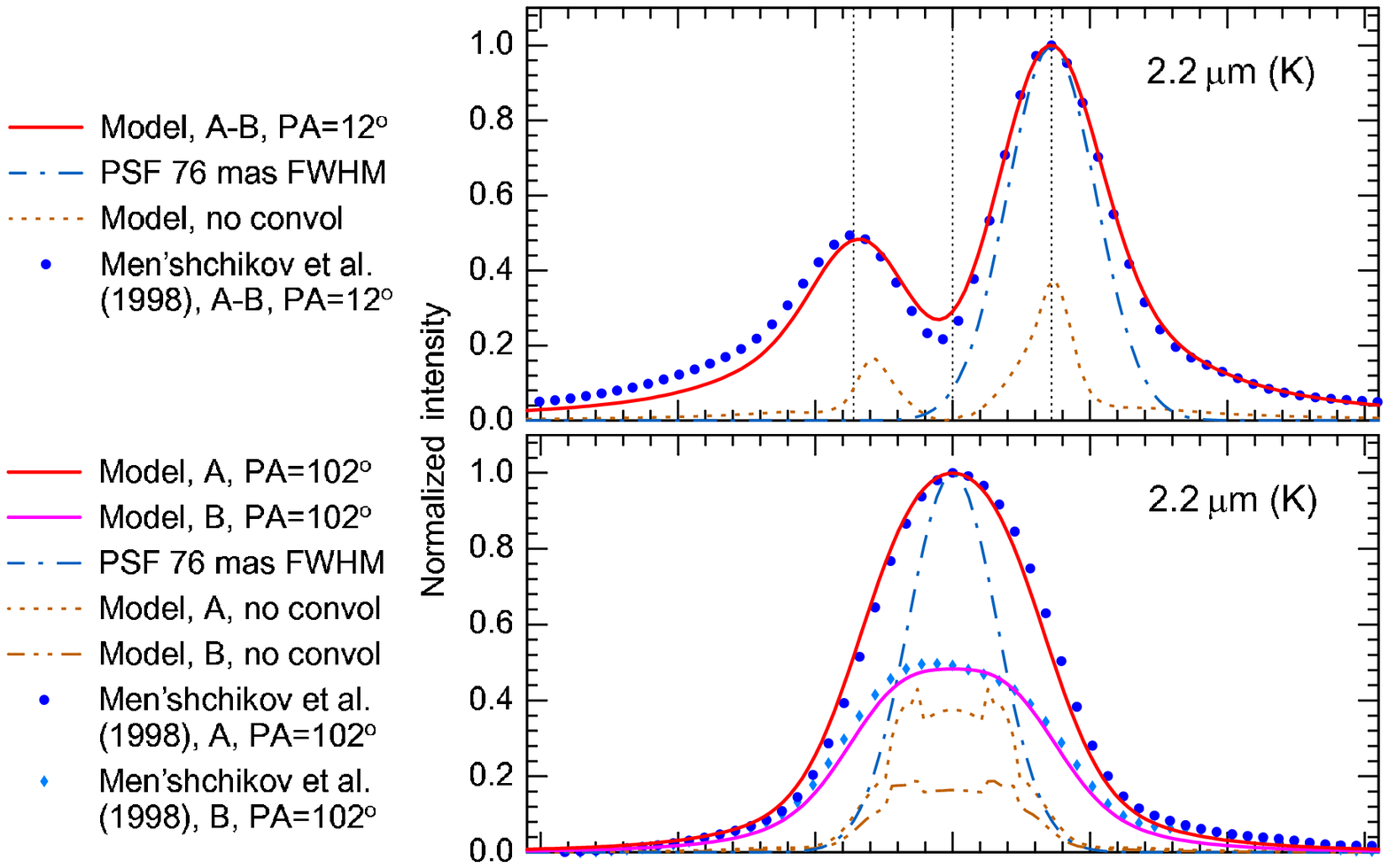}
    \hspace{2.15mm}
    \includegraphics{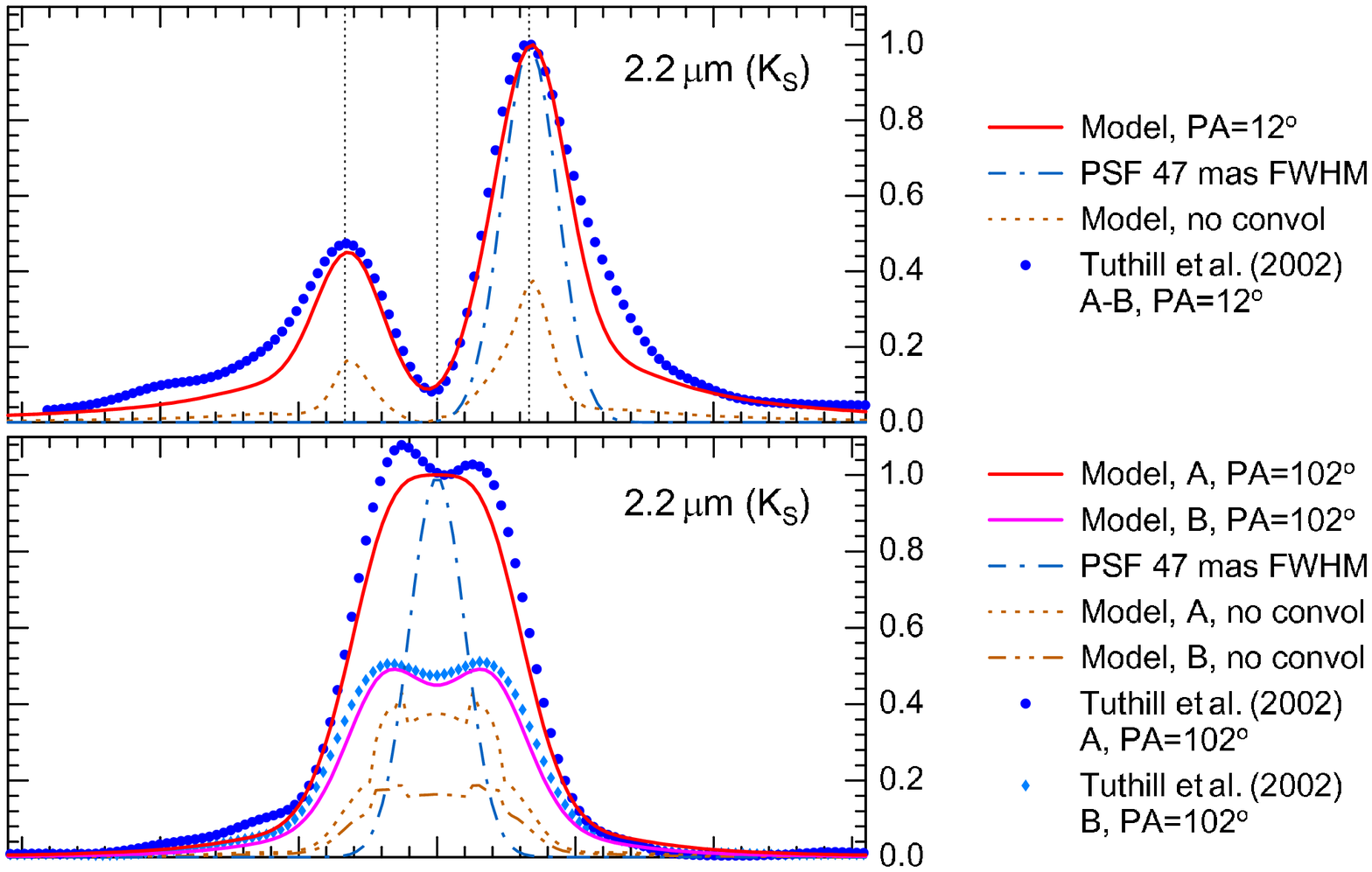}
   }
\resizebox{0.945\hsize}{!}
   {
    \includegraphics{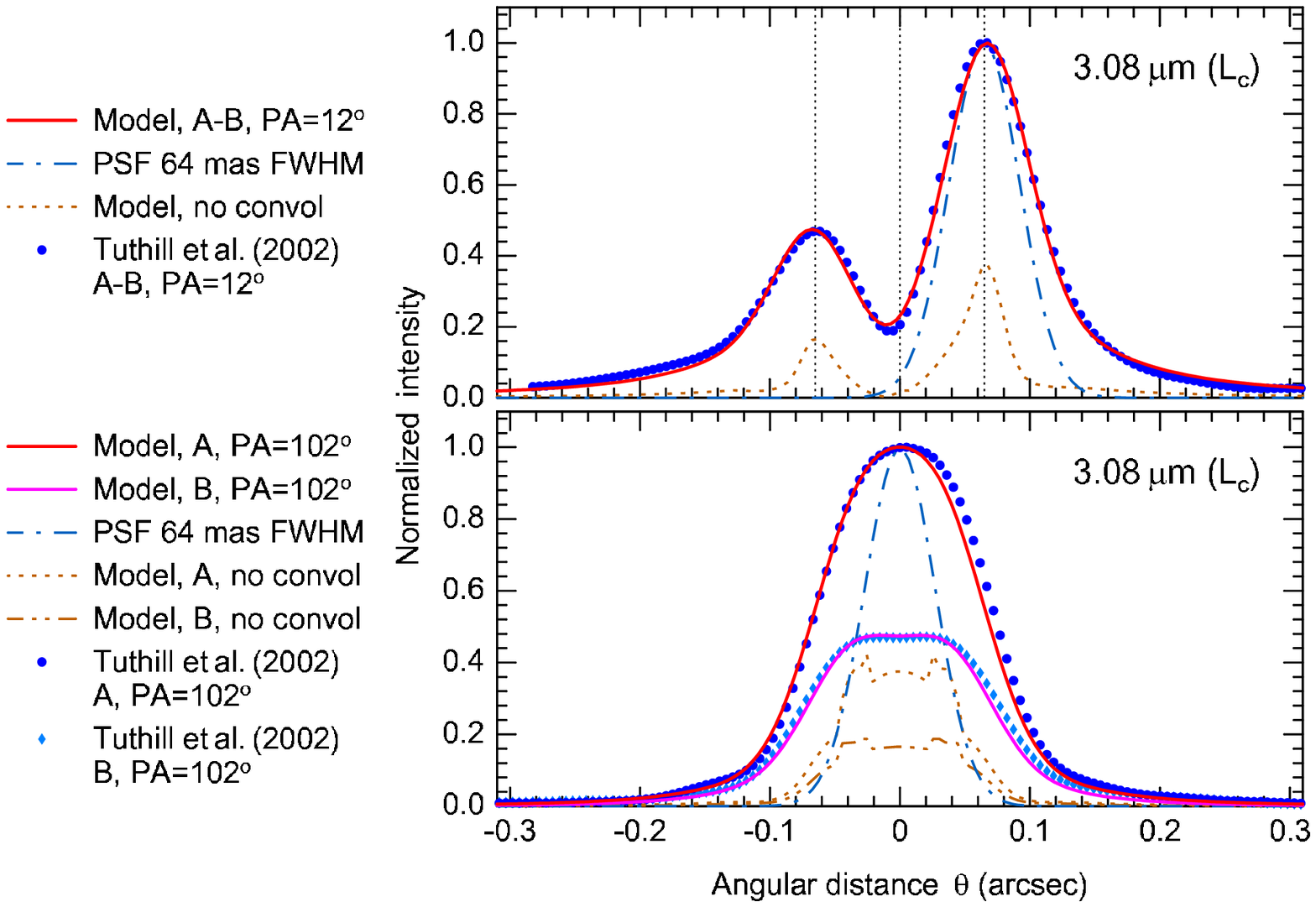}
    \hspace{-3mm}
    \includegraphics{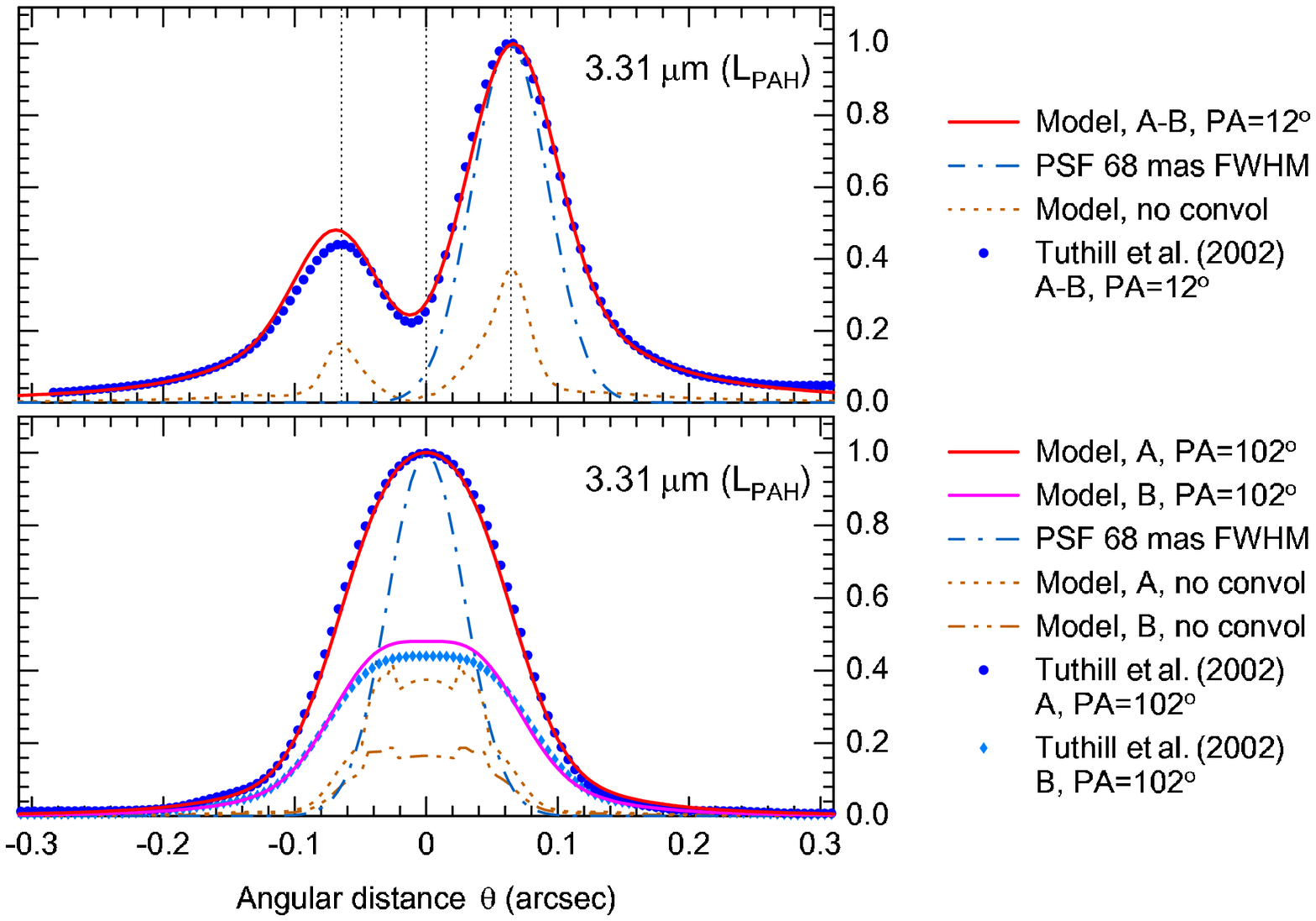}
   }
\caption
{
Comparison of the model intensity profiles with those from our optical and
near-IR speckle images of the {\rr}. Also shown are the profiles of the
point-spread functions and true (unconvolved) model intensities. Vertical
dotted lines enable easier comparison of the relative offsets of the bright
peaks (75\,mas in the visible, 72\,mas in $H$ and $K$, 64.5\,mas in {\Ks},
65\,mas in {\Lc}, and 66.5\,mas in {\Lpah}) from the center of the nebula.
}
\label{Intens1}
\end{center}
\end{figure*}


\subsection{Intensity profiles}
\label{IntensProf}

Figure~\ref{Intens1} presents a quantitative comparison of the model images
with available high-resolution optical and near-IR images of the {\rr} in terms
of the intensity profiles in two directions, parallel (PA = 12{\degr}) and
perpendicular (PA = 102{\degr}) to the projected symmetry axis of the nebula.
The profiles (intensity cuts) were taken from the images of
\citetalias{Osterbart_etal1997} (RG\,656, RG\,780),
\citetalias{Men'shchikov_etal1998} (RG\,715, $H$, $K$), and
\citetalias{Tuthill_etal2002} ({\Ks}, {\Lc}, {\Lpah}).

An inspection of the intensity profiles shows that the model is in good
agreement with the observed intensity distributions of the {\rr}. There are
almost no changes in the intensity profiles over all the near-IR wavelengths
except for those caused by slightly different resolutions. Such an invariance
of the intensity profiles indicates that very large grains dominate the
scattered and emitted radiation in the very dense circumbinary torus of the
{\rr}.

There are considerable discrepancies between the mo\-del and observed images at
optical wavelengths, where the bright lobes start to appear more diffuse and
extended in the direction perpendicular to the symmetry (outflow) axis
\citepalias[see also images in][]{Osterbart_etal1997,Men'shchikov_etal1998}. In
part, these deviations may be due to the approximations we used to describe the
model geometry. The model density distribution is assumed to be independent of
the polar angle and this assumption might become more important at short
wavelengths. We believe, however, that this morphological change can be
naturally explained by additional contribution of the extended red emission,
known to be present in the {\rr} (but not in our model) at the optical
wavelengths (0.55--0.75\,{\mic}). This emission veils the {\sf X}-shaped
spikes, making them appear relatively fainter in the visible than at the
near-IR wavelengths, where the extended red emission does not affect the
images.


\subsection{Comparison with previous models}
\label{Comparison}

Our results are consistent with the optical images of the {\rr} simulated by
\citet{Yusef-Zadeh_etal1984}, although the density distribution predicted by
our model is more complex than a simple $\rho \propto r^{-2}$ profile assumed in
their calculations. The two models have derived a similar bipolar geometry of the
dusty envelope with density distributions (almost) independent of the polar angle
up to a limiting latitude, where the density drops to zero (or very small
values) inside the biconical cavities. The authors assumed a full opening angle
of 70{\degr} for the cavities, derived from low-resolution optical images
\citep[e.g.,][]{Cohen_etal1975}. In our model, a smaller opening angle $\omega
= 50${\degr} was adopted, which is consistent with our new high-resolution images.

The model constructed by \citet{Lopez_etal1997} differs from our model, in part
due to the factor of 4 greater resolutions of our new speckle images containing
much more detailed information about the physical properties of the nebula. The
differences can also be attributed to more extensive exploration of the
parameter space in our modeling. The overall geometry of their envelope with
bipolar cavities is very similar to the one adopted in
\citet{Yusef-Zadeh_etal1984} and in the present modeling, although the density
is decreasing almost linearly with the latitude. The radial density
distribution $\rho \propto r^{-2}$ adopted by \citet{Lopez_etal1997} (with an
order-of-magnitude jump upward beyond $\sim 100$ AU) is relatively flat,
containing most of the mass at the outer boundary. This would have a pronounced
effect (by a factor of $\sim$ 2--3) on the beam-matched fluxes at long
wavelengths, which their fit of the observed SED did not take into account.
Very low densities of the innermost regions in the model are inconsistent with
the bipolar appearance of the {\rr} in the high-resolution near-IR speckle
images shown in Fig.~\ref{Images1}.

Our previous model \citepalias{Men'shchikov_etal1998} is very similar to the
new model presented in this paper. As explained in Sect.~\ref{Introduction}, the
old model was unable to reproduce the shape of the bright lobes of the {\rr} in
the new Keck telescope images at the wavelengths of 3.1--3.3\,{\mic}. The model
was also not very successful in fitting the observed SED over the UV to optical
wavelength range and in explaining the flat distribution of the {\HmK} color
image over the bright lobes. All this stimulated us to reconsider all existing
observations and construct a more realistic model described in this paper.


\section{Discussion}
\label{Discussion}

Below we interpret the results of our modeling of the {\rr} and discuss
the properties and evolution of its close binary and dense circumbinary
environment.


\subsection{Physical parameters of the binary}
\label{Binary}

The radio emission of a small H\,II region observed in the {\rr} (cf.
Sects.~\ref{EffTem}, \ref{SpEnDi}) indicates that there is a hot source of
ionizing photons in the nebula. However, IUE spectroscopy \citep{Sitko1983}
shows no signature of any hot $T \ga 2.5 \times 10^{4}$\,K companion in the
far-UV spectrum of the {\rr}, where one would expect to see the peak of its
radiation. This fact suggests that the hot invisible component has much lower
luminosity than {\Lstar} $\approx 6050$\,{\Lsun} of the observed post-AGB
star. Taking into account interstellar extinction in our model (see
Sect.~\ref{SpEnDi}), we can estimate that the well-observed SED sets an upper
limit for the hot companion's luminosity of $\sim$ 100\,{\Lsun}.

To illustrate this, in Fig.~\ref{SED} we have added a hot ($6 \times 10^{4}$
K) blackbody to the stellar continuum of the post-AGB object, normalized so as
to maximize its luminosity and yet to preserve our model fit of the observed
SED. It is this hot additional component that produces rising far-UV fluxes
toward the shortest wavelengths ($\lambda \la 0.14$\,{\mic}) in the stellar
continuum and in the model SED with no interstellar reddening. Although the
increasing interstellar extinction at these wavelengths (3 mag at
0.12\,{\mic}) forces the model fluxes to decline, a bump in the SED due to the
hot companion remains (Fig.~\ref{SED}). Our normalization implies that the
invisible hot component's luminosity is a factor of at least $\sim$ 60 less
luminous than the post-AGB object, giving the upper limit of $\sim$
100\,{\Lsun}. Together with the inferred effective temperature of $T \ga 2.5
\times 10^{4}$\,K, this limit marks an area in the Hertzsprung-Russell (HR)
diagram populated by cooling white dwarfs with masses $M \ga 0.23$\,{\Msun}
\citep{Driebe_etal1998}. This can be interpreted as the evidence that the close
binary indeed consists of a post-AGB object and a relatively hot white dwarf.

This idea is further supported by the spectroscopic orbital elements of the
binary's observed component. \cite{VanWinckel_etal1995} determined the orbital
period $P$ = 298 days and the eccentricity $e$ = 0.45, whereas
\cite{Waelkens_etal1996} gave $P$ = 318 days, and \cite{Waters_etal1998} cited
$e$ = 0.38. We have recomputed the full set of the latest orbital elements from
original measurements \citep{VanWinckel_etal1995,Waelkens_etal1996} using the
computer code of \cite{Tokovinin1992}.

The new orbital elements of {\hd} are presented in Table~\ref{Orbit}. With the
viewing angle ${\theta_{\rm v}} \approx$ 11{\degr} (Fig.~\ref{Geometries}) and
the current mass of the post-AGB star of {\Mstar} $\approx 0.57$\,{\Msun}
(see for details Sect.~\ref{Evolution}), the mass function $f(M)$ =
0.049\,{\Msun} implies that the invisible companion is a low-mass object with
{\Mwd} $\approx 0.35$\,{\Msun}. If one adopts a lower-limit effective
temperature of {\Twd} $\approx 2.5 \times 10^{4}$\,K (Sect.~\ref{EffTem}), one
gets from the evolutionary tracks of white dwarfs \citep{Driebe_etal1998} a
very low luminosity of {\Lwd} $\approx 0.3$\,{\Lsun} for the star. In this
paper we assume, however, that the hot component's luminosity is just at the
upper limit of {\Lwd} $\approx 100$\,{\Lsun} and, therefore,
significantly hotter, with {\Twd} $\approx 6 \times 10^{4}$\,K.

\begin{table}
\caption{Orbital elements of the close binary {\hd}.}
\label{Orbit}
\smallskip
\begin{tabular}{lllll}
\hline
\noalign{\smallskip}
Parameter              & Symbol         & Value  & Error & Units        \\
\noalign{\smallskip}
\hline
\noalign{\smallskip}
Period                 & $P$            & 322    & 2.0   & days         \\
Periastron\,(JD\,244+) & $T_0$          & 8075   & 7.0   & days         \\
System velocity        & $V_0$          & 19.8   & 0.4   & km\,s$^{-1}$ \\
Semiamplitude          & $K_1$          & 12.2   & 0.5   & km\,s$^{-1}$ \\
Eccentricity           & $e$            & 0.37   & 0.03  &              \\
Semimajor axis         & $a_1\,\sin{i}$ & 0.34   & 0.01  & AU           \\
Mass function          & $f(M)$         & 0.049  & 0.006 & {\Msun}      \\
\noalign{\smallskip}
\hline
\end{tabular}
\end{table}


\subsection{The bipolar geometry}
\label{Bipolar}

Despite the narrowness of the dark lane, the circumbinary material we see in
the images is definitely {\em not} confined to a flat disk. As discussed by
\cite{Men'shchikovHenning2000} and \cite{Men'shchikov_etal2001}, the
appearance of optically thick structures with bipolar outflow cavities may
easily be misleading. The narrow dark lane is a result of the radiation
transfer effects, with optical depths along different lines of sight greatly
varying in such an axisymmetric density distribution. The narrow dark lane
appears quite naturally in bipolar envelopes like the one shown in
Fig.~\ref{Geometries} even when the ``torus'' density is independent of the
polar angle. This is demonstrated by the model images in Figs.~\ref{Images1},
\ref{Colors} which show very good agreement with observations.

A thin flat disk observed edge-on would appear as an ellipsoidal structure
highly elongated at PA $\approx$ 102{\degr} (by a factor of ${\sim\,}$5)
perpendicular to the symmetry axis of the nebula, with a narrow and long dark
lane \citep[see, e.g., $L$-band model images of a circumstellar disk
in][]{SuttnerYorke2001}. The expected appearance of a thin disk sharply
contrasts the observed bipolar brightness distribution extended {\em along} the
symmetry axis. This rules out the flat disk geometry as a model for the
interpretation of the observed properties of the {\rr} and other similar
bipolar nebulae. Plausible distributions of dusty material may include either
a strongly flaring, geometrically very thick disk or a toroidal structure
(envelope) with biconical outflow cavities.

We believe that it would be more accurate to describe the circumstellar
environment of the {\rr} as an envelope with conical cavities, since the term
``disk'' is usually associated with a thin flat geometry. Many such dense
envelopes are known among both young and evolved low-mass stars \citep[{\irs},
{\hlt}, {\irc} are just a few of them; e.g.,][] {Men'shchikovHenning1997,
Men'shchikov_etal1999, White_etal2000, Men'shchikov_etal2001}. In young stellar
objects this geometry is naturally produced by the (radial) collapse of an
initially spherical protostar, which eventually creates powerful jets and
excavates bipolar (conical) outflows. In old stars this geometry is naturally
produced by the high (radial) mass loss they experience on the top of the AGB
and by the fast bipolar winds expected at the later stages of their evolution.
Close binarity of the stars plays an important role in the creation of such
extremely dense, opaque bipolar envelopes, and the {\rr} is an outstanding
example of such a structure.


\subsection{The gas component}
\label{GasProp}

Although our model suggests a self-consistent picture of the {\rr} explaining
a variety of observational data, the modeling was done using only the constraints
from the dust continuum radiation (Sect.~\ref{GeomAss}). Since the modeling
directly predicts only the properties of the binary and its bipolar dusty
envelope, discussion of a wealth of information on the molecular and atomic
species obtained during the past decades of observations is beyond the scope of
this paper. Below we briefly discuss, however, a few aspects related to the
circumbinary gas component.

Where is the H\,II region located in the {\rr}? Our model requires such high
densities in central regions of the nebula ($n_{\rm H} \approx 2.5 \times
10^{12}$\,cm$^{-3}$, Sect.~\ref{DensTemp}) that the relatively
low-luminosity white dwarf would not be able to ionize hydrogen within a radius
of $\sim$ 17--170\,AU, as seems to be required by the observed radio
fluxes \citep{Jura_etal1997}. This is true, however, only for a spherical
distribution of circumstellar matter. The {\rr} has biconical outflow cavities
with much lower densities of $\sim 5 \times 10^{5}$\,cm$^{-3}$
(Fig.~\ref{Geometries}, Sect.~\ref{DensTemp}). An estimate of the Str{\"o}mgren
sphere radius $s_{0}$ \citep{Lang1974} for the parameters of the hot white
dwarf (Sect.~\ref{Binary}) and electron temperature of $10^{4}$\,K, gives
$s_{0} \approx 160$\,AU. This suggests that the H\,II region fills in the
outflow cavity of the dense circumbinary torus of the {\rr}.

Optical spectroscopy of the {\rr} by \citet{Jura_etal1997} revealed a
two-component H$\alpha$ emission consisting of a narrow spike of a full width
of 20 km\,s$^{-1}$ at half-maximum, on top of a broad plateau with a width of
200 km\,s$^{-1}$ at zero intensity. Their analysis attributed the slow gas to
the extended H\,II region, whereas the fast-moving ionized gas is likely to be
confined to a much smaller zone that fits into the the binary's orbit. The
ultraviolet spectra of CO and C\,I presented by \citet{Glinski_etal1997}
displayed narrow emission lines and wide absorption bands with full widths of
25--30 km\,s$^{-1}$ and 230 km\,s$^{-1}$, respectively. The narrow emission and
wide absorption lines of C\,I seem to be related to the center of mass of the
binary, although the absorption could be assigned to a binary component, too.
They noted a similarity of the lines with the two-component H$\alpha$ profile
observed by \citet{Jura_etal1997}, suggesting that the respective narrow and
wide lines arise in the same distinct environments.

Clearly, new observations are needed to verify these findings of the
high-velocity gas in the {\rr} and to place more precise constraints on the
spatial distribution of the observed species, before any realistic model could
be able to interpret all existing data. Without any possibility to explicitly
and self-consistently incorporate the gas components in our dust continuum
radiative transfer modeling, there is no way to extract more information on
the gases in the {\rr} from the model presented in this paper.


\subsection{Stability of the thick torus}
\label{Stability}

What keeps the gas and dust components of our model at high latitudes above
the midplane, if the configuration is considered as stationary and long-lived, as
proposed by \citet{Jura_etal1995}? The torus-like (Fig.~\ref{Geometries}) dense
circumbinary envelope (Fig.~\ref{Images1}) of the {\rr} does not resemble a
flat disk (Sect.~\ref{Bipolar}) and it may not necessarily be a stationary,
centrifugally-supported configuration. Our model implies that it is a massive,
selfgravitating, slowly evolving envelope ejected by the binary components, in
which cavities were carved out presumably by a faster bipolar outflow.

The idea of a stable disk in the {\rr} was invoked by \citet{Waters_etal1992}
to explain an unusual chemical composition of the spectroscopically observed
post-AGB star \citep{VanWinckel_etal1995} that suggested separation of the
chemical elements due to the formation of dust and reaccretion of the depleted
gas back onto the photosphere \citep{MathisLamers1992}. There is no good
reason, however, why the circumbinary material should be necessarily
distributed in a thin disk for this mechanism, which is based just on the
radiation pressure on dust, to work. A long-lived disk was also proposed by
\citet{Jura_etal1995} to explain unusual properties of dust in the {\rr} and
possibly growth of dust particles to very large sizes of at least 200\,{\mic}.
In a (stationary) disk, the dust grains would have much more time to grow than
in the transient conditions of a normal wind of an AGB star, whose density
$\rho \propto r^{-2}$ quickly drops as the envelope expands.

The growth rates due to coagulation of grains and accretion of gas atoms
on the grain surface is proportional to the rate of collisions and, therefore,
to the gas and dust densities. As an example, the time scale (in years) for a
grain to grow by accretion to a radius $a$ can be written as
\begin{equation}
t_{\rm acc} \approx 150
\left(\frac{a}{0.2\,{\rm cm}}\right)\!\!
\left(\!\frac{\rho_{\rm gr}}{2\,{\rm g\,cm}^{-3}}\!\!\right)\!\!
\left(\!\frac{10^{12}\,{\rm cm^{-3}}}{n_{\rm H}}\!\right)\!\!
\left(\!\frac{10^{3}\,{\rm K}}{T}\!\right)^{\!\!1/2}\!\!\!,
\label{growth}
\end{equation}
where $\rho_{\rm gr}$ is the grain material density, $n_{\rm H}$ is the number
density of H atoms, $T$ is the kinetic gas temperature, and a sticking
probability of 0.1 and an abundance of $10^{-3}$ of the accreted atoms were
adopted. The high densities in the torus of the {\rr} (Sect.~\ref{DensTemp})
imply $t_{\rm acc} \sim 60$ years for large dust particles to grow by
accretion. This illustrates the possibility of a very rapid coagulation and
growth of grains, on a crossing time scale of the dense toroidal envelope
($\sim$ 100 years, Fig.~\ref{DenTem}). Thus, there seems to be no need in a
long-lived, rotating disk in the {\rr} to grow millimeter-sized dust particles.

Although the dense environment is probably not supported by the centrifugal
forces, the toroidal envelope may well be quasi-static, balanced against the
gravity by the gas pressure and by the radiation pressure on dust grains.
The free-fall time scale for the gas component can be estimated from
\begin{equation}
t_{\rm ff} \approx 118
\left(\!\frac{R_{\rm c}}{100\,{\rm AU}}\!\right)^{\!3/2}\!
\left(\!\frac{2.12\,{M_{\odot}}}{M_{\rm c}}\!\right)^{\!1/2}
{\rm yr},
\label{FreeFall}
\end{equation}
where $R_{\rm c}$ is the outer radius of the dense torus and $M_{\rm c} \approx
M_{\star} + M_{\rm WD} + M$ is the mass inside it (as before, $M$ is the mass
of the torus). Substituting the model values, we find a free-fall time $t_{\rm
ff} \approx 120$ years. The dynamical expansion time scale of the torus due to
the gas pressure can be expressed as
\begin{equation}
t_{\rm exp} \approx 170
\left(\!\frac{R_{\rm c}}{100\,{\rm AU}}\!\right)\!
\left(\!\frac{10^3\,{\rm K}}{T}\!\right)^{\!1/2}
{\rm yr}.
\label{Explosion}
\end{equation}
As before, assuming that the gas temperature is equal to the dust temperature
of $\sim 10^{3}$ K, we find $t_{\rm exp} \approx 170$ years. The
order-of-magnitude comparison suggests that the gas component may be in a
quasi-hydrostatic equilibrium.

Estimating the maximum radius of dust particles $a_{\rm max}$ that can be
removed by the radiation pressure of the central star, we get an upper limit of
\begin{equation}
a_{\rm max} \approx 0.2
\left(\!\frac{L_{\star}}{6050\,L_{\odot}}\!\right)\!\!
\left(\!\frac{M_{\odot}}{M_r}\!\right)\!\!
\left(\!\frac{2\,{\rm g\,cm}^{-3}}{\rho_{\rm gr}}\!\right)
{\rm cm},
\label{a_max}
\end{equation}
where $M_r$ is the total mass within the radius $r$. For the high luminosity
of the {\rr} derived in this modeling, the condition for dust grains to be
gravitationally bound requires particles with $a \ga 0.2$ cm at the inner dust
boundary ($r = 14$ AU), where $M_r \approx M_{\star} + M_{\rm WD}$. Since
$M_{\rm c}/(M_{\star} + M_{\rm WD}) \approx 2.3$, the same condition of
gravitationally bound grains at $r = R_{\rm c}$ requires particles with $a \ga
0.09$ cm. In our model, the radiation pressure cannot remove dust particles
with $a = 0.2$ cm from the dense torus. The gravitation and radiation pressure
on dust grains seem to balance each other in the envelope, causing its very
slow quasi-static evolution.

The dust grains outside the dense torus ($r > 100$ AU) have a normal size
distribution in the model, therefore they are more efficiently accelerated by
the radiation pressure and can drive a bipolar outflow. This can explain the
observation by \citet{Jura_etal1995} that the CO emission in the {\rr} is
associated with the center of the mass of the binary and that it resembles a
bipolar outflow with a speed of $\sim$ 7 km\,s$^{-1}$.


\subsection{Evolution of the binary}
\label{Evolution}

The presence of a large amount of matter in the close vicinity of the binary
nucleus of the {\rr} suggests that the origin of the nebula may be associated
with an unstable mass loss which accompanied formation of the post-AGB star, as
noticed by \citet{Osterbart_etal1997}. This ``unstable mass loss'' may be
related to the ejection of a common envelope in a close binary
\citep{BondLivio1990, Yungelson_etal1993}. Thus we may consider an evolutionary
scenario in which one component of a binary first forms a low-mass He or CO
white dwarf, while another becomes an AGB star and then a CO dwarf, ejecting
the currently observed envelope. In our analysis we followed the general outlines
of the work of \citet{Nelemans_etal2000} on reconstruction of the evolution of
close binary white dwarfs. For the evaluation of the lifetimes, radii,
luminosities, and core masses of the AGB stars we used the SSE package
\citep{Hurley_etal2000}. These quantities slightly depend on the specific
prescriptions for mass loss adopted by \citet{Hurley_etal2000}.

In our discussion, the following notation is adopted. The initial masses of the
components (on the main sequence) are $M_{10}$ and $M_{20}$ ($M_{10} > M_{20}$)
and their initial separation is $a_0$. The mass of the primary after Roche
lobe overflow (RLOF) is $M_{\rm WD1}$ and the separation of the components is
$a_1$. The mass of the secondary prior to RLOF is $M^{\prime}_2$, its radius is
$R_2$, and the components' separation prior to the overflow is $a_2$. After the
RLOF, the mass of the secondary component is $M_{\rm WD2}$ (we consider the
present post-AGB star as a nascent white dwarf) and the final separation of the
components is $a_{\rm f}$. We attempted to reconstruct the formation of the
binary of the {\rr} starting from the present configuration and ``evolving''
the system back in time toward the main sequence, to derive the initial
parameters of the components.

Changes of the components' separation in the common envelope (CE) formed after
the RLOF by the AGB star may be described by the equation of the balance of the
binding energy of the giant's envelope and the difference of the orbital
energies of the pre- and post-CE binary
\citep{Webbink1984}:
\begin{equation}
\label{eq:ce}
\frac{M^{\prime}_2 (M^{\prime}_2 \! - \! M_{\rm WD2})}{\lambda\,R_2} = \alpha \!
\left(
\frac{M_{\rm WD1} M_{\rm WD2}}{2\,a_{\rm f}} \! - \! \frac{M_{\rm WD1} M^{\prime}_2}{2\,a_2}
\right)\!.
\end{equation}
Here $\lambda$ is a dimensionless factor which allows one to express the
binding energy of the envelope as a function of $M^{\prime}_2$, $M_{\rm WD1}$,
and $R_2$ \citep{deKool1990}. It depends on the structure of the stellar
envelope, being a function of the evolutionary stage of the donor
\citep[see][]{DewiTauris2000,TaurisDewi2001}. The parameter $\alpha$ represents
the efficiency of the orbital energy deposition into the common envelope,
depending also on the structure of the donor and the nature of the accretor. At
the present state of the art in stellar evolution theory, one cannot
disentangle these two parameters and it is reasonable to treat the product
$\alpha\lambda$ as a single parameter.

The effective temperature {\Tstar} $\approx 7800$\,K of the post-AGB object
suggests that it is now on the horizontal part of the white dwarf cooling
track, therefore we can assume that the present {\Lstar} $\approx
6000$\,{\Lsun} is close to its luminosity before the CE ejection. Using
``tracks'' constructed by the SSE package, we estimate the mass $M^{\prime}_2$,
core mass $M_{\rm c}$, and radius $R_{2}$ of the star at the moment when its
luminosity is equal to the present value. We make then the usual assumption
that $M_{\rm WD2} = M_{\rm c}$. The analysis shows that the initial mass
$M_{20}$ of the main-sequence progenitor of the post-AGB object had to be
greater than 1.5\,{\Msun}, since stars with lower masses do not reach {\Lstar}
$= 6000$\,{\Lsun}.

One can also obtain an upper limit for the initial mass of the post-AGB object.
Since the r.h.s. of Eq. (\ref{eq:ce}) must be positive, we obtain the
condition $a_2 > (M^{\prime}_2/M_{\rm WD2})\,a_f$. Assuming that the present
mean separation of the components in the {\rr} represents the post-CE value of
$a_f$, we derive $M_{20} \la 1.9$\,{\Msun}. The narrow range $1.5 \la
M_{20}/M_{\sun} \la 1.9$ of possible initial masses for the present post-AGB
object implies that now its mass is $M_{\rm WD2} \approx 0.57$\,{\Msun}. Then
the mass function and the inclination of the orbit provide the mass $M_{\rm
WD1} \approx 0.35$\,{\Msun} of the first-formed white dwarf.

The secondary component lost an amount of mass $M_{20} - M^{\prime}_2$ as
stellar wind, predominantly at the AGB stage. This widened the orbit, as the
product of the separation and total mass was conserved. At the onset of the
RLOF, the stellar radius $R_2$ was equal to the radius of the Roche lobe, for
which one can write \citep[e.g.,][]{Paczynski1967}
\begin{equation}
\label{eq:rl}
R_{\rm L} \approx 0.46\,a \left( \frac {m_{\rm d}}
{m_{\rm d}+m_{\rm a}} \right)^{1/3},
\end{equation}
where $a$ is the separation of components, $m_{\rm d}$ and $m_{\rm a}$ are the
masses of the donor and accretor, respectively. Substituting $R_2$,
$M^{\prime}_2$, and $M_{\rm WD1}$ in Eq.~(\ref{eq:rl}), one can obtain an
estimate of the separation of the components $a_2 \approx$ 670--590\,{\Rsun}
(3.1--2.7\,AU) prior to the RLOF, for the initial masses of $M_{20} \approx$
1.5--1.9\,{\Msun}.

\begin{figure}
\resizebox{\hsize}{!}
   {
    \includegraphics{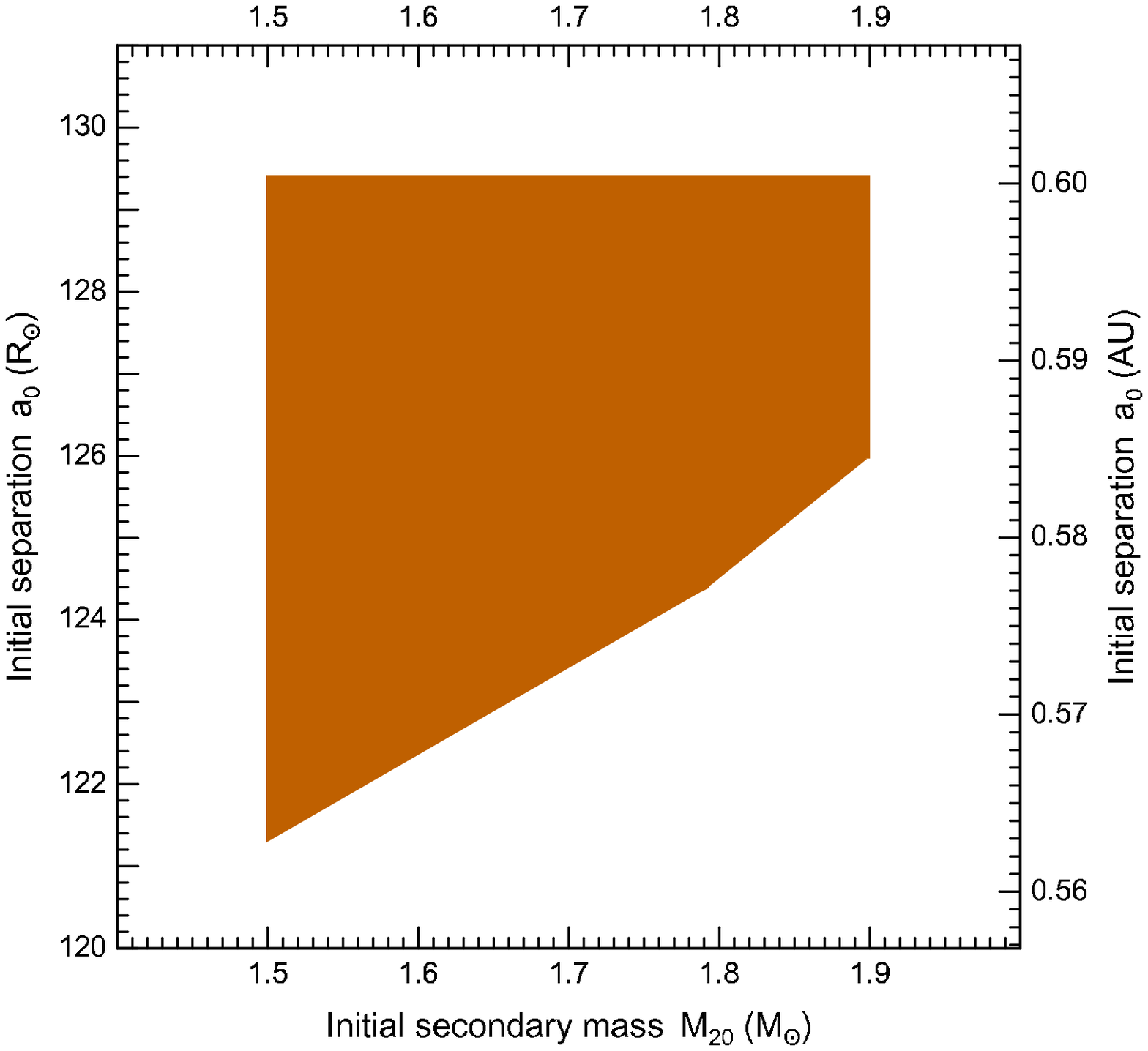}
   }
\caption
{
The range of possible initial separations $a_{0}$ of the components in the
progenitor of the {\rr} binary depending on the initial mass $M_{20}$ of the
secondary component (the progenitor of the present post-AGB star). For the
given $M_{20}$, the initial separation depends on the parameter $\gamma$ in
Eq.~(\ref{eq:achange}). The upper edge of the shaded region corresponds to
the initial mass $M_{10} = 2.3$\,{\Msun} of the primary component, whereas
the lower edge corresponds to equal masses of the components.
}
\label{Separations}
\end{figure}

The mass $M_{\rm WD1} \approx 0.35$\,{\Msun} of the first white dwarf suggests
that it is a helium white dwarf descending from a red giant with $M_{10} \la
2.3$\,{\Msun}, which had a degenerate helium core. The radii of such stars
depend only on the core mass \citep{RefsdalWeigert1970}. As for the secondary,
we assume that the dwarf's mass $M_{\rm WD1}$ is equal to the mass of the
helium core of its progenitor. Then one can estimate the radius of the primary
at the onset of the RLOF using, e.g., the core mass - stellar radius relation
for giants of solar chemical composition \citep{IbenTutukov1985}: $R/R_{\odot}
\approx 10^{3.5} (M_{\rm WD1}/M_{\odot})^4$. The radius of the giant with a
0.35\,{\Msun} helium core was $R_1 \approx 47.5$\,{\Rsun} and the star should
have had a deep convective envelope \citep[more than 50\,{\%} of the mass in
the outer convective zone, see the models of][]{Mengel_etal1979}. Then the RLOF
results in an unstable mass loss on a time scale intermediate between the
thermal and dynamical ones. \citet{Nelemans_etal2000} have shown that, for
comparable masses of the donor and accretor, the unstable loss of the stellar
envelope most probably happens without spiral-in.

The effect of the mass loss can be described in terms of the
balance of the angular momentum. Following \citet{PaczynskiZiolkowski1967}, one
may assume that the losses of mass and orbital angular momentum are related in
a linear way, $J_0{-}J_1 = \gamma J_0 {\Delta M}/{M_{\rm t0}}$. Here $J_0$ and
$J_1$ are the orbital angular momenta of the binary before and after the mass
loss, $\Delta M$ is the mass lost from the system, and $M_{\rm t0} = M_{10} +
M_{20}$ is the total initial mass of the binary. The separation of components
changes as
\begin{equation}
\label{eq:achange}
\frac{a_1}{a_0} = \left(\frac{M_{10} M_{20}}{M_{\rm WD1} M_2}\right)^{2} \!\!
\left(\frac{M_{\rm WD1} + M_2}{M_{\rm t0}}\right) \!\!
\left(1 - \gamma\frac{\Delta M}{M_{\rm t0}}\right)^{2}\!,
\end{equation}
where $M_2$ is the mass of the secondary after the first mass-loss episode. For
the stars in the mass range under consideration, a difference of, e.g., only
$\sim 0.1$\,{\Msun} in initial mass is sufficient for the primary to complete
its evolution in the red giants branch, while the secondary is still a
main-sequence star. Therefore we can expect that the amount of accreted matter
in the first mass loss episode in the system was negligible, i.e. that $\Delta
M \approx M_{10} - M_{\rm WD1}$, $M_2 \approx M_{20}$. Equation~(\ref{eq:achange})
allows us to find the separation of components $a_0$ of the initial binary if
$\gamma$ is known. On the other hand, $a_0$ can be found for any $M_{20}$ using
the core mass - radius relation and Eq.~(\ref{eq:rl}), since $M_{20} \le M_{10}
\le 2.3$\,{\Msun}. This enables us to estimate the range of the parameter
$\gamma$. The solutions were obtained for $\gamma =$ 1.0--1.3, which are
marginally compatible with the estimates derived by \citet{Nelemans_etal2000}
for low-mass close double white dwarfs. The range of $a_0$ for different
$M_{20}$ is shown in Fig.~\ref{Separations}.

Several cautionary notes should be made on the analysis. In our scenario, the
combined parameter $\alpha\lambda$ in Eq.~(\ref{eq:ce}) has to be $\sim 10$.
This may suggest that the release of the internal thermodynamic energy may be
involved in unbinding the envelope \citep{Han_etal1994,Han_etal1995}. Although
the values of the structural parameter $\lambda$ have never been derived for
stars with $M < 3$\,{\Msun}, large values of $\lambda$ may be characteristic
for the tip of the AGB \citep[see][]{DewiTauris2000}. In addition, on the AGB
the effect of the orbital shrinkage may be somewhat offset by the wind outflow,
as has been noticed by \citet{Iben2000}. Strictly speaking, Eq.~(\ref{eq:ce})
becomes then invalid, but effectively this would imply an increase of
$\alpha\lambda$.

Our estimate of $M_{\rm WD1}$ is compatible as well with an assumption that
this dwarf is a descendant of a star with an initial mass of about
2.5--3\,{\Msun}, which overfilled the Roche lobe prior to the helium ignition
in the core and produced a helim star that later became a CO dwarf
\citep{IbenTutukov1985,Han_etal2000}. However, for the formation of a
0.35\,{\Msun} dwarf the RLOF had to occur when $R_2 \la 20$\,{\Rsun} and
the resulting separation of components $a_2$ would not be sufficient to
harbor an AGB star with a 0.57\,{\Msun} core.

\begin{figure}
\resizebox{\hsize}{!}
   {
    \includegraphics{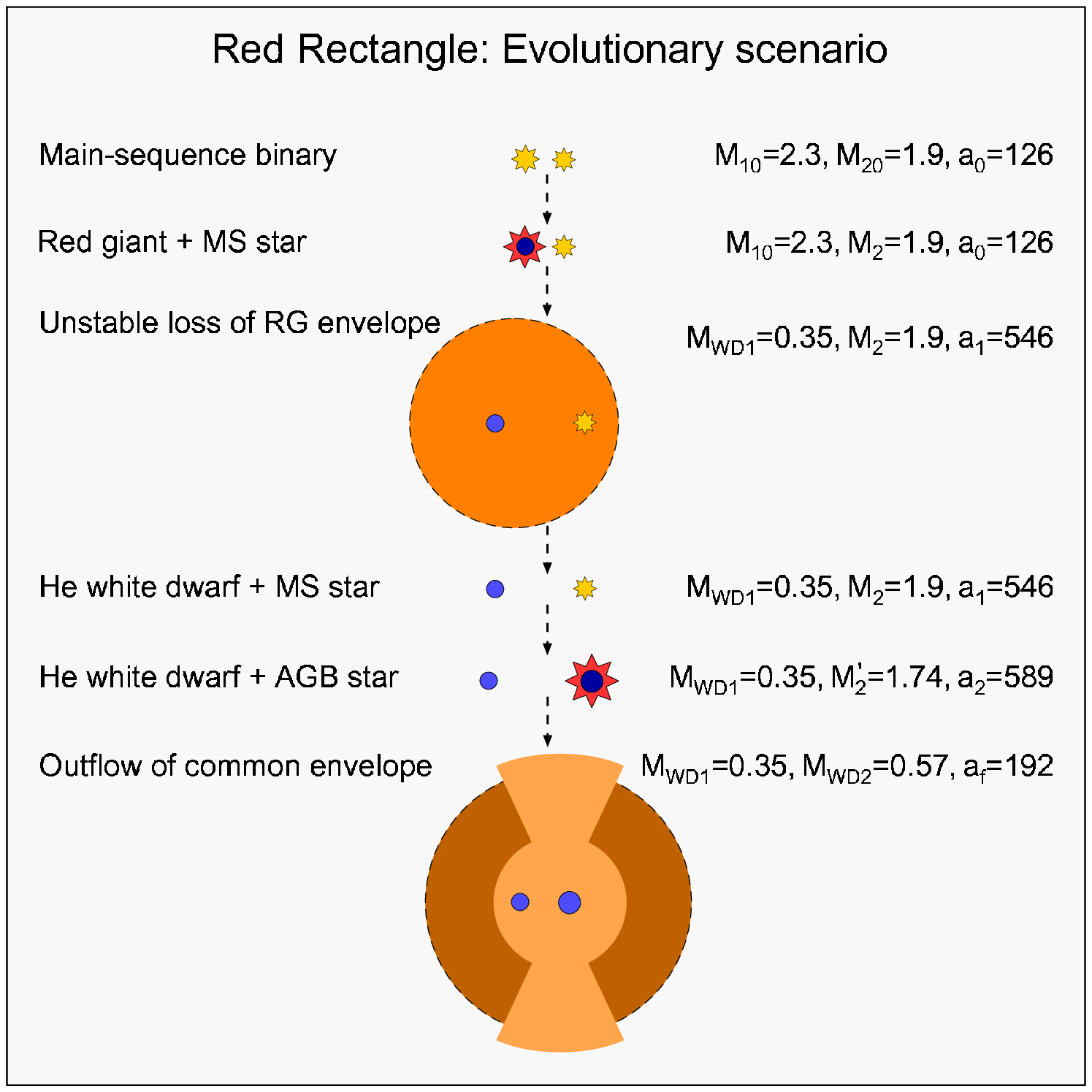}
   }
\caption
{
The sequence of evolutionary stages of the binary resulting in the formation of
the {\rr} nebula. The binary parameters are given in solar units.
}
\label{Scenario}
\end{figure}

The RLOF occurs close to the tip of AGB and some mass at this evolutionary
stage has already been lost by stellar wind. For the initial masses of
1.5--1.9\,{\Msun} of the post-AGB object in the {\rr}, this amounts to the loss
of 0.35--0.16\,{\Msun}. The mass of the common envelope is therefore
constrained to $M =$ 0.58--1.17\,{\Msun} in our evolutionary scenario. This
value derived from the above analysis allowed us to constrain the
dust-to-gas mass ratio {\dustgas} in a way similar to the approach used in our
model of {\irc} \citep{Men'shchikov_etal2001}. In fact, from our continuum
radiative transfer modeling we know the total mass of dust $M_{\rm d} \approx
0.015$\,{\Msun} in the {\rr}, which is contained in the dense torus. The
estimated range of possible masses of the circumbinary torus implies that
{\dustgas} = 0.021--0.01, rather high values compared to the usual {\dustgas}
$\sim$ 0.003 in the winds of single mass-losing AGB stars. The higher value
seems to be too high considering the abundances of elements that can make up
dust grains. We therefore adopted in this study the lower end of the interval,
{\dustgas} = 0.01, which corresponds to the total torus mass $M \approx
1.2$\,{\Msun} and the initial mass $M_{20} \approx 1.9$\,{\Msun} of the star.

Figure~\ref{Scenario} illustrates and summarizes the evolutionary scenario
presented above. The initial masses of the components in the progenitor binary
system of the {\rr} are very similar to each other, but not equal. Whereas a
2.3\,{\Msun} star completes its evolution to the tip of the red-giant branch in
about $\sim$ 800\,Myr, a 1.9\,{\Msun} star evolves to the tip of AGB in $\sim$
1600\,Myr. Thus, there may be a difference of about 800\,Myr in the ages of the
binary components of the {\rr}. Within this time a helium white dwarf would
cool down to $\sim 0.01$\,{\Lsun} \citep{Driebe_etal1998}. If its present
luminosity has indeed a much higher value of {\Lwd} $\la 100$\,{\Lsun}, this
might mean that the He white dwarf has accreted some hydrogen from the AGB star
wind and experienced one or several hydrogen-burning flashes. Then the latter
brought it close to the high luminosity and temperature ``knee'' of the cooling
track again.


\subsection{Eccentricity of the orbit}
\label{Eccentricity}

It is well known that tidal interactions in close binaries tend to quickly
circularize their orbits. The large eccentricity $e$ = 0.37 of the close binary
in the {\rr} has been interpreted in two different ways.
\cite{VanWinckel_etal1995} suggested that increased mass loss, when the star
passes periastron, may be the process responsible for the increased
eccentricity, whereas in a later paper \citep{Waelkens_etal1996}, the authors
suggested that gravitational interaction of the binary with the circumbinary
disk may be the reason. Recently, \cite{Soker2000} argued that it is the mass
loss at periastron which increases the binary eccentricity and that the inner
disk mass is too small for the other mechanism to be efficient in this and
other similar objects.

In our analysis, the initial orbit of the binary is circular. It is usually
assumed that the unstable mass loss by Roche lobe filling stars does not
produce eccentric orbits. However, the dynamical time scale for the AGB star
was shorter than the orbital period of the binary and we may speculate that the
relatively fast loss of the envelope made the orbit elliptic. Later, the
eccentricity can increase as a result of the binary-disk interaction.

Our modeling has shown that the torus' mass in the {\rr} may be large
enough to account for the eccentricity changes. The mass in question resides
inside a radius of $r \approx 6\,a_{1}$ and it must be sufficiently large for
the circumbinary disk to efficiently drive the increasing eccentricity
\citep{Artymowicz_etal1991}. A lower bound on the model torus mass within this
radius can be found by extrapolating the density at the inner dust boundary
($4.2 \times 10^{-12}$ g\,cm$^{-3}$) into the dust-free zone. This gives
a mass of $\sim$ 0.001\,{\Msun} and a disk-to-binary mass ratio $q_{\rm d}$ of
the same order. Assuming an initial eccentricity of 0.1, we find from $\dot{e}
\approx 0.002\,q_{\rm d}\,2\pi /P$ \citep{Artymowicz_etal1991} a high rate of
changes on a time scale of $e/\dot{e} \sim 7 \times 10^3$\,yr. We note,
however, that this estimate is just an illustration and that it remains far
from being certain that this specific mechanism, assumptions, and formulae are
applicable for the binary's environment existing in the {\rr}.


\section{Conclusions}
\label{Conclusions}

Recent diffraction-limited near-IR speckle images of the {\rr} in the
wavelength range 2.1--3.3\,{\mic} with angular resolutions of 44--68 mas
\citepalias{Tuthill_etal2002} as well as the previous optical and near-IR speckle
images at 0.7--2.2\,{\mic} \citepalias{Osterbart_etal1997,Men'shchikov_etal1998}
revealed a geometrically thick circumbinary torus with bipolar outflow cones
and {\sf X}-shaped spikes originating deep inside the cavities. This
multiwavelength set of high-resolution images enabled us to reanalyze most
of the existing observations using our two-dimensional radiative transfer code.
Results of this study are summarized below.

{\sc Distance.}
An important by-product of the modeling of the {\rr} is a new
determination of its distance $D \approx 710$\,pc (with model
uncertainties of about 10\,{\%}), which is twice as large as the commonly
adopted value of 330\,pc (Sect.~\ref{Distance}). The new distance, based on the
account for interstellar extinction in our model (Sects.~\ref{Distance},
\ref{SpEnDi}), is consistent with high luminosities expected from the stellar
evolution theory. Based on this distance, we reconstructed physical
parameters of the binary and its circumbinary torus (Table~\ref{ModelParams}).

{\sc Close binary.}
In our model, the observed component of the spectroscopic binary is a luminous,
low-mass post-AGB star with {\Mstar} $\approx 0.57$\,{\Msun}, {\Tstar}
$\approx 7750$\,K, and {\Lstar} $\approx 6050$\,{\Lsun}. It is a product
of the mass loss by the secondary component in a close binary system. We
identified the now invisible descendant of the primary component with a
relatively hot white dwarf with {\Mwd} $\approx$ 0.35\,{\Msun}, {\Twd}
$\approx 6 \times 10^{4}$\,K, and {\Lwd} $\approx$ 100\,{\Lsun}.
The presence and parameters of the compact degenerate star in the close binary
{\hd} have been deduced (Sect.~\ref{Binary}) from (1) the spectroscopic mass
function $f(M)$ = 0.049\,{\Msun}, (2) presence of a compact H\,II region
ionized by a hot source of radiation, (3) upper limit of the contribution of
the hot object to the continuum in the far UV.

{\sc Circumbinary structure.}
The intensity distribution of the high-resolution images is definitely
inconsistent with the flat disk geometry frequently used to visualize bipolar
nebulae (Sect.~\ref{Bipolar}). A geometrically very thick density
distribution of a compact, dense torus with biconical outflow cavities
(Fig.~\ref{Geometries}) is best suitable for reproducing the observed images.
The opening angle of the cavities is 50{\degr} and the observer's viewing angle
is 11{\degr} below the midplane.

{\sc Density distribution.}
Although the nebula extends to at least $R_2 \approx 4 \times 10^{4}$ AU
from the central binary, most of its mass of $M \approx 1.2$\,{\Msun} is
contained in the extremely dense torus having an outer radius of 100\,AU
(Sect.~\ref{DensTemp}). The inner {\em dust} boundary of the torus is located
at a distance $R_1 \approx 14$\,AU from the center, where gas densities
reach values of $\rho \approx 4.2 \times 10^{-12}$ g\,cm$^{-3}$ ($n_{\rm H}
\approx 2.5 \times 10^{12}$\,cm$^{-3}$). The density of the outflow
cavities is many orders of magnitude lower than the torus density in the
region dominated by the dense torus (Fig.~\ref{DenTem}), whereas in the outer
regions ($r \ge$ 800 AU) the outflow cones are denser than the rest of the
toroidal envelope.

{\sc Dust properties.}
Our model of dust in the circumbinary torus of the {\rr} has two distinct
components whose chemical composition is not constrained by observations.
Amorphous carbon grains with radii $a$ in the range of 0.005--600\,{\mic} and
typical interstellar size distribution ${\rm d}n/{\rm d}a \propto a^{-3.5}$
exist mainly in the outer regions of the toroidal envelope ($r > 100$\,AU).
Most of the dust mass is contained in very large ($a$ = 0.2\,cm)
particles of the massive, dense torus, which produce an almost gray optical
depth of $\tau \approx 47$. Although there exist more dust components somewhere
in the nebula, including crystalline silicates and PAHs, their mass must be
small compared to the mass of the very large particles, which dominate the
observed appearance of the {\rr}.

{\sc Evolution of the binary.}
Based on our estimates of the masses of the {\rr} binary components, their
separation and luminosities, we suggest an evolutionary scenario for the
formation of the nebula, in which components had initial masses of about
2.3 and 1.9\,{\Msun} and a separation of $\sim 130\,${\Rsun}. The
scenario associates the formation of the {\rr} nebula with the ejection of the
common envelope upon the Roche lobe overflow by the present post-AGB star.


\begin{acknowledgement}
We are grateful to Viktor Malanushenko for the assistance in computing
the orbital parameters, to Anatoly Miroshnichenko for his help in the
derivation of the observational estimate of the interstellar reddening, and to
Jarrod Hurley for providing the SSE package. LRY was supported by the Russian
``Astronomy and Space Research'' program. This research has made use of
the SIMBAD database operated at CDS, Strasbourg, France, and of the data
products from the Two Micron All Sky Survey, which is a joint project of the
University of Massachusetts and the Infrared Processing and Analysis Center,
funded by the National Aeronautic and Space Administration and the National
Science Foundation. We thank the referee, Michael Barlow, for his very useful
comments.
\end{acknowledgement}


\bibliographystyle{aa}
\bibliography{aamnem99,h3541}
\end{document}